\documentclass[authoryear,3p,12pt,fleqn]{elsarticle}

\usepackage{amsmath,amssymb}
\usepackage{mathrsfs}
\usepackage{mathtools}
\usepackage{amssymb}
\usepackage{amsfonts}
\usepackage{array}
\usepackage{color}
\usepackage{comment}
\usepackage{enumitem}
\usepackage[export]{adjustbox}
\usepackage{float}
\usepackage{lineno}
\usepackage{algorithm,xpatch}
\usepackage{algpseudocode}

\newcommand{\OO}[1]{}

\newdimen{\algindent}
\algnewcommand{\LineComment}[2]{\Statex \hspace{#1\algindent} \hspace{-0.75em} \(\triangleright\) #2 \hfill}

\DeclareDocumentCommand{\setequal}{m m O{$:=$} m}{\rlap{#2} \hspace{#1em} #3 #4}

\journal{arXiv} 

\begin{document}
	
	\begin{frontmatter}
		
		
		
		\title{A predictive model for fluid-saturated, brittle granular materials during high-velocity impact events}
		
		
		\author[hemi]{Aaron S. Baumgarten\corref{cor1}}
		\cortext[cor1]{Corresponding Author: Aaron S. Baumgarten, abaumg11@jhu.edu}
		\author[hemi]{Justin Moreno}
		\author[meche]{Brett Kuwik}
		\author[meche]{Sohanjit Ghosh}
		\author[hemi,meche]{Ryan Hurley}
		\author[hemi,meche]{K.T. Ramesh}
		
		\affiliation[hemi]{organization={Hopkins Extreme Materials Institute, Johns Hopkins University},
			city={Baltimore},
			state={MD},
			postcode={21218}, 
			country={USA}}
		
		\affiliation[meche]{organization={Department of Mechanical Engineering, Johns Hopkins University},
			city={Baltimore},
			state={MD},
			postcode={21218}, 
			country={USA}}
		
		\vspace*{-4\baselineskip}
		
		\begin{abstract}
			Granular materials --- aggregates of many discrete, disconnected solid particles --- are ubiquitous in natural and industrial settings. Predictive models for their behavior have wide ranging applications, e.g. in defense, mining, construction, pharmaceuticals, and the exploration of planetary surfaces. In many of these applications, granular materials mix and interact with liquids and gases, changing their effective behavior in non-intuitive ways. Although such materials have been studied for more than a century, a unified description of their behaviors remains elusive.
			
			In this work, we develop a model for granular materials and mixtures that is usable under particularly challenging conditions: high-velocity impact events. This model combines descriptions for the many deformation mechanisms that are activated during impact --- particle fracture and breakage; pore collapse and dilation; shock loading; and pore fluid coupling  --- within a thermo-mechanical framework based on poromechanics and mixture theory. This approach allows for simultaneous modeling of the granular material and the pore fluid, and includes both their independent motions and their complex interactions.
			A general form of the model is presented alongside its specific application to two types of sands that have been studied in the literature. The model predictions are shown to closely match experimental observation of these materials through several GPa stresses, and simulations are shown to capture the different dynamic responses of dry and fully-saturated sand to projectile impacts at 1.3 km/s.
		\end{abstract}

		\begin{keyword}
			constitutive behavior \sep granular material \sep porous material \sep mixture model \sep impact testing
			
		\end{keyword}
		
	\end{frontmatter}
	
	
	\section{Introduction}
	\label{sec:introduction}
	The severity of landslides, subterranean explosions, earthquakes, and high-velocity impact events is significantly influenced by the strength and dynamic response of granular materials --- e.g., sands, soils, snow, and lunar dust. Despite their relatively simple composition, this class of materials exhibits surprisingly complex behaviors, especially during dynamic loading events. These behaviors include phenomena such as pore collapse \citep{mandl1977}, dilation or bulking \citep{rudnicki1975}, liquefaction \citep{lade1994,sawicki2006}, and material ejection \citep{melosh1989, housen2011}; each of which directly influences the motion and deformation of these materials as well as the loads they place on surrounding structures.
	
	A key component of their behavior, particularly during dynamic loading, is the interaction of the bulk granular material with the interstitial (or pore) fluids that fill the space between individual particles \citep{lade1994, jackson2000, coussy2004, boyer2011}. Although granular materials are frequently considered in isolation, most granular sediments are porous, with a significant fraction of their apparent volume occupied by pore liquids and gases. Under quasi-steady loading conditions of laboratory-scale specimens, the presence of these interstitial fluids and their effects on the material response can be easily accounted for. However, under dynamic loading conditions, the motion and deformation of these pore fluids can have large, non-intuitive effects on the behavior of the bulk material \citep[e.g., see][]{pailha2009, baumgarten2019a}.
	
	For more than a century, researchers have been developing models for the behavior of granular materials and fluid--sediment mixtures, producing many analytical and mathematical descriptions for simple loading conditions. For example, there are models for the pressure dependent yield strength of these materials during shear and tri-axial loading \citep[e.g.,][]{drucker1952,lade1975,jop2006}; for the tendency of granular materials to dilate toward a critical (steady) state while shearing \citep[e.g.,][]{roux1998,pailha2009}; for the drag force acting against pore fluids as they move through packed granular beds \citep[e.g.,][]{darcy1856,carman1937}; and for the evolution of material stresses within settling sediments \citep[e.g.,][]{biot1941, terzaghi1943, terzaghi1996}. Despite the predictive power of these models, they are generally limited to specific geometries and regimes of material motion: frequently providing limited information to engineers about the complex dynamics of these materials.
	
	In recent years, significant research has focused on the development of models for more general, dynamic loading conditions. Models of this type are constructed using the framework of continuum mechanics: where the bulk material is described using smoothly varying, aggregated fields (e.g., stress, density, and velocity) rather than modeling the motion and deformation of each individual grain and pore. Such models combine constitutive descriptions of the stress--strain response of the granular media with numerical simulation frameworks capable of solving the continuum equations of motion --- namely, conservation of mass, momentum, and energy. Examples of such models include low-pressure granular flow models \citep[e.g.,][]{dunatunga2015,dunatunga2022}; high-pressure granular breakage and compaction models \citep[e.g.,][]{rubin2011,cil2020a,herbold2020}, as well as poromechanics and mixture theory models \citep[e.g.,][]{bandara2015,gao2018,baumgarten2019b}. 
	%
	%
	In the current literature, however, there are no models for the dynamic behavior of brittle granular materials --- including their complex interaction with pore fluids --- through their transition from dense, compacted, high-pressures states all the way to gaseous, stress-free states.
	
\begin{figure}[!h]
		\centering
		\includegraphics[width=0.9\linewidth]{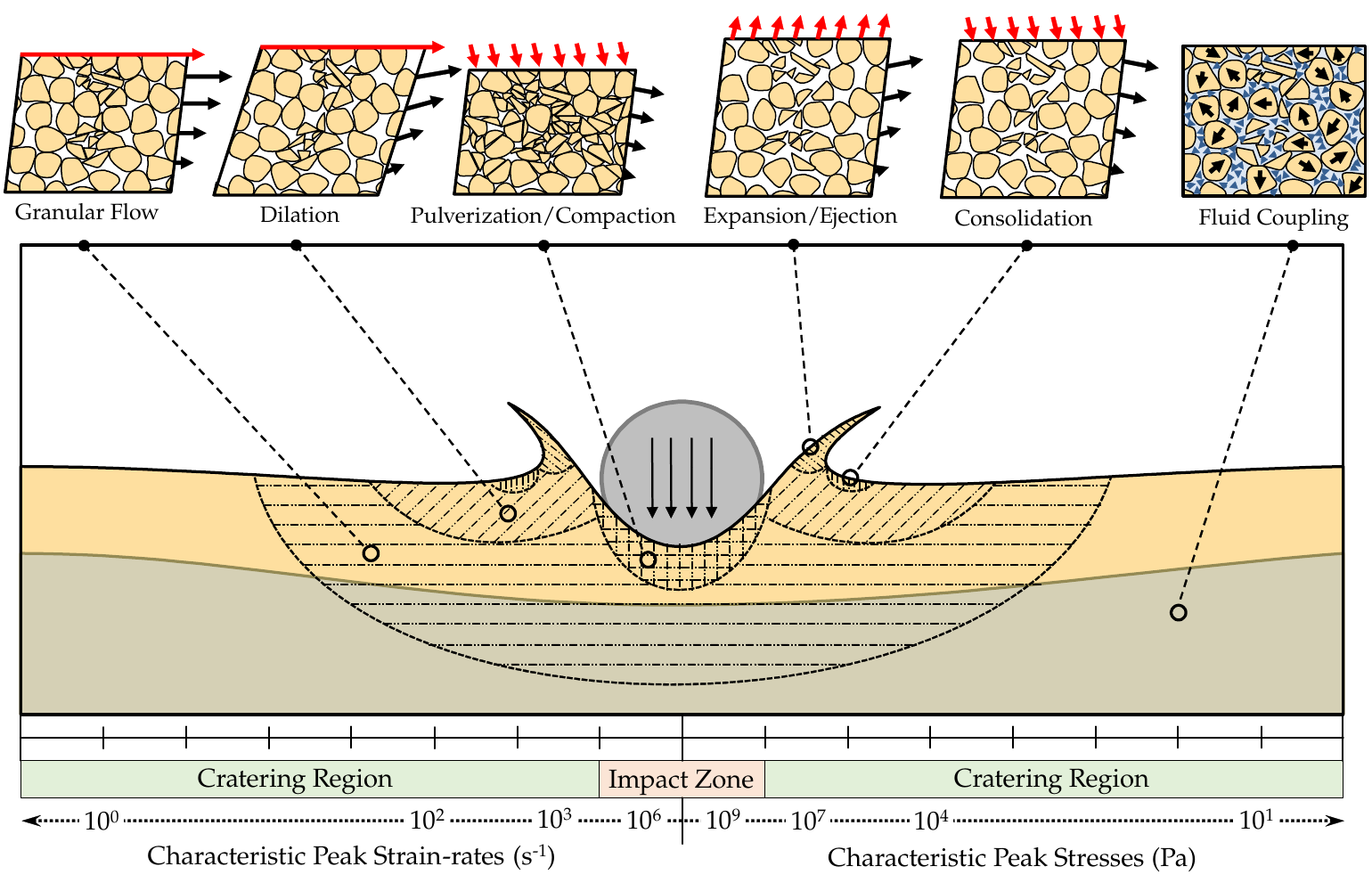}
		\caption[]{Illustration of dominant deformation mechanisms during high-velocity impact into fluid-saturated granular material. In the impact zone, stresses and strain-rates may exceed 10$^9$ Pa and 10$^6$ s$^{-1}$, respectively, leading to pulverization and compaction of the granular particles. (Although impact heating and thermal effects are modeled in this work, melting and solid phase changes are not.) In the cratering region, the magnitude of stresses and strain-rates decreases significantly, leading to granular flow; shear dilation; material ejection (expansion and re-consolidation); and importantly, coupling with interstitial (pore) fluids.
		}
		\label{fig:mechanisms}
	\end{figure}

	In this work, we are particularly interested in models that can be applied under conditions relevant to high-velocity impact events ($<$ 3 km/s; see \citealp{signetti2022}). This condition presents a unique challenge in the field of granular material modeling due to the number of grain-scale mechanisms that become important --- shown in Figure \ref{fig:mechanisms}. At the point of impact, pressures and strain-rates may exceed $10^9$ Pa and $10^6$ s$^{-1}$, respectively, leading to rapid compaction; particle fracture and fragmentation; and pore fluid compression. Moving outward from the impact site, within a radius of $\sim$10--20 times the size of the impactor, the process of crater formation leads to extreme material deformations at much lower stresses and strain-rates: $10^1$--$10^7$ Pa and $10^0$--$10^3$ s$^{-1}$, respectively. These deformations are accommodated by frictional granular shearing, dilation, pore fluid flow, and eventually material ejection. The transition between these different regimes is determined, in part, by the microscopic elastic deformations of the individual particles, which allow the transmission of stresses through the particle contact network \citep{radjai1999} and drive the formation of the stress concentrations that lead to particle fracture \citep{hurley2018}.
	
	The present work develops a predictive model for granular materials that can be used to study high-velocity impact events. This model combines elements from a range of disciplines in mechanics, including classical theories in soil mechanics \citep[e.g.,][]{carman1937}, poromechanics \citep[e.g.,][]{terzaghi1943}, and granular flow \citep[e.g.,][]{drucker1952}, with more contemporary theories for mixtures \citep[e.g.,][]{bedford1983} and shock physics \citep[e.g.,][]{drumheller1998}.
	Further, we discuss the grain-scale deformation mechanisms that are activated during these impact events, and we build mathematical descriptions of these mechanisms into this multi-component continuum model.

	\section{Governing Equations}
	\label{sec:governing_equations}
	The model developed in this work is constructed within the framework of continuum mechanics, which describes the motion and deformation of materials using continuous fields rather than following individual particles and pores. Within this broad framework, we are particularly interested in the governing equations derived from the theory of mixtures and poromechanics. In this section, we present the kinematic rules and balance laws associated with these theories, and apply these governing equations to fluid-saturated sediments.
	
	
	
\begin{figure}[!h]
		\centering
		\includegraphics[width=0.9\linewidth]{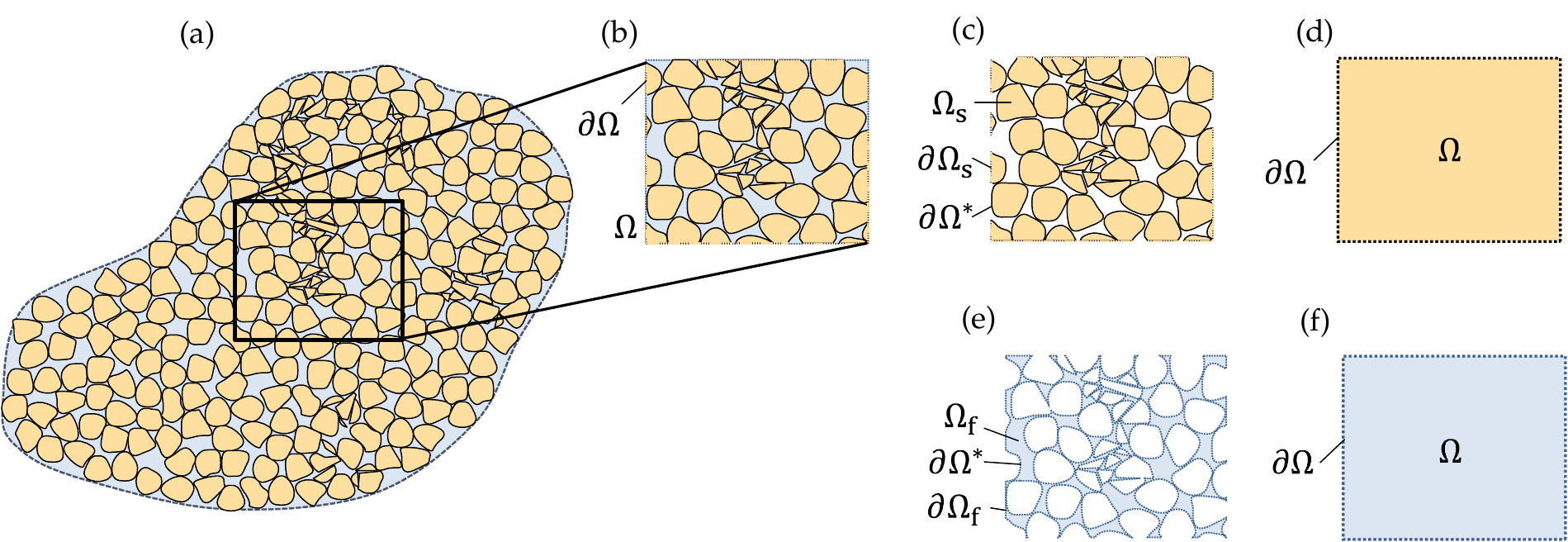}
		\caption[]{Illustration of fluid-saturated, granular material and associated representative volume elements (RVEs). The representative volume $\Omega$ shown in (b) is separable into the solid component $\Omega_s$ shown in (c) and the fluid component $\Omega_f$ shown in (e). These two components of the RVE, their external boundaries $\partial \Omega_s$ and $\partial \Omega_f$, and their interior surface $\partial \Omega^*$ are used to construct \emph{effective} material fields, which are continuous in the volume $\Omega$ and shown in (d) and (f).
		}
		\label{fig:rve}
	\end{figure}

	Consider a representative volume of granular media --- e.g. Figure \ref{fig:rve}a,b. This volume consists of two primary components: (i) a volume occupied by the solid material that composes the individual particles and (ii) a volume occupied by the interstitial fluid that fills the pore spaces between particles. In mixture theories, these two components are considered immiscible and are homogenized into separate, overlapping continua \citep[e.g., see][and Figure \ref{fig:rve}c--f]{baumgarten2019a}, each with their own density, velocity, stress, and internal energy fields.
	
	The total volume $V$ within a representative volume element or RVE is the sum of the volumes occupied by either solid material $V_s$ or interstitial fluid $V_f$: $V = V_s + V_f$. The fraction of the total volume occupied by each material represents the respective volume fraction: $\phi_s = V_s/V$, the solid volume fraction; and $\phi_f = V_f/V$, the fluid volume fraction or \emph{porosity} (so that $\phi_s + \phi_f = 1$).
	These volume fractions fundamentally connect the \emph{effective} material fields of the continua shown in Figures \ref{fig:rve}d,f with the \emph{true} material fields shown in Figures \ref{fig:rve}c,e of the constituents we are interested in modeling. For example, we relate the effective densities ($\bar{\rho}_s$ and $\bar{\rho}_f$) of the continua in Figures \ref{fig:rve}d,f with the true densities ($\rho_s$ and $\rho_f$) of the constituents in Figures \ref{fig:rve}c,e:
	\begin{equation}
		\label{eqn:porosity}
		\qquad \bar{\rho}_s = \phi_s \rho_s, \qquad \bar{\rho}_f = \phi_f \rho_f.
	\end{equation}

	Thus the solid particles shown in Figure \ref{fig:rve}c are homogenized into the continuous material called the \emph{solid} (or \emph{granular}) \emph{phase} shown in Figure \ref{fig:rve}d, with an effective density, $\bar{\rho}_s$; velocity, $\boldsymbol{v}_s$; and specific (per unit mass) internal energy, $\varepsilon_s$. Similarly, the interstitial fluid shown in Figure \ref{fig:rve}e is homogenized into the continuous material shown in Figure \ref{fig:rve}f --- called the \emph{fluid phase} --- which has an effective density, $\bar{\rho}_f$; velocity, $\boldsymbol{v}_f$; and specific internal energy, $\varepsilon_f$.
	
	
	
	
	Following the fluid--sediment mixture theories of \citet{bedford1983} and \citet{jackson2000}, we express the conservation of mass, momentum, and energy at an arbitrary \emph{spatial} point $\boldsymbol{x}$ within the two overlapping continua 
	with the following governing equations. These equations are expressed in the \textit{material} reference frame for each continua and make use of their respective material time derivatives:
	\begin{subequations}
		\label{eqn:material_time_derivative}
		\begin{align}
			d^s/dt &\equiv \partial / \partial t + \boldsymbol{v}_s \cdot \nabla,\\
			d^f/dt &\equiv \partial / \partial t + \boldsymbol{v}_f \cdot \nabla,
		\end{align}
	\end{subequations}
	with $\nabla$ the \textit{spatial} gradient operator (i.e., $\nabla \equiv \partial/\partial \boldsymbol{x}$). Conservation of mass is therefore expressed as follows and defines the time-rate of change of the effective mass densities:
	\begin{subequations}
		\label{eqn:mass_conservation}
		\begin{align}
			\frac{d^s \bar{\rho}_s}{dt} &= -\bar{\rho}_s\ \text{div}(\boldsymbol{v}_s),\\
			\frac{d^f \bar{\rho}_f}{dt} &= -\bar{\rho}_f\ \text{div}(\boldsymbol{v}_f),
		\end{align}
	\end{subequations}
	with $\text{div}()$ the \emph{spatial} divergence operator (i.e., $\text{div}(\boldsymbol{v}_s) \equiv \nabla \cdot \boldsymbol{v}_s$). 
	
	Conservation of momentum defines the time-rate of change of the effective velocities:
	\begin{subequations}
		\label{eqn:momentum_conservation}
		\begin{align}
			\bar{\rho}_s \frac{d^s \boldsymbol{v}_s}{dt} & = \text{div}(\boldsymbol{\sigma}_s) + \bar{\rho}_s \boldsymbol{g} - \boldsymbol{f}_d - \phi_s \nabla p_f,\\
			\bar{\rho}_f \frac{d^f \boldsymbol{v}_f}{dt} & = \text{div}(\boldsymbol{\tau}_f) + \bar{\rho}_f \boldsymbol{g} + \boldsymbol{f}_d - \phi_f \nabla p_f.
			%
			%
		\end{align}
	\end{subequations}
	In these equations, $\boldsymbol{\sigma}_s$ represents the \emph{effective granular stress} tensor, $\boldsymbol{\tau}_f$ denotes the \emph{effective fluid shear stress} tensor, and $p_f$ is the \emph{fluid pore pressure}. These effective stresses are generally analogous to the Cauchy stress in {classical} continuum mechanics; however, unlike the Cauchy stress, the pore fluid pressure contributes to the motion of \emph{both} materials, not only the fluid continuum. Interactions between the two constituent materials along their shared interfaces ($\partial \Omega^*$ in Figure \ref{fig:rve}c,e) give rise to internal interaction forces --- namely, the buoyant force \citep[e.g., see][]{drumheller2000} and the \emph{inter-phase drag force}, $\boldsymbol{f}_d$. In addition to these internal forces, the motion of each material is affected by  $\boldsymbol{g}$, the gravitational acceleration vector.
	
	Conservation of energy defines the time-rate of change of the internal energies:
	\begin{subequations}
		\label{eqn:energy_conservation}
		\begin{align}
			\bar{\rho}_s \frac{d^s \varepsilon_s}{dt} & = \boldsymbol{\sigma}_s : \nabla \boldsymbol{v}_s + \frac{\phi_s p_f}{\rho_s} \bigg(\frac{d^s \rho_s}{dt}\bigg) - \text{div}(\boldsymbol{q}_s) + q_s - q_i,\\
			\bar{\rho}_f \frac{d^f \varepsilon_s}{dt} &= \boldsymbol{\tau}_f : \nabla \boldsymbol{v}_f + \frac{\phi_f p_f}{\rho_f} \bigg(\frac{d^f \rho_f}{dt}\bigg) - \text{div}(\boldsymbol{q}_f) + q_f + q_i + \boldsymbol{f}_d \cdot(\boldsymbol{v}_s - \boldsymbol{v}_f).
			%
			%
		\end{align}
	\end{subequations}
	Finally, we may propose expressions for the imbalance of entropy, which defines the time-rate of change of the specific internal entropies, $s_s$ and $s_f$, within each continua:
	\begin{subequations}
		\label{eqn:entropy_imbalance}
		\begin{align}
			\bar{\rho}_s \frac{d^s s_s}{dt} & \geq -\text{div}(\boldsymbol{q_s}/T_s) + q_s/T_s - q_i/T_f,\\
			\bar{\rho}_f \frac{d^f s_f}{dt} & \geq -\text{div}(\boldsymbol{q}_f/T_f) + q_f/T_f + q_i/T_s.
			%
			%
		\end{align}
	\end{subequations}
	%
	

	%
	
	%
	\noindent Equations \eqref{eqn:energy_conservation} and \eqref{eqn:entropy_imbalance} include the true internal temperatures $T_s$ and $T_f$ of the constituent materials; the heat flow vectors, $\boldsymbol{q}_s$ and $\boldsymbol{q}_f$; scalar rates of internal heat generation, $q_s$ and $q_f$; and the \emph{inter-phase heat flow} per unit volume, $q_i$.
	
	Together, \eqref{eqn:mass_conservation}--\eqref{eqn:entropy_imbalance} define a system of governing equations for modeling the motion and deformation of fluid--saturated granular materials. Solving this system of equations is only possible when they are coupled with specific constitutive models for the effective granular stress $\boldsymbol{\sigma}_s$, the effective fluid shear stress $\boldsymbol{\tau}_f$, the fluid pore pressure $p_f$, the inter-phase drag force $\boldsymbol{f}_d$, and the rates of heat flow and heat generation --- $\boldsymbol{q}_s$, $\boldsymbol{q}_f$, $q_s$, $q_f$, and $q_i$. Further discussion and derivation of these equations can be found in {\ref{sec:appendix_theory}}, as well as in Chapter 2 of \citet{baumgarten2021b}.

	
	\section{Granular Constitutive Model}
	\label{sec:model}
	The constitutive model proposed in this section is formulated using the theory of breakage mechanics \citep[][]{einav2007a}, and  incorporates the effects of non-linear elasticity \citep{nguyen2009}; dilation and compaction \citep{rubin2011,cil2020a}; critical state behavior \citep{pailha2009,tengattini2016}; shock compression \citep{herbold2020}; and importantly, coupling with pore fluids \citep{baumgarten2019a}.
	
	We use a thermodynamic formulations founded on the \emph{specific Helmholtz free energies}, $\psi_s = \varepsilon_s - T_s s_s$ and $\psi_f = \varepsilon_f - T_f s_f$, which describe the amount of available energy --- or \emph{strain energy} --- stored in each material. Combining constitutive equations for $\psi_s$ and $\psi_f$ with the first and second law of thermodynamics in \eqref{eqn:energy_conservation} and \eqref{eqn:entropy_imbalance}, we formulate models for $\boldsymbol{\sigma}_s$, $\boldsymbol{\tau}_f$, $p_f$, and $\boldsymbol{f}_d$ that are thermodynamically sound across the range of loading conditions experienced during high-velocity impact events. 
	
	\subsection{Kinematics}
	Consider the motion of the overlapping continua shown in Figure \ref{fig:rve}d,f. In poromechanics and mixture theory, each continuum material moves through space following its own independent velocity field --- here, $\boldsymbol{v}_s$ and $\boldsymbol{v}_f$ (in general, $\boldsymbol{v}_s \neq \boldsymbol{v}_f$). We define the \emph{effective mesoscopic distortion rates}, $\boldsymbol{L}_s$ and $\boldsymbol{L}_f$, of each continuum material in terms of the gradient of their respective velocity fields:
	\begin{subequations}
		\label{eqn:strain_rates}
		\begin{align}
			\boldsymbol{L}_s & \equiv \nabla \boldsymbol{v}_s, \quad \text{and}\\
			\boldsymbol{L}_f & \equiv \nabla \boldsymbol{v}_f,
		\end{align}
	\end{subequations}
	\noindent with $\nabla \boldsymbol{v}_s$ and $\nabla \boldsymbol{v}_f$ the second-order, velocity gradient tensors. The corresponding \emph{effective mesoscopic strain-rates} $\boldsymbol{D}_s$ and $\boldsymbol{D}_f$ are the symmetric parts of $\boldsymbol{L}_s$ and $\boldsymbol{L}_f$.
	
	This mesoscopic picture of material deformation is identical to the standard picture of deformation in continuum mechanics. However, in poromechanics, soil mechanics, and breakage mechanics, we are also interested in \emph{microscopic} strains and distortions, which must be inferred from the \emph{mesoscopic} states of both continua.

	\subsection{Granular Micromechanics}
	\label{sec:granular_micromechanics}
\begin{figure}[!h]
		\centering
		\includegraphics[width=0.98\linewidth]{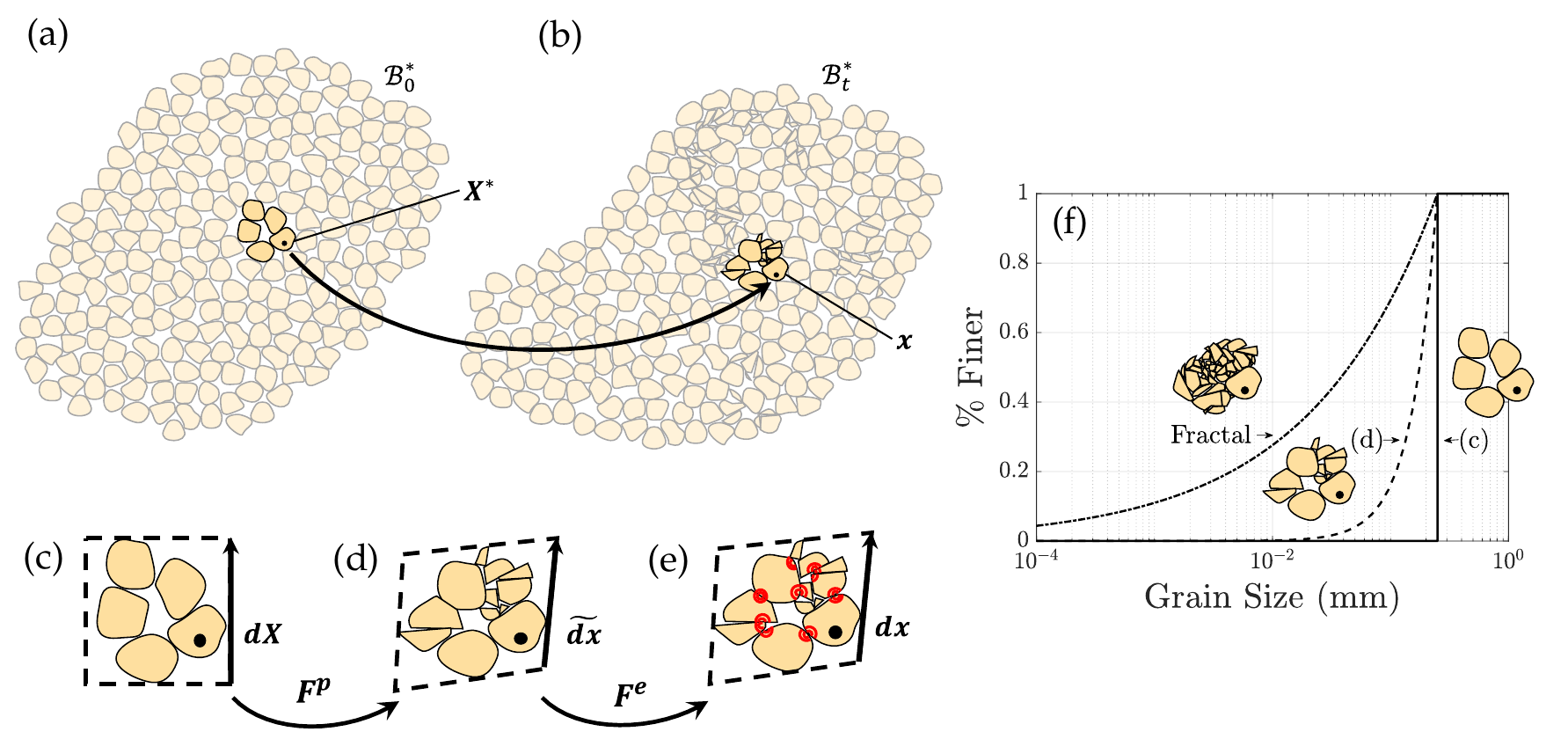}
		\caption[]{Illustration of a granular material in (a) the stress-free reference configuration $\mathcal{B}_0^*$ and (b) the current, deformed configuration $\mathcal{B}_t^*$. Highlighted region shows the material point $\boldsymbol{X}^*$ in $\mathcal{B}_0^*$, which maps to the spatial point $\boldsymbol{x}$ in $\mathcal{B}_t^*$.
			The illustrations in (c)--(e) show the proposed elasto-plastic mapping of vectors in the reference neighborhood around $\boldsymbol{X}^*$ in (c) to an intermediate, stress-free neighborhood in (d) to the current, deformed neighborhood in (e). The plot in (f) shows a way to characterize the evolving shapes of particles shown in (c)--(d) using breakage mechanics theory: the \emph{cumulative particle size distribution}.
		}
		\label{fig:kinematics}
	\end{figure}

	This microscopic-to-mesoscopic connection is developed by considering the behavior of the \emph{material neighborhoods} that surround individual particles.
	%
	%
	Figure \ref{fig:kinematics} illustrates one such material neighborhood as it is taken from an (assumed) initially stress-free reference state in the body $\mathcal{B}_0^*$ to its current deformed state in the body $\mathcal{B}_t^*$.
	We characterize the deformation of this material neighborhood using the \emph{deformation gradient} tensor $\boldsymbol{F}$, which admits the local multiplicative decomposition $\boldsymbol{F} = \boldsymbol{F}^e \boldsymbol{F}^p$ \citep{lee1968}, and obeys the following evolution rule: $d^s \boldsymbol{F}/dt = \boldsymbol{L}_s \boldsymbol{F}$.
	
	The \emph{elastic} deformation gradient tensor $\boldsymbol{F}^e$ describes the part of the total mesoscopic deformation that stores strain energy in the material. In principle, this deformation is completely reversible and associated with microscopic, elastic strains concentrated at particle--particle contact points (Figure \ref{fig:kinematics}d--e).
	The \emph{plastic} deformation gradient tensor $\boldsymbol{F}^p$, on the other hand, describes the mesoscopic deformations that are \emph{inelastic} --- i.e., that do not store strain energy in the material. These deformations can be significant and are associated with inelastic mechanisms such as granular rearrangement, pore dilation and compaction, and particle fragmentation (Figure \ref{fig:kinematics}c--d).

	Constitutive models for granular materials are frequently expressed in terms of strains and volume ratios \citep[e.g.,][]{nguyen2009, rubin2011, tengattini2016, cil2020a} which can be easily defined in terms of the deformation gradient tensor.
	For example, the \emph{elastic volume ratio} $J^e$, and a mesoscopic elastic strain tensor
	$\boldsymbol{E}^e$ and its primary strain invariants --- the \emph{elastic volumetric strain} $\epsilon_v^e$, and the \emph{elastic shear strain} $\epsilon_s^e$ --- may be obtained from the elastic deformation gradient $\boldsymbol{F}^e$:
	\begin{equation}
		\label{eqn:elastic_strain}
		\boldsymbol{E}^e \equiv \tfrac{1}{2}(\boldsymbol{F}^{e\top} \boldsymbol{F}^e - \boldsymbol{1}), \quad
		\epsilon_v^e \equiv \text{tr}(\boldsymbol{E}^e),
		\quad
		\epsilon_s^e \equiv \sqrt{\tfrac{2}{3} \boldsymbol{E}^e_0:\boldsymbol{E}^e_0},
		\quad \text{and} \quad
		J^e \equiv \text{det}(\boldsymbol{F^e}).
	\end{equation}
	Here, the trace of a tensor $\boldsymbol{A}$ is denoted by $\text{tr}(\boldsymbol{A})$, its transpose by $\boldsymbol{A}^\top$, its determinant by $\text{det}(\boldsymbol{A})$, and its deviatoric part by $\boldsymbol{A}_0$.
	Assuming an additive decomposition of the mesoscopic strain-rates in \eqref{eqn:strain_rates}, $\boldsymbol{F}^e$ and $\boldsymbol{E}^e$ evolve according to the following rule: $d^s \boldsymbol{F}^e / dt = \boldsymbol{L}^e \boldsymbol{F}^e$ with $\boldsymbol{L}^e = \boldsymbol{L}_s - \boldsymbol{\tilde{D}}^p$,
	where $\boldsymbol{\tilde{D}}^p$ denotes the \emph{inelastic deformation rate} tensor, which is defined later in this section.
	
	The final component of this micromechanical picture of granular materials is highlighted in Figure \ref{fig:kinematics}f: the \emph{cumulative particle size distribution}.
	In breakage mechanics theory \citep[][]{einav2007a}, this is characterized by the auxiliary variable $B$, the \emph{relative breakage}.
	Originally introduced in \citet{hardin1985}, $B$ measures the pulverization of granular particles by comparing the current distribution of particle sizes with the hypothetical initial and ultimate distributions shown in Figure \ref{fig:kinematics}f (see {\ref{sec:appendix_theory}}).
	%
	%
	In addition to the mesoscopic picture of inelastic deformations provided by $\boldsymbol{F}^p$ and $\boldsymbol{\tilde{D}}^p$, the relative breakage $B$ allows us to incorporate information about the evolving sizes of the individual particles into our model. 

	All together, this micromechanical picture of granular materials allows us to predict the amount of elastic deformation experienced within the individual grains --- through $B$, $\epsilon_v^e$, $\epsilon_s^e$, and $J^e$ --- by measuring the rate at which the material is deforming ($\boldsymbol{L}_s$ and $\boldsymbol{D}_s$), determining how much of that deformation is inelastic ($\boldsymbol{\tilde{D}}^p$), and connecting this inelastic deformation with changes in the particle size distribution ($d^s B/ dt$). 
	A thorough discussion of these strains, strain-rates, and auxiliary variables can be found in {\ref{sec:appendix_theory}}.
	
	%
	%
	%
	
	%
	%
	
	\subsection{Helmholtz Free Energy}
\begin{figure}[!h]
		\centering
		\includegraphics[width=0.75\linewidth]{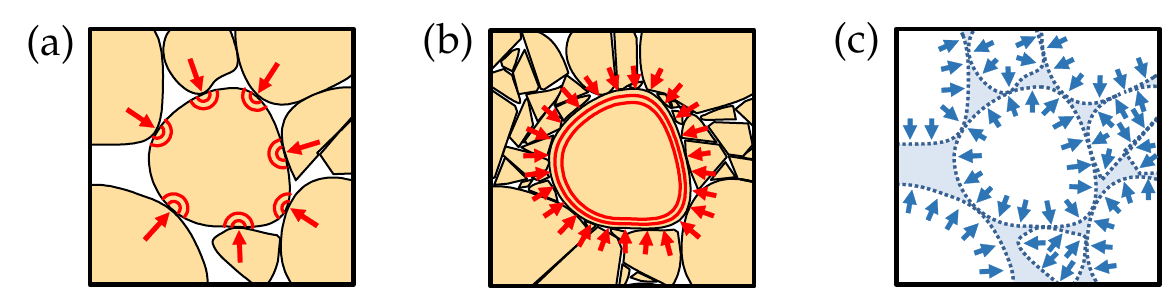}
		\caption[]{
			Illustration of strain energy storage mechanisms in fluid-saturated granular materials. (a) Under low--moderate confining stresses, highly porous granular materials store strain energy near particle--particle contact points. (b) After crushing, low porosity granular materials store energy more uniformly through direct compression of the solid material that composes the individual grains. (c) Within a  connected pore network, the interstitial fluid primarily stores strain energy through bulk compression and internal temperature changes.
		}
		\label{fig:strain_energy}
	\end{figure}

	Following the thermomechanical models developed in \citet{herbold2020} and \citet{baumgarten2021a}, we propose expressions for the mass-specific Helmholtz free energies, $\psi_s$ and $\psi_f$. These are assumed to be functions of the mesoscopic elastic deformations as described by $\epsilon_v^e$ and $\epsilon_s^e$; the distribution of particle sizes, as captured by $B$; the true densities of the constituent materials $\rho_s$ and $\rho_f$; and their absolute temperatures, $T_s$ and $T_f$. We assume that these Helmholtz free energies can be written as
	\begin{subequations}
		\label{eqn:helmholtz_free_energy}
		\begin{align}
			\psi_s &= \hat{\psi}_c(\epsilon_v^e, \epsilon_s^e, B) + \hat{\psi}_g(\rho_s, T_s),\\
			\psi_f &= \hat{\psi}_f(\rho_f, T_f).
		\end{align}
	\end{subequations}
	Here $\psi_c$ represents the component of $\psi_s$ that is associated with strain energy stored at particle--particle contact points \citep[Figure \ref{fig:strain_energy}a; e.g., see][]{hiramatsu1966}, while $\psi_g$ and $\psi_f$ represent the mechanically distinct \emph{densification} strain energies associated with volume and temperature changes within constituent grains and interstitial fluid (Figures \ref{fig:strain_energy}b,c).
	%
	
	Together with the assumed expressions for $\psi_s$ and $\psi_f$ in \eqref{eqn:helmholtz_free_energy}, Equations  \eqref{eqn:energy_conservation} and \eqref{eqn:entropy_imbalance} provide a set of thermodynamic constraints for the constitutive model presented here. First, the free energy functions $\psi_s$ and $\psi_f$ must satisfy,
	%
	%
	\begin{subequations}
		\label{eqn:free_energy_imbalance}
		\begin{align}
			\bar{\rho}_s \frac{d^s \psi_s}{dt} &= \boldsymbol{\sigma}_s : \nabla \boldsymbol{v}_s + \frac{\phi_s p_f}{\rho_s} \bigg(\frac{d^s \rho_s}{dt}\bigg) - \bar{\rho}_s s_s \bigg(\frac{d^s T_s}{dt}\bigg) - D_s,
			\label{subeqn:solid_free_energy_imbalance}\\
			\bar{\rho}_f \frac{d^f \psi_f}{dt} &= \boldsymbol{\tau}_f : \nabla \boldsymbol{v}_f + \frac{\phi_f p_f}{\rho_f} \bigg(\frac{d^f \rho_f}{dt}\bigg) -  \bar{\rho}_f s_f \bigg(\frac{d^f T_f}{dt}\bigg) - D_f,
			%
			%
		\end{align}
	\end{subequations}
	{with $D_s \geq 0$ and $D_f \geq 0$ the positive rates of \emph{mechanical dissipation}.}
	Second, the second law requires that heat flows from hot to cold regions in both materials and that drag forces act against relative motions --- i.e.,
	\begin{equation}
		\label{eqn:heat_flow}
		\boldsymbol{q}_s \cdot \nabla T_s \leq 0, \quad  \boldsymbol{q}_f \cdot \nabla T_f \leq 0, \quad  q_i (T_s - T_f) \geq 0, \quad \text{and} \quad \boldsymbol{f}_d \cdot (\boldsymbol{v}_s - \boldsymbol{v}_f) \geq 0.
	\end{equation}
	%
	These conditions constrain the form of the constitutive equations for $\boldsymbol{\sigma}_s$, $\boldsymbol{\tau}_f$, $p_f$, $\boldsymbol{f}_d$, $\boldsymbol{q}_s$, $\boldsymbol{q}_f$, and $q_i$ --- ensuring that our model predictions are physically reasonable and dissipative across the full range of conditions experienced during high-velocity impact events. 
	The derivation of these equations can be found in {\ref{sec:appendix_theory}} together with further discussions.

	\subsection{Mechanical Model for Porosity}
	The porosity is constrained according to \eqref{eqn:porosity} with $\phi_s \in [0,1]$; $\phi_f \in [0,1]$; and $\phi_s + \phi_f = 1$. Here, we discuss two useful notations for porosity: (i) the \emph{inelastic porosity}, $\phi_p$; and (ii) the \emph{true porosity}, $\phi_f = 1 - \phi_s$.
	
	The inelastic porosity, $\phi_p$, defines the porosity of the granular sediment in the intermediate, inelastically deformed space shown in Figure \ref{fig:kinematics}d. It represents the porosity of the material in the absence of significant elastic deformations --- which tend to squeeze particles into open pore spaces. This measure of porosity is primarily used to model dilation and critical state behavior in the breakage mechanics literature \citep[e.g.,][]{tengattini2016,cil2020a} and follows the deformation theory from \citet{collins2010}: 
	\begin{equation}
		\label{eqn:inelastic_porosity}
		d^s \phi_p / dt = - (1 - \phi_p)\ \text{tr}(\boldsymbol{\tilde{D}}^p).
	\end{equation}
	%
	Although broadly useful for characterizing how inelastic deformations change the available pore space, the inelastic porosity is mathematically distinct from the true porosity $\phi_f$ that appears in \eqref{eqn:porosity}--\eqref{eqn:entropy_imbalance}.
	
	
	The true porosity $\phi_f$, on the other hand, is primarily used in the geomechanics literature and defines the porosity of the granular material in its current deformed space (shown in Figure \ref{fig:kinematics}e). This measure of porosity is frequently determined using the solid volume fraction $\phi_s$, which is calculated from the effective density $\bar{\rho}_s$ and true solid density $\rho_s$
	\citep[e.g., see][]{danielson1986}: 
	\begin{equation}
		\label{eqn:porosity_explicit}
		\phi_f = 1 - \phi_s, \quad \text{and} \quad \phi_s = \bar{\rho}_s / \rho_s.
	\end{equation}
	%
	%
	%
	%
	Here, the effective density $\bar{\rho}_s$ is easily determined by bulk kinematics in \eqref{eqn:mass_conservation} and \eqref{eqn:strain_rates}; however, the true solid density $\rho_s$ must be determined using a constitutive equation. At low--moderate confining stresses (Figure \ref{fig:strain_energy}a), the solid constituent is generally treated as \emph{incompressible} --- i.e., $\rho_s$ has a constant value $\rho_0$ --- greatly simplifying the calculation in \eqref{eqn:porosity_explicit}. At high confining stresses (Figure \ref{fig:strain_energy}b), on the other hand, this assumption is no longer valid, and the constituent solid must be treated as \emph{compressible}.
	
	To complete the model for $\phi_f$ in \eqref{eqn:porosity_explicit}, we propose the following mechanical model for the solid density $\rho_s$:
	\begin{equation}
		\label{eqn:solid_density}
		\rho_s/\rho_0 = 1 + \hat{\alpha}(\phi_s)(J^{e-1} - 1),
	\end{equation}
	with $\rho_0$ the stress-free solid reference density (at room temperature) and $\hat{\alpha}(\phi_s) \in [0,1]$ a constitutive function that depends on the solid volume fraction.
	%
	At relatively high porosities (Figure \ref{fig:strain_energy}a), the solid constituent is reasonably modeled as incompressible with $\hat{\alpha}(\phi_s) \approx 0$. 
	On the other hand, at relatively low porosities (Figure \ref{fig:strain_energy}b), the solid constituent must compress to accommodate mesoscopic deformations with $\hat{\alpha}(\phi_s) \approx 1$. 
	In this work, we assume a simple power law model for $\hat{\alpha}$:
	\begin{equation}
		\label{eqn:alpha}
		\hat{\alpha}(\phi_s) = \phi_s^b,
	\end{equation}
	with $b$ a fitting parameter. Together with \eqref{eqn:porosity} and \eqref{eqn:solid_density}, this mechanical model uniquely determines the solid volume fraction --- and thus the porosity --- in terms of the effective density $\bar{\rho}_s$, the mesoscopic volume ratio $J^e$, and the reference density $\rho_0$ according to the implicit equation: 
	\begin{equation}
		\label{eqn:solid_volume_fraction}
		\phi_s = \bar{\rho}_s / \rho_0 - \phi_s^{b+1} (J^{e-1} - 1).
	\end{equation}
	
	Note that this purely mechanical model for porosity does not consider thermal expansion or the influence of extreme pore fluid pressures \citep[e.g.,][]{biot1941}. 
	
	\subsection{Effective Granular Stress}
	The constitutive equation for the effective granular stress $\boldsymbol{\sigma}_s$ can be deduced from \eqref{eqn:strain_rates}, \eqref{eqn:elastic_strain}, \eqref{eqn:helmholtz_free_energy}, \eqref{eqn:free_energy_imbalance}, and \eqref{eqn:solid_density} following the Coleman--Noll procedure \citep{coleman1963}:
	\begin{subequations}
		\label{eqn:effective_granular_stress}
		\begin{align}
			\boldsymbol{\sigma}_s &= \frac{\bar{q}}{3 \epsilon_s^e} \boldsymbol{B}^e_0 \boldsymbol{B}^e - \bar{p} \boldsymbol{B}^e - \phi_s p^* \hat{A}(\phi_s, J^e) \boldsymbol{1},
			\label{subeqn:stress}\\
			\bar{p} &= -\bar{\rho}_s \frac{\partial \hat{\psi}_c}{\partial \epsilon_v^e}, \quad
			\bar{q} = \bar{\rho}_s \frac{\partial \hat{\psi}_c }{ \partial \epsilon_s^e}, \quad \text{and} \quad
			p^* = \rho_s^2 \frac{\partial \hat{\psi}_g}{\partial \rho_s},
		\end{align}
	\end{subequations}
	with $\boldsymbol{B}^e = \boldsymbol{F}^e \boldsymbol{F}^{e\top}$, and $\hat{A}(\phi_s, J^e) = -(J^e/\rho_s) (d \rho_s / d J^e)|_{\boldsymbol{\tilde{D}}^p = \boldsymbol{0}}$. An explicit expression for $\hat{A}(\phi_s, J^e)$ is provided in {\ref{sec:appendix_theory}}. In \eqref{eqn:effective_granular_stress}, $\bar{p}$ and $\bar{q}$ denote the pressure and shear stresses associated with elastic deformations at the particle--particle contact points.
	
	We assume the nonlinear free energy function, $\psi_c$ \citep[e.g., see][]{nguyen2009} to be in the form:
	\begin{subequations}
		\label{eqn:contact_stresses}
		\begin{align}
			\hat{\psi}_c(\epsilon_v^e, \epsilon_s^e, B) &= (1 - \theta B) \frac{p_r}{\rho_0} \bigg(\frac{-\bar{K}^2 \epsilon_v^{e3}}{12} - \frac{3 \bar{G}\bar{K} \epsilon_v^e \epsilon_s^{e2}}{ 4} + \frac{\bar{G} \sqrt{3\bar{G}\bar{K}} \epsilon_s^{e3}}{4}\bigg),\\
			\bar{p} &= \frac{\bar{\rho}_s}{\rho_0} (1 - \theta B) p_r \bigg(\frac{\bar{K}^2 \epsilon_v^{e2}}{4} + \frac{3\bar{G}\bar{K}\epsilon_s^{e2}}{4}\bigg),
			\label{subeqn:pbar}\\
			\bar{q} &= \frac{\bar{\rho}_s}{\rho_0} (1 - \theta B)  3 \bar{G} p_r \bigg(\frac{-\bar{K} \epsilon_v^e \epsilon_s^e}{2} + \frac{\sqrt{3 \bar{G} \bar{K}} \epsilon_s^{e2}}{4}\bigg).
			\label{subeqn:qbar}
			%
		\end{align}
	\end{subequations}
	where $\theta$ denotes the constant grading index from \citet{einav2007a}; $p_r$ denotes the non-linear reference pressure; $\bar{K}$ and $\bar{G}$ denote the dimensionless, reference bulk modulus and shear modulus from \citet{nguyen2009}; and the remaining variables have already been defined. 
	
	The third term in $\boldsymbol{\sigma}_s$ in \eqref{eqn:effective_granular_stress}a is the solid pressure term, $p^*$, which is associated with strain energy stored during direct compression of the constituent solid material and defined by the constituent free energy function, $\psi_g$ \citep[e.g., see][]{herbold2020}. Any thermodynamically valid equation of state (EOS) may be used, but here we use the Mie--Gr{\"u}neisen EOS to define $\psi_g$ \citep[][]{mie1903,gruneisen1912}:
	\begin{subequations}
		\label{eqn:solid_pressure}
		\begin{align}
			\hat{\psi}_g(\rho_s, T_s) &= \hat{e}_c(\rho_s) + c_v T_s - T_s \bigg[c_v \text{ln}\bigg(\frac{T_s}{T_0}\bigg) - c_v \Gamma_0 \bigg(1 -  \frac{\rho_0}{\rho_s}\bigg)\bigg],\\
			p^* &= \hat{p}_H(\rho_s) \bigg[1 - \frac{\Gamma_0}{2} \bigg(1 - \frac{\rho_0}{\rho_s}\bigg)\bigg] + \rho_0 \Gamma_0 \big(\hat{e}_c(\rho_s) + c_v (T_s - T_0)\big),\label{subeqn:pstar}
			%
		\end{align}
	\end{subequations}
	where \eqref{subeqn:pstar} is computed using 
	the solid heat capacity $c_v$; the reference temperature $T_0$; the Gr{\"u}neisen parameter $\Gamma_0$; and the constitutive functions, $\hat{e}_c(\rho_s)$ and $\hat{p}_H(\rho_s)$, which define the cold energy and shock Hugoniot curves, respectively.
	%
	%
	In the first-order Mie--Gr{\"u}neisen EOS, these functions are determined by the reference sound speed, $C_0$, and the slope of the $U_s$--$U_p$ curve, $S_0$. Explicit expressions for these functions are provided in {\ref{sec:appendix_explicit_eos}}.
	
	%
	%
	
	
	
	\subsection{Yielding and Dissipation}
	The inelastic response of the granular material is considered in terms of the inelastic deformation rate $\boldsymbol{\tilde{D}}^p$, and the rate of change of the relative breakage, $d^s B / dt$. The inelastic flow rules that define $\boldsymbol{\tilde{D}}^p$ and $d^s B / dt$ are \emph{non-associative} --- i.e., they are defined by both a \emph{yield function} and by a \emph{flow direction}. 
	
	As in \citet{baumgarten2019a}, we define the flow direction for $\boldsymbol{\tilde{D}}^p$ using the \emph{stress conjugate to yielding}, $\boldsymbol{\sigma}_y$, and the stress invariants $p_y$ and $q_y$ as follows:
	\begin{equation}
		\label{eqn:inelastic_deformation_rate}
		\boldsymbol{\tilde{D}}^p = \frac{3 \xi_s^p}{2 q_y} \boldsymbol{\sigma}_{y0} + \tfrac{1}{3}\big(\xi_v^p + \xi_2^p + \xi_3^p\big) \boldsymbol{1},
		\quad \text{with} \quad
		p_y = -\tfrac{1}{3} \text{tr}(\boldsymbol{\sigma}_y), \quad
		q_y = \sqrt{\tfrac{3}{2} (\boldsymbol{\sigma}_{y0}:\boldsymbol{\sigma}_{y0})}.
	\end{equation}
	Here, {$\boldsymbol{\sigma}_{y0}$ denotes the deviatoric component of $\boldsymbol{\sigma}_y$ and}  
	the scalar rates $\xi_s^p$, $\xi_v^p$, $\xi_2^p$, and $\xi_3^p$ describe the four dominant, inelastic deformation mechanisms shown in Figure \ref{fig:mechanisms}: the \emph{granular shear rate}, $\xi_s^p$; the \emph{dilation/compaction rate}, $\xi_v^p$; the \emph{free expansion rate}, $\xi_2^p$; and the \emph{consolidation rate}, $\xi_3^p$.
	
	In \eqref{eqn:inelastic_deformation_rate}, the stress conjugate to yielding, $\boldsymbol{\sigma}_y$, defines the stress components that dissipate energy during inelastic flow ($\boldsymbol{\tilde{D}}^p \neq \boldsymbol{0}$). This stress measure is nearly identical to the effective granular stress, $\boldsymbol{\sigma}_s$, and
	is defined as follows:
	\begin{equation}
		\label{eqn:yield_stress}
		\boldsymbol{\sigma}_y = \frac{\bar{q}}{3 \epsilon_s^e} \boldsymbol{B}^e_0 \boldsymbol{B}^e - \bar{p} \boldsymbol{B}^e - \phi_s p^* \hat{C}(\phi_s, J^e) \boldsymbol{1},
	\end{equation}
	with $\bar{q}$, $\bar{p}$, and $p^*$ defined in \eqref{eqn:effective_granular_stress} and $\hat{C}(\phi_s, J^e) = -(J^e/\rho_s) (d \rho_s / d J^e)|_{\boldsymbol{D}_s = \boldsymbol{0}}$. 
	The distinction between the stress conjugate to yielding $\boldsymbol{\sigma}_y$ and the effective granular stress $\boldsymbol{\sigma}_s$ is required to satisfy the second law of thermodynamics and is discussed further in {\ref{sec:appendix_theory}}.
	
	To complete the flow rule in \eqref{eqn:inelastic_deformation_rate}, we define a set of yield functions that uniquely determine the scalar inelastic rates, $\xi_s^p$, $\xi_v^p$, $\xi_2^p$, $\xi_3^p$, and $d^s B / dt$, subject to the following dissipation condition from \eqref{subeqn:solid_free_energy_imbalance}:
	\begin{equation}
		\label{eqn:solid_dissipation}
		D_s = q_y \xi_s^p - p_y (\xi_v^p + \xi_2^p + \xi_3^p) + E_B \frac{d^s B}{dt},
		\quad \text{with} \quad
		D_s \geq 0, \quad
		E_B = -\bar{\rho}_s \frac{\partial \hat{\psi}_c}{\partial B},
	\end{equation}
	with $E_B$ the \emph{breakage energy} that is dissipated through particle pulverization \citep[][]{einav2007a}.

	\subsection{Onset of Yielding}
	First, we determine the granular shear rate $\xi_s^p$, the dilation/compaction rate $\xi_v^p$, and the rate of breakage $d^s B/ dt$ using the scalar multiplier $\lambda_1$ (i.e., $\xi_s^p$, $\xi_v^p$, and $d^s B/ dt \propto \lambda_1$) according to the yield function proposed in \citet{rubin2011}:
	\begin{equation}
		\label{eqn:y1}
		y_1 = \frac{E_B (1 - B)^2}{E_c} + \frac{q_y^2}{(M p_y)^2} - 1,
		\quad \text{with} \quad
		y_1 \leq 0, \quad \lambda_1 \geq 0, \quad y_1 \lambda_1 = 0.
	\end{equation}
	Here, $E_c$ denotes the \emph{critical breakage energy}, and $M$ denotes the \emph{internal friction coefficient}.
	This form of the yield function combines mathematical models for frictional granular flow \citep[$q_y = Mp_y$;][]{drucker1952} and particle fragmentation \citep[$E_B(1-B)^2 = E_c$;][]{einav2007a} into a single rule for dense, inelastic deformation.
	
	The second yield function determines the rate of free expansion $\xi_2^p$ using the scalar multiplier $\lambda_2$ (i.e., $\xi_2^p \propto \lambda_2$) according to the yield function proposed in \citet{baumgarten2019a}:
	\begin{equation}
		\label{eqn:y2}
		y_2 = -p_y,
		\quad \text{with} \quad
		y_2 \leq 0, \quad \lambda_2 \geq 0, \quad y_2 \lambda_2 = 0.
	\end{equation}
	This form of the second yield function ensures pressure positivity and defines a granular material in which the particles are free to separate and are unable to support tension.
	
	Finally, the third yield function determines the consolidation rate $\xi_3^p$ using the scalar multiplier $\lambda_3$ (i.e., $\xi_3^p \propto \lambda_3$) according to a modified yield function from \citet{baumgarten2019a}:
	\begin{equation}
		\label{eqn:y3}
		y_3 =
		\begin{cases}
			\phi_f - \phi_\text{max} & \text{if}\quad \phi_f \leq \phi_\text{max},\\
			0 & \text{if}\quad \phi_f > \phi_\text{max},
		\end{cases}
		\quad \text{with} \quad
		y_3 \leq 0, \quad \lambda_3 \geq 0, \quad y_3 \lambda_3 = 0,
	\end{equation}
	with $\phi_{\text{min}} = \phi_l (1 - B)^l$ and $\phi_{\text{max}} = \phi_u (1 - B)^u$ the limiting inelastic porosities proposed in \citet{rubin2011}. This form of the third yield function ensures that disconnected granular sediments ($\phi_f \geq \phi_\text{max}$) are unable to support any stresses.
	
	
	\subsection{Inelastic Flow}
	At the onset of dense, inelastic yielding (i.e., $y_1 = 0$), $d^s B/dt$, $\xi_v^p$, and $\xi_s^p$ are allowed to have non-zero values obeying the modified flow rules from \citet{tengattini2016} and \citet{cil2020a}. These flow rules are determined by the scalar multiplier $\lambda_1 \geq 0$, which ensures that $y_1 = 0$ while the material is yielding:
	\begin{subequations}
		\label{eqn:y1_rates}
		\begin{align}
			\label{subeqn:dBdt}
			\frac{d^s B}{dt} &= \lambda_1 \frac{E_B (1 - B)^2}{E_B E_c} \text{cos}^2(\omega),\\
			\label{subeqn:devdt}
			\xi_v^p &= \lambda_1 \frac{E_B (1 - B)^2}{E_c} \frac{-p_y}{(p_y^2 + q_y^2)}  \text{sin}^2(\omega) + \lambda_1 M_d \frac{q_y}{(M p_y)^2},\\
			\label{subeqn:desdt}
			\xi_s^p &=  \lambda_1 \frac{E_B (1 - B)^2}{E_c} \frac{q_y}{(p_y^2 + q_y^2)} \text{sin}^2(\omega) + \lambda_1 \frac{q_y}{(M p_y)^2}.
		\end{align}
	\end{subequations}
	Here, $\omega$ is the \emph{coupling angle} from \citet{einav2007a}, and $M_d$ is the \emph{dilation coefficient}.
	The first terms in \eqref{subeqn:dBdt}, \eqref{subeqn:devdt}, and \eqref{subeqn:desdt} define the components of inelastic deformation associated with particle fragmentation {and are coupled together by the coupling angle $\omega$. These terms capture how fragmentation simultaneously changes the distribution of particle sizes ($d^s B/dt$) \emph{and} relaxes stress concentrations at particle--particle contact points ($\xi_v^p$ and $\xi_s^p$).}
	%
	The second terms in \eqref{subeqn:devdt} and \eqref{subeqn:desdt}, on the other hand, are associated with frictional granular rearrangement and shear dilation ($\xi_v^p$ and $\xi_s^p$).
	
	An important component of \eqref{eqn:y1_rates} is the incorporation of the critical state theories of \citet{pailha2009}; \citet{tengattini2016}; and \citet{cil2020a}.
	In particular, $\omega$, $M$, and $M_d$ are all functions of the \emph{relative density}, $\tau$, and its critical value, $\tau_{cs}$:
	\begin{equation}
		\label{eqn:critical_state}
		\tau = \frac{\phi_\text{max} - \phi_p}{\phi_\text{max} - \phi_\text{min}},
		\quad \text{and} \quad
		\tau_{cs} = \sqrt{\frac{E_B}{E_c}} \frac{(1 - B)}{\gamma}
	\end{equation}
	with $\phi_\text{min}$ and $\phi_\text{max}$ defined in \citet{rubin2011}, and $\gamma \in [0,1]$ the constant dilation parameter from \citet{tengattini2016}. From these, we define:
	\begin{equation}
		\label{eqn:friction_and_dilation}
		M = M_0 + M_d,
		\quad
		M_d = \gamma (\tau - \tau_{cs}) \bigg(\frac{6\ \text{sin}(\theta_p)}{3 - \text{sin}(\theta_p)} - M_0\bigg),
		\quad \text{and} \quad 
		\omega = \frac{\pi}{2}(1 - \tau),
	\end{equation}
	%
	%
	where $M_0$ is the \emph{critical state friction coefficient} and $\theta_p$ is the \emph{peak dilation angle}. Here, $\theta_p = \pi/15 + \text{sin}^{-1}(3 M_0 / (6 + M_0))$ is assumed from \citet{cil2020a}.
	
	At the onset of tensile yielding (i.e., $y_2 = 0$), $\xi_2^p$ is allowed to have a non-zero value obeying the flow rule from \citet{baumgarten2019a}. This flow rule is determined by the scalar multiplier $\lambda_2 \geq 0$, which ensures that $y_2 = 0$ while the material is yielding:
	\begin{equation}
		\label{eqn:y2_rates}
		\xi_2^p = \lambda_2.
	\end{equation}
	Similarly, at the onset of disconnected yielding (i.e., $y_3 = 0$), $\xi_3^p$ is allowed to have a non-zero value determined by the scalar multiplier $\lambda_3 \geq 0$, which ensures that $y_3 = 0$ while the material is yielding:
	\begin{equation}
		\label{eqn:y3_rates}
		\xi_3^p = - \lambda_3.
	\end{equation}

	All together, \eqref{eqn:inelastic_deformation_rate}--\eqref{eqn:y1_rates}, \eqref{eqn:y2_rates}, and \eqref{eqn:y3_rates} define the inelastic behavior of the granular material model proposed in this work. Further discussion of these flow rules, along with proof that the dissipation inequality in \eqref{eqn:solid_dissipation} is satisfied, is provided in {\ref{sec:appendix_theory}}.
	
	\subsection{Fluid Equation of State and Viscous Stresses}
	The behavior of the interstitial fluid is dominated by the pore fluid stresses as captued by the effective fluid shear stress, $\boldsymbol{\tau}_f$, and the pore fluid pressure, $p_f$.
	The constitutive equations for these stresses 
	are deduced from \eqref{eqn:strain_rates}, \eqref{eqn:helmholtz_free_energy}, and \eqref{eqn:free_energy_imbalance} following the Coleman--Noll procedure:
	\begin{equation}
		\label{eqn:effective_fluid_stresses}
		p_f = \rho_f^2 \frac{\partial \psi_f}{\partial \rho_f}, \quad \text{and} \quad
		\boldsymbol{\tau}_f : \boldsymbol{D}_s \geq 0.
	\end{equation}
	For the effective fluid shear stress, $\boldsymbol{\tau}_f$, there are many admissible constitutive equations available in the literature \citep[e.g.,][]{eilers1941, krieger1959, morris1999}, and here, we adopt the simple model proposed in \citet{einstein1906}:
	\begin{equation}
		\label{eqn:effective_fluid_shear_stress}
		\boldsymbol{\tau}_f = 2 \eta_0 (1 + 5 \phi_s / 2) \boldsymbol{D}_{f0},
	\end{equation}
	with $\eta_0$ the \emph{fluid viscosity} and $\boldsymbol{D}_{f0}$ the deviator of the effective fluid strain-rate $\boldsymbol{D}_f$. For the fluid pore pressure $p_f$ --- which depends on the density and temperature of the fluid constituent, $\rho_f$ and $T_f$ --- we use the Tillotson EOS \citep[][]{tillotson1962} as implemented in \citet{brundage2013}; however any thermodynamically valid EOS may also be used {\citep[e.g., the Mie--Gr{\"u}neisen EOS or the Sackur--Tetrode EOS; see][]{gruneisen1912,sackur1913}. Note that the choice of the Tillotson EOS here is motivated by the possibility of vaporization in the fluid phase during high-velocity impact.}
	
	This model constructs the fluid pressure using constitutive equations that depend on 
	the specific internal energy $\varepsilon_f$, which is defined as follows:
	\begin{equation}
		\label{eqn:fluid_internal_energy}
		\varepsilon_f = \hat{e}_{cf}(\rho_f) + c_{vf} (T_f - T_0),
	\end{equation}
	where $\hat{e}_{cf}(\rho_f)$ defines the fluid cold energy curve and $c_{vf}$ denotes the \emph{specific fluid heat capacity}. The fluid pore pressure $p_f$ is then defined piece-wise in energy--density space. The full form of this EOS is provided in the {\ref{sec:appendix_explicit_eos}}; however, for compressed states, the following expression may be used:
	\begin{equation}
		\label{eqn:tillotson_eos_p1}
		p_{f1} = 
		\bigg[a_f + \frac{b_f}{\varepsilon_f/(E_0 \eta^2) + 1}\bigg]\rho_f \varepsilon_f + A_f \mu + B_f \mu^2,
		\quad \text{for} \quad \rho_f \geq \rho_{f0},
	\end{equation}
	with $\eta = \rho_f / \rho_{f0}$ and $\mu = \eta - 1$. In this equation, $a_f$, $b_f$, $A_f$, $B_f$, and $E_0$ are constant fitting parameters, while $\rho_{f0}$ denotes the stress-free, reference density at $T_0$. The full form of the model includes the additional fitting parameters $\alpha_f$ and $\beta_f$, along with the density of \emph{incipient vaporization}, $\rho_\text{IV}$; the energy of incipient vaporization, $E_\text{IV}$; the energy of \emph{complete vaporization} $E_\text{CV}$; and the \emph{cavitation pressure} $P_\text{cav}$.
	Further discussion of the model and its implementation may be found in \citet{brundage2013}.
	
	\subsection{Inter-phase Drag Force}
	The flow of a viscous fluid through the interstitial space between particles produces a volumetric drag force which is represented in this model using the inter-phase drag force vector, $\boldsymbol{f}_d$.
	The constitutive equation for this force 
	is constrained by the thermodynamic restrictions in \eqref{eqn:heat_flow}; in particular: $\boldsymbol{f}_d \cdot (\boldsymbol{v}_s - \boldsymbol{v}_f) \geq 0$. There are many admissible constitutive equations available in the literature \citep[e.g.,][]{darcy1856,vanderhoef2005,baumgarten2021b}, and here we adopt the Carman--Kozeny drag model from \citet{carman1937}:
	\begin{equation}
		\label{eqn:interphase_drag}
		\boldsymbol{f}_d = \frac{180 \phi_s^2 \eta_0}{d_{50}^2 (1 - \phi_s)} (\boldsymbol{v}_s - \boldsymbol{v}_f).
	\end{equation}
	In this equation, $\eta_0$ is the pore fluid viscosity, and $d_{50}$ is a characteristic grain size. Here, we let $d_{50}$ denote the sieve diameter that allows 50\% (by mass) of the granular particles to pass through.
	
	\subsection{Adiabatic and Isothermal Behavior}
	The final constitutive equations that must be defined in order to solve the governing equations in \eqref{eqn:mass_conservation}--\eqref{eqn:energy_conservation} are the heat flow rates $\boldsymbol{q}_s$, $\boldsymbol{q}_f$, and $q_i$. 
	Any model for these heat flows is constrained by the thermodynamic restrictions in \eqref{eqn:heat_flow}: namely, $\boldsymbol{q}_s \cdot \nabla T_s \leq 0$; $\boldsymbol{q}_f \cdot \nabla T_f \leq 0$; and $q_i (T_s - T_f) \geq 0$. Although there are many admissible models that could be chosen, here we consider two limiting conditions: \emph{isothermal} conditions with $T_s = T_f = T_0$; and \emph{adiabatic} conditions with $\boldsymbol{q}_s = \boldsymbol{q}_f = \boldsymbol{0}$ and $q_i = 0$.
	
	For many laboratory tests performed on granular media (e.g., oedometer testing, triaxial loading, etc.), the loading rate is relatively slow, which allows temperatures to equilibrate during an experiment. For model fitting and comparison with laboratory data, the isothermal limit should be used with $T_s = T_f = T_0$. In these cases, $\boldsymbol{q}_s$, $\boldsymbol{q}_f$, and $q_i$ are defined implicitly to satisfy the conditions in \eqref{eqn:energy_conservation}, \eqref{eqn:solid_pressure}, and \eqref{eqn:fluid_internal_energy}.
	
	On the other hand, during the high-velocity impact events considered in this work, the loading rate is relatively high, and likely does not allow for significant heat flow or thermal dissipation. For predicting the material response under high-velocity impact conditions, the adiabatic limit should be used with $\boldsymbol{q}_s = \boldsymbol{q}_f = \boldsymbol{0}$ and $q_i = 0$. In these cases, $T_s$ and $T_f$ must be computed by integrating \eqref{eqn:energy_conservation} and solving for the thermal component of the internal energies: $\varepsilon_s = \hat{e}_c(\rho_s) + c_v T_s$, and $\varepsilon_f = \hat{e}_{cf}(\rho_f) + c_{vf} (T_f - T_0)$.
	
	\section{Model Summary and Example Parameters}
	\label{sec:summary}
	The model described in this work provides a unified constitutive description of fluid-saturated granular materials that can be used for impact applications. This model is governed by the system of physical conservation laws in \eqref{eqn:mass_conservation}--\eqref{eqn:entropy_imbalance} and is formulated to capture the complex, coupled dynamics of granular sediments and interstitial fluids during impact.
	
	The constitutive equations for the effective granular stress $\boldsymbol{\sigma}_s$ are based on a multiplicative decomposition of deformation (i.e., $\boldsymbol{F} = \boldsymbol{F}^e \boldsymbol{F}^p$) with an elastic stress response based on a strain-energy formulation --- i.e., $\psi_s = \hat{\psi}_c(\epsilon_v^e, \epsilon_s^e, B) + \hat{\psi}(\rho_s, T_s)$. 
	To determine the elastic deformations used in the model, the corresponding additive decomposition of the deformation rates (i.e., $\boldsymbol{L}_s = \boldsymbol{L}^e + \boldsymbol{\tilde{D}}^p$) is used, which incorporates analytical models for the four dominant, inelastic deformation mechanisms: granular shearing; dilation and compaction; free granular separation; and disconnected reconsolidation. The corresponding model parameters for two sands (Dog's Bay and Ottawa) are presented in Table \ref{tab:parameters}.
	
	\begin{table}[h!]
		\centering\footnotesize
		\caption{List of constitutive model parameters, with values presented for Dog's Bay and Ottawa sands.}
		\label{tab:parameters}
		\begin{tabular}{l p{1.6cm} p{1.6cm} l l}
			\noalign{\smallskip}
			\hline\noalign{\smallskip}
			Parameter & Dog's Bay & Ottawa & Units & Description\\
			\hline\noalign{\smallskip}
			\multicolumn{5}{l}{Elastic Stress Parameters}\\
			\hline\noalign{\smallskip}
			$\theta$ & 0.83 & 0.83 & -- & grading index \citep[][]{einav2007a}\\
			$p_r$ & 1000 & 1000 & Pa & reference pressure \citep[][]{nguyen2009}\\
			$\bar{K}$ & 16330 & 15340 & -- & dimensionless, reference bulk modulus\\
			$\bar{G}$ & 25710 & 9200 & -- & dimensionless, reference shear modulus\\
			$b$ & -- & 10 & -- & density--elasticity coupling parameter\\
			$\rho_0$ & 2710 & 2650 & kg/m$^3$ & reference solid density\\
			$T_0$ & 298 & 298 & K & reference temperature\\
			$c_v$ & -- & 736 & J/kg$\cdot$K & solid specific heat capacity\\
			$\Gamma_0$ & -- & 0.67 & -- & reference Gr{\"u}neisen parameter \citep[][]{boehler1979}\\
			$C_0$ & -- & 3630 & m/s & reference solid bulk sound speed\\
			$S_0$ & -- & 0.89 & -- & $U_s$--$U_p$ slope for Mie--Gr{\"u}neisen EOS \citep[][]{wackerle1962}\\
			\hline\noalign{\smallskip}
			\multicolumn{5}{l}{Inelastic Deformation Parameters}\\
			\hline\noalign{\smallskip}
			$E_c$ & 280 & 5.0$\times 10^5$ & J/m$^3$ & critical breakage energy \citep[][]{einav2007a}\\
			$M_0$ & 1.65 & 1.02 & -- & critical state friction coefficient\\
			$\gamma$ & 0.95 & 0.15 & -- & dilation parameter \citep[][]{tengattini2016}\\
			$\phi_l$ & 0.80 & 0.31 & -- & lower porosity at $B = 0$ \citep[][]{rubin2011}\\
			$\phi_u$ & 0.90 & 0.45 & -- & upper porosity at $B = 0$ \citep[][]{rubin2011}\\
			$l$ & 0.26 & 0.22 & -- & lower porosity power law parameter\\
			$u$ & 0.21 & 0.17 & -- & upper porosity power law parameter\\
			\hline\noalign{\smallskip}
			\multicolumn{5}{l}{Fluid EOS Parameters \citep[Water;][]{brundage2013}}\\
			\hline\noalign{\smallskip}
			$\eta_0$ & -- & 8.9$\times 10^{-4}$ & Pa$\cdot$s & dynamic fluid viscosity\\
			$\rho_{f0}$ & -- & 998 & kg/m$^3$ & reference fluid density\\
			$c_{vf}$ & -- & 3690 & J/kg$\cdot$K & fluid specific heat capacity\\
			$E_0$ & -- & 7$\times10^6$ & J/kg & reference internal energy\\
			$a_f$ & -- & 0.7 & -- & Tillotson EOS parameter\\
			$b_f$ & -- & 0.15 & -- & Tillotson EOS parameter\\
			$A_f$ & -- & 2.18$\times 10^9$ & Pa & Tillotson EOS parameter\\
			$B_f$ & -- & 1.325$\times 10^{10}$ & Pa & Tillotson EOS parameter\\
			$\alpha_f$ & -- & 10 & -- & Tillotson EOS parameter\\
			$\beta_f$ & -- & 5 & -- & Tillotson EOS parameter\\
			$\rho_\text{IV}$ & -- & 958 & kg/m$^3$ & density of incipient vaporization\\
			$E_\text{IV}$ & -- & 4.2$\times10^5$ & J/kg & energy of incipient vaporization\\
			$E_\text{CV}$ & -- & 2.5$\times10^6$ & J/kg & energy of complete vaporization\\
			$P_\text{cav}$ & -- & -2.5$\times10^7$ & Pa & cavitation pressure \citep[][]{herbert2006}\\
			\hline\noalign{\smallskip}
			\multicolumn{5}{l}{Inter-phase Drag Parameters}\\
			\hline\noalign{\smallskip}
			$d_{50}$ & -- & 3$\times10^{-4}$ & m & characteristic grain size\\
			\hline\noalign{\smallskip}
		\end{tabular}
	\end{table}
	
	The remaining constitutive equations for the interstitial fluid are based on a dissipative, energy-density formulation --- i.e., $\psi_f = \hat{\psi}_f(\rho_f, T_f)$ with $\boldsymbol{\tau}_f:\boldsymbol{D}_f \geq 0$ and $\boldsymbol{f}_d \cdot (\boldsymbol{v}_s - \boldsymbol{v}_f) \geq 0$. These constitutive equations incorporate analytical models for the compression of the pore fluid and its viscous interaction with surrounding particles, which are primarily based on available models from the literature. The pore fluid pressure $p_f$, the viscous shear stress $\boldsymbol{\tau}_f$, and the drag force $\boldsymbol{f}_d$ are characterized for water using the model parameters presented in Table \ref{tab:parameters}.
	
	\subsection{Example: Uniaxial Strain Compression of Ottawa Sand}
	\label{sec:example_uniaxial}
\begin{figure}[!h]
		\centering
		\includegraphics[width=0.32\linewidth]{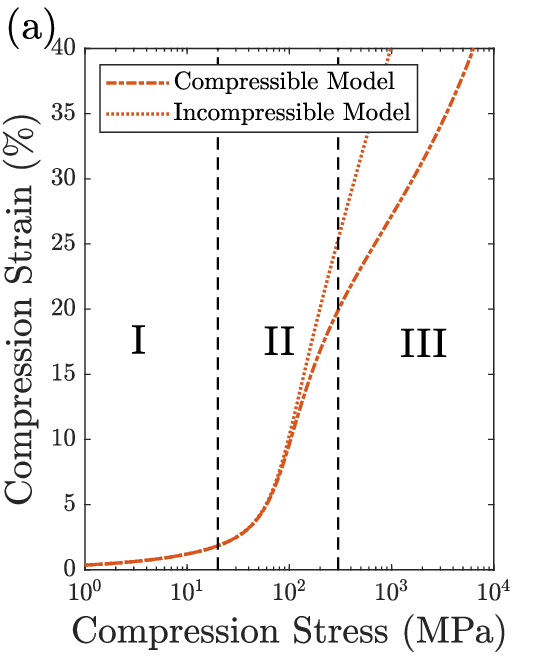}
		\includegraphics[width=0.32\linewidth]{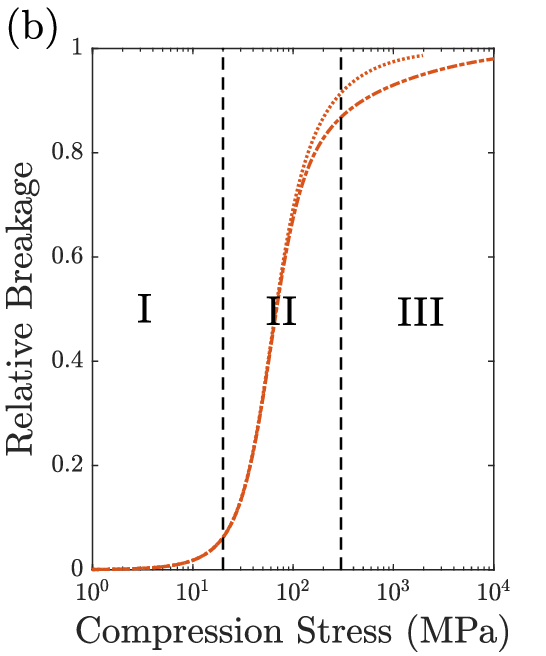}
		\includegraphics[width=0.32\linewidth]{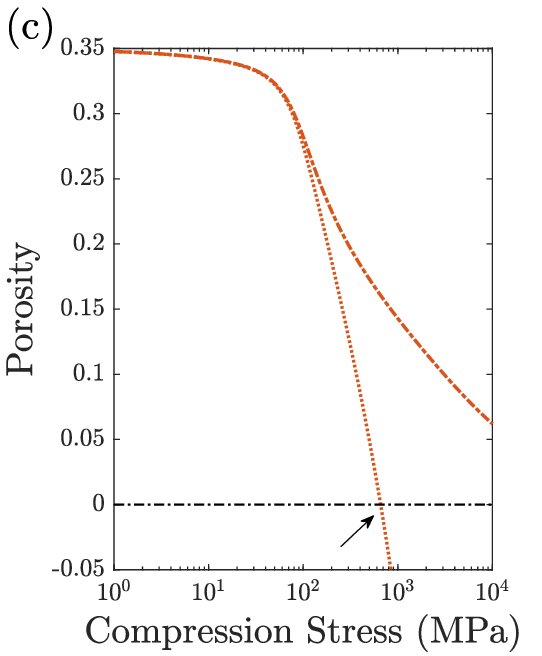}
		\caption[]{
			Example model predictions for uniaxial compression of an initially monodisperse Ottawa sand sample ($\phi_{s0} = 0.65$, $B_0 = 0.0$). (a) Comparison of predicted stress-strain response for incompressible and complete model implementations; black-dashed lines separate three canonical compaction stages from \citet{reed1995}. (b) Comparison of predicted stress-breakage response for incompressible and complete model implementations. (c) Comparison of predicted stress-porosity response for both model implementations; black-dashed line denotes theoretical minimum value. The incompressible model predicts a non-physical porosity above 700 MPa.
		}
		\label{fig:example_uniaxial}
	\end{figure}

	The basic capabilities of the model are demonstrated through an example compression test on dry Ottawa sand using the model parameters from Table \ref{tab:parameters}. Here, an initially mono-disperse sample of Ottawa sand ($d_{50} = 300$ $\mu$m, $\rho_0 = 2650$ kg/m$^3$, $\phi_s = 0.65$, $B_0 = 0.0$) is compressed along one axis while the other two axes remain fixed --- i.e., uniaxial strain compression, a condition that is typically attained in plate impact experiments. 
	For this example, we consider two possible implementations of the constitutive model: a \emph{compressible} version and an \emph{incompressible} version. The compressible version is the complete model, including both strain-energy contributions, $\psi_c$ and $\psi_g$ from \eqref{eqn:helmholtz_free_energy}. The incompressible version is a simplified implementation which only considers the contribution of $\psi_c$ to the effective granular stress $\boldsymbol{\sigma}_s$.
	The predicted material response to this loading condition is shown in Figure \ref{fig:example_uniaxial}, and highlights important characteristic behaviors of the model.
	
	Figure \ref{fig:example_uniaxial}a,b show that the model exhibits three distinct stages of compaction, which are analogous to the canonical stages of powder compaction described in \citet{reed1995}. Stage I is the low-pressure stage, dominated by granular rearrangement and the growth of contact stresses (see Figures \ref{fig:strain_energy}a and \ref{fig:example_uniaxial}a). In this stage, the model predicts almost no evolution of the relative breakage, $B$, or compaction of pore space (Figure \ref{fig:example_uniaxial}c). Stage II is the compaction stage, dominated by particle fracture, fragmentation, and pore collapse (Figure \ref{fig:example_uniaxial}b,c). Almost all of the particle fracture predicted by the model occurs in this stage. Finally, stage III is the fully compacted stage, dominated by elastic compression of the constituent solid (Figures \ref{fig:strain_energy}b and \ref{fig:example_uniaxial}a,b).
	
	In this final stage, we begin to see important differences between the compressible and incompressible model implementations.
	Below 100 MPa, there is negligible difference between the predictions of these two models, and the individual sand grains may be reasonably modeled as elastically incompressible --- i.e., 
	$\rho_s = \rho_0$ and $p^* = 0$.
	However, at higher confining stresses, the model predictions begin to deviate, and the simplified incompressible implementation becomes non-physical (Figure \ref{fig:example_uniaxial}c).
	Under these conditions the full compressible model should be used --- i.e., 
	$\rho_s = \rho_0 + \rho_0\ \hat{\alpha}(\phi_s)(J^{e-1} - 1)$ and $p^* \geq 0$ --- to avoid erroneous predictions of available pore space between individual particles.
	
	Although the uniaxial strain compression curves presented in this section highlight key features of the model, they of course do not capture the full range of potential applications of the model. In Section \ref{sec:results}, we implement this constitutive model in the full system of governing equations from Section \ref{sec:governing_equations} to predict the response of fluid-saturated granular materials along multi-stage loading paths and under the complex, dynamic conditions of impact.
	
	\section{Materials and Methods}
	\label{sec:methods}
	Laboratory data for two sands (Dog's Bay and Ottawa) are considered to calibrate {and validate} this constitutive model. Dog's Bay sand is a weak, biogenic carbonate sand from western Ireland that consists of foraminifera and mollusc shells \citep[][]{coop1990}, and Ottawa sand is a quartz-based sand mined from deposits located in Ottawa, Illinois \citep[][]{erdogan2017}. The model parameters that characterize these two sands are presented in Table \ref{tab:parameters} and calibrated using the procedure described in {\ref{sec:appendix_fitting}}.
	
	The primary data sources for Dog's Bay sand are the isotropic compression and bender element tests performed in \citet{jovicic1997}, which are used to calibrate $p_r$, $\bar{K}$, and $\bar{G}$; the compression tests performed in \citet{coop1990}, used to calibrate $E_c$; and the multi-stage triaxial tests performed in \citet{bandini2011}, used to calibrate $M_0$ and $\gamma$. Calibration of $\theta$, $\phi_u$, $\phi_l$, $u$, and $l$ is based on previous modeling work from \citet{tengattini2016}. { There is insufficient data to calibrate model parameters for Dog's Bay sand in the high-pressure, low-porosity regime --- namely, $b$, $\rho_0$, $c_v$, $\Gamma_0$, $C_0$, and $S_0$. Future uniaxial strain compression experiments or plate impact experiments could be used for this purpose.
		
		As discussed in Section \ref{sec:example_uniaxial}, the individual sand grains may be modeled as elastically incompressible for pressures up to 100 MPa. Although the model parameter $b$ could not be calibrated for Dog's Bay sand, we apply the model to problems within the domain of calibration and assume that $\rho_s = \rho_0$.
		Several triaxial loading experiments reported in \citet{bandini2011} for Dog's Bay sand are used to validate the model in this low-pressure, high-porosity regime. (Note that select data points from these experiments were also used to calibrate $M_0$ and $\gamma$; here the complete loading path is used for model validation.)}
	
	
	The primary data sources for Ottawa sand are the grading tests performed in \citet{youd1973}, which are used to calibrate $\phi_u$, $\phi_l$, $u$, and $l$; the bender element tests performed in \citet{robertson1995}, used to calibrate $p_r$ and $\bar{G}$; the uniaxial compression tests performed in \citet{kuwik2022}, used to calibrate $\bar{K}$, $E_c$, and $b$; the ring shear tests performed in \citet{wijewickreme1986} and \citet{dakoulas1992}, used to calibrate $M_0$; the ring shear tests performed in \citet{sadrekarimi2011}, used to calibrate $\gamma$; and the quartz crystal characterization tests performed in \citet{wackerle1962}, \citet{boehler1979}, \citet{lyzenga1983}, and \citet{heyliger2003}, used to calibrate $\rho_0$, $c_v$, $\Gamma_0$, $C_0$, and $S_0$.
	
	To validate the application of this model to {Ottawa sand} in conditions relevant for high-velocity impact events, additional experimental data are considered --- including original data collected in this study. { The uniaxial compression experiments reported in \citet{kuwik2022} are used to calibrate the stress--strain response of the model and validate the stress--breakage ($B$) behavior. The triaxial loading experiments reported in \citet{shahin2022} are used to validate the model along a two-stage loading path in the high-pressure, low-porosity regime. Additional projectile penetration and cratering data were collected for several high-velocity impact experiments performed on Ottawa sand at the Hopkins Extreme Materials Institute HyFIRE (Hypervelocity Facility for Impact Research Experiments) facility.}
	
	
	
	
	\subsection{Experimental Methods}
	\label{sec:experimental_methods}
\begin{figure}[!h]
		\centering
		\includegraphics[width=0.92\linewidth]{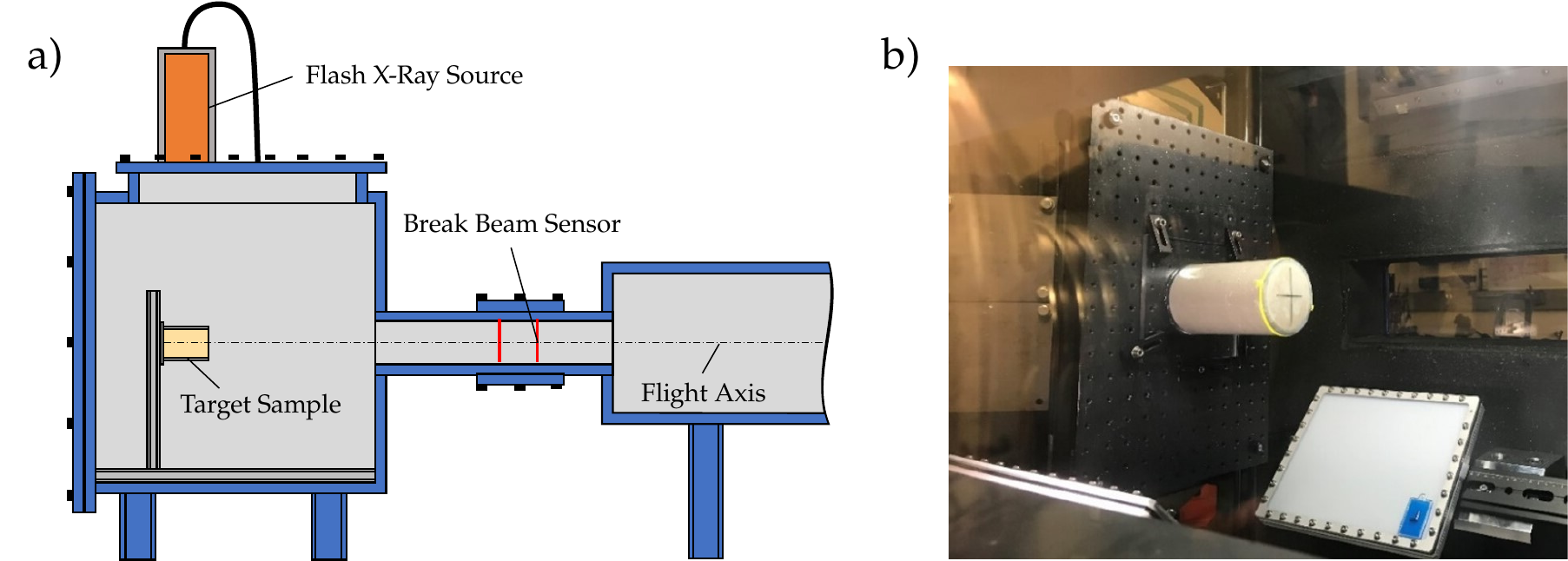}
		\caption[]{
			(a) Illustration of the experimental configuration for the tests listed in Table \ref{tab:shots} at the Hopkins Extreme Materials Institute (HEMI) Hypervelocity Facility for Impact Research Experiments (HyFIRE). (b) Digital photograph of Ottawa sand target sample mounted in target chamber; X-ray image plates are shown mounted in casings on the lower left and right of image.
		}
		\label{fig:hyfire}
	\end{figure}

	The high-velocity impact experiments reported in this study were performed using a two-stage light gas gun in the HyFIRE facility (shown in Figure \ref{fig:hyfire}). Target specimens of Ottawa sand \citep[$d_{50} = 300$ $\mu$m, $\rho_0 = 2650$ kg/m$^3$, $\phi_s = 0.37$;][]{shahin2022} were prepared in 7.62 cm diameter polycarbonate tubes measuring 15.24 cm in length. Dry samples were prepared with 1.16--1.17 kg of material and sealed with Parafilm$^\text{TM}$ M wax film. Saturated samples were prepared with 1.16--1.17 kg of material added to 0.26--0.27 liters of water and sealed with Parafilm$^\text{TM}$ M wax film. To remove trapped gases, the saturated samples were also vibrated for 30 minutes and allowed to settle for 48 hours before the experiment.
	Each specimen was mounted horizontally in the target chamber and impacted by a 3 mm, 440C stainless steel sphere at between 1.16 and 1.70 km/s (average 1.28 km/s). A summary of the impact test conditions is presented in Table \ref{tab:shots}, {including experiments previously reported in \citet{kuwik2023}.}
	
	\begin{table}[h!]
		\centering\footnotesize
		\caption{List of impact tests for Ottawa sand performed at the Hopkins Extreme Materials Institute.}
		\label{tab:shots}
		\begin{tabular}{l l c l l l}
			\noalign{\smallskip}
			\hline\noalign{\smallskip}
			Test \# & Condition & Chamber Pressure (Torr) & \multicolumn{2}{l}{Right, Left X-Ray Flash ($\mu$s)} & Velocity (km/s)\\
			\hline\noalign{\smallskip}
			1$^*$ & Dry & 100 & 50, & 50 & 1.22\\
			2$^*$ & Dry & 100 & 150, & 150 & 1.32\\
			3$^*$ & Dry & 500 & 50, & 50 & 1.21\\
			4$^*$ & Dry & 760 & 89, & 89 & 1.17\\
			5$^*$ & Saturated & 760 & 54, & 54 & 1.25\\
			6$^*$ & Saturated & 760 & 90, & 90 & 1.70\\
			7$^*$ & Saturated & 760 & 150, & 150 & 1.16\\
			8$^\dagger$ & Saturated & 500 & 54, & 154 & 1.26\\
			9$^\dagger$ & Saturated & 500 & 23, & 98 & 1.28\\
			10$^\dagger$ & Dry & 500 & 28, & 103 & 1.27\\
			11$^\dagger$ & Dry & 500 & 29, & 154 & 1.29\\
			\hline\noalign{\smallskip}
		\end{tabular}
		\\{$^*$Samples interspersed with 2mm lead spheres to visualize deformations \citep[see][]{kuwik2023}.}
		\\$^\dagger$Asynchronous X-ray flashes used for multiple depth and crater measurements.
	\end{table}
	
	During the impact event, synchronous and asynchronous flash X-ray images were captured using a two-channel, dual-output X-ray system from L3 Communications, with images captured on Carestream Industrex digital imaging plates. These plates were scanned using a ScanX Discover HC scanner. Simultaneously, high-speed imaging of the impact was captured using a Shimadzu HPV-X2 high-speed video camera, which was synchronized with an AMOtronics Saturn System test sequencer and a Physics Applications International break beam sensor. An illustration of the experimental configuration within HyFIRE is shown in Figure \ref{fig:hyfire}.
	
	\subsection{Numerical Methods}
	\label{sec:numerical_methods}
	The model predictions presented in this work are numerical approximations of solutions to the system of governing equations in \eqref{eqn:mass_conservation}--\eqref{eqn:energy_conservation}. For the uniaxial and triaxial tests simulated in Section \ref{sec:example_uniaxial} and Sections \ref{sec:dogs_bay_triaxial}--\ref{sec:ottawa_triaxial}, isothermal, quasi-steady solutions are computed for single material points using a MATLAB implementation of the stress update algorithm discussed in {\ref{sec:appendix_algorithm}}. For the dynamic simulations presented in Section \ref{sec:ottawa_impact}, we use a custom implementation of the material point method (MPM) based on the numerical algorithm previously published in \citet{baumgarten2019a}. 

	\section{Results}
	\label{sec:results}
	In this section, we use the numerical methods summarized in Section \ref{sec:numerical_methods} to validate the model and study its application in conditions relevant to high-velocity impact events. Using the model parameters presented in Table \ref{tab:parameters}, we simulate the triaxial compression and shearing of Dog's Bay sand from \citet{bandini2011} as well as the uniaixial compression and triaxial shearing of Ottawa sand from \citet{kuwik2022} and \citet{shahin2022}, respectively. Additionally, we evaluate the predictive capabilities of the model by simulating both {projectile dynamics and crater development} during 1.3 km/s impacts into dry and water-saturated Ottawa sand. 
	
	\subsection{Triaxial Loading of Dog's Bay Sand}
	\label{sec:dogs_bay_triaxial}
	In \citet{bandini2011}, a series of triaxial tests are performed on Dog's Bay sand ($d_{50}=200$ $\mu$m, $\rho_0 = 2710$ kg/m$^3$, $\phi_s = 0.35$--0.39) to study the the combined crushing and critical state response of the material. During these tests, samples were isotropically compressed to pressures between 500 kPa and 4 MPa, leading to significant inelastic compaction of the bulk material. After this initial compression, samples were sheared, unloaded, reconfigured, and sheared a second time. This multi-stage loading path allows for a unique {evaluation} of model predictions: starting from the same initial state, we assess how well the model follows experimental observations during the entire loading cycle. Although we fit the parameters $M_0$ and $\gamma$ to data presented in \citet{bandini2011}, no additional fitting is performed to match the multi-stage loading path reported here.
	
\begin{figure}[!h]
		\centering
		\includegraphics[width=0.49\linewidth]{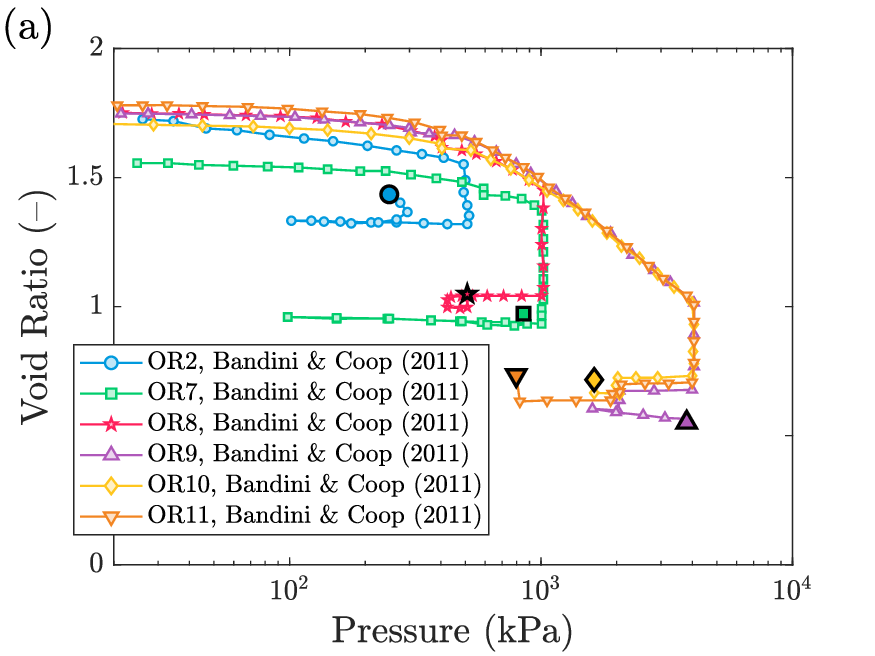}
		\includegraphics[width=0.49\linewidth]{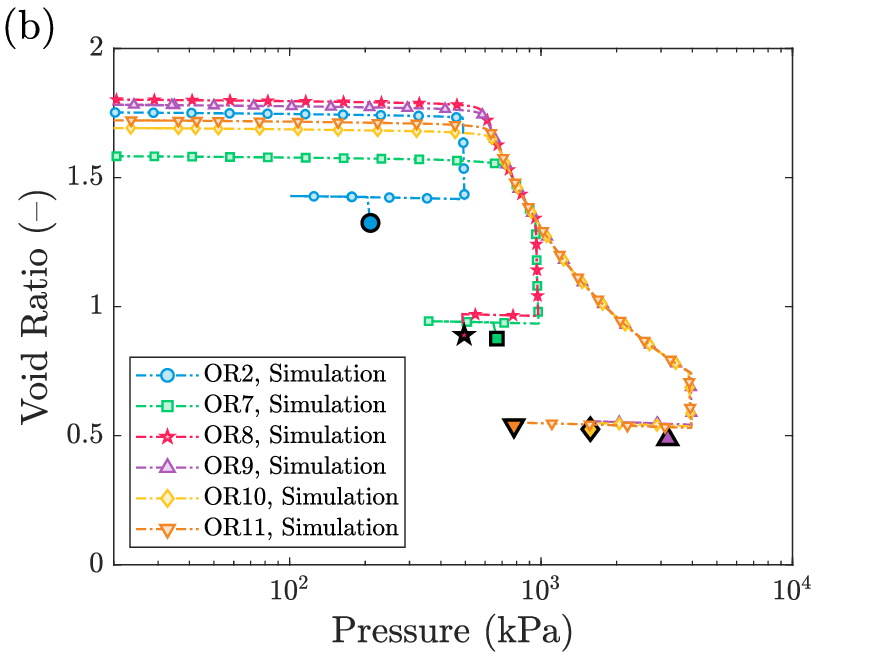}
		\newline \vskip4pt
		\includegraphics[width=0.49\linewidth]{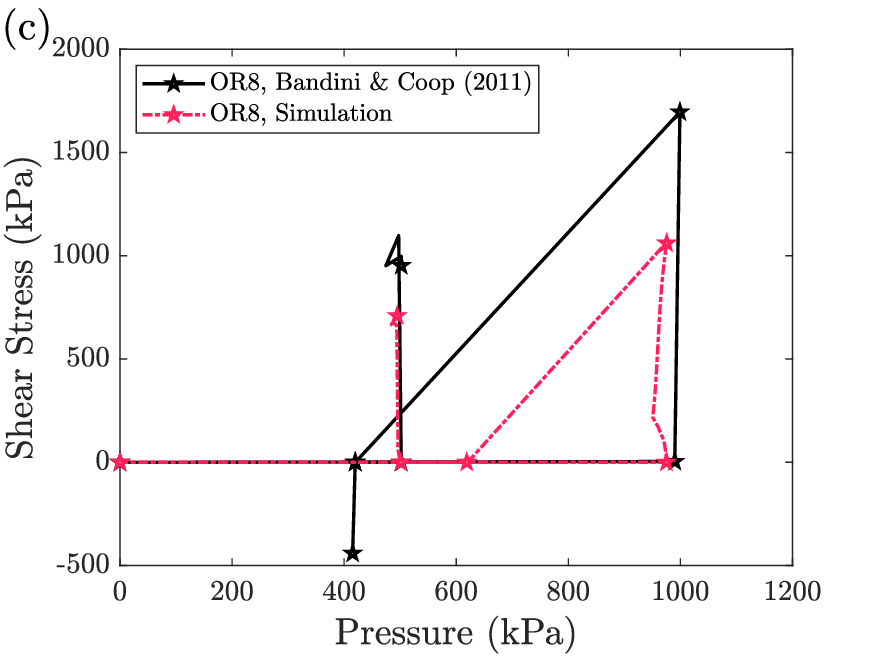}
		\caption[]{
			(a) Experimental compression curves for the Dog's Bay sand tests listed in Table \ref{tab:dogs_bay_triaxial} from \citet{bandini2011}. (b) Simulated compression curves following experimental loading paths. 
			(c) Comparison of experimental (black, solid line) and simulated (red, dashed line) stress path for test OR8 --- shearing stages are strain controlled at a fixed pressure. {The measured and simulated values of relative breakage ($B$) at the end of each test in (a),(b) are listed in Table \ref{tab:dogs_bay_breakage}.}
		}
		\label{fig:dogs_bay_triaxial}
	\end{figure}

	
	In this work, we are particularly interested in six of the drained tests reported in \citet{bandini2011}: OR2, OR7, OR8, OR9, OR10, and OR11, which are summarized in Table \ref{tab:dogs_bay_triaxial}. Each of these tests involves an initial compression followed by two shearing stages: one at the initial confining pressure and a second at a reduced confining stress. This multi-stage loading path is used to highlight the general applicability {and robustness} of the model, which is able to closely follow the compression, shearing, and volumetric changes observed in the experimental data (see Figure \ref{fig:dogs_bay_triaxial}). 
	
	\begin{table}[!h]
		\centering\footnotesize
		\caption{List of simulated, triaxial compression tests for Dog's Bay sand, following \citet{bandini2011}.}
		\label{tab:dogs_bay_triaxial}
		\begin{tabular}{l p{1.8cm} p{2.3cm} p{2.6cm} p{2.4cm} p{2.6cm}}
			\noalign{\smallskip}
			\hline\noalign{\smallskip}
			Test & Initial Void Ratio (--) & Initial Loading (kPa) & Axial Strain (\%), First Shearing & Second Loading (kPa) & Axial Strain (\%), Second Shearing\\
			\hline\noalign{\smallskip}
			OR2 & 1.758 & 500 & 21.07 & 100$^\dagger$ & 19.71\\
			OR7 & 1.591 & 1000 & 29.19 & 350$^\dagger$ & 14.75\\
			OR8 & 1.807 & 1000 & 25.07 & 500 & 21.79\\
			OR9 & 1.787 & 4000 & 19.56 & 1600$^\dagger$ & 19.49\\
			OR10 & 1.695 & 4000 & 22.28 & 1600 & 15.15\\
			OR11 & 1.737 & 4000 & 22.59 & 800 & 14.29\\
			\hline\noalign{\smallskip}
		\end{tabular}
		\newline
		$^\dagger$ Second shearing performed at fixed radial stress.
	\end{table}

	\begin{table}[!h]
		
		\centering\footnotesize
		\caption{Comparison of the breakage, $B$, measured before the triaxial compression tests reported in \citet{bandini2011}, $B_0$; after the completion of these tests, $B_\text{exp}$; and as simulated using the proposed model.}
		\label{tab:dogs_bay_breakage}
		\begin{tabular}{l l l l}
			\noalign{\smallskip}
			\hline\noalign{\smallskip}
			Test & Initial Breakage, $B_0$ & Final Breakage, $B_\text{exp}$ & Final Breakage, Model\\
			\hline\noalign{\smallskip}
			OR2 & 0.52 & 0.61 & 0.85 \\
			OR7 & 0.52 & 0.74 & 0.94 \\
			OR8 & 0.52 & 0.75 & 0.94 \\
			OR9 & 0.52 & 0.88 & 0.99 \\
			OR10 & 0.52 & 0.68 & 0.98 \\
			OR11 & 0.52 & 0.89 & 0.98 \\
			\hline\noalign{\smallskip}
		\end{tabular}
	\end{table}
	
	To simulate these six experiments, we implement the stress update algorithm described in {\ref{sec:appendix_algorithm}} and model a single material point under drained loading conditions --- i.e., without confined, interstitial fluid. The deformation rate $\boldsymbol{L}_s$ is assigned to follow the loading histories summarized in Table \ref{tab:dogs_bay_triaxial}, using a numerically computed stress-gradient to satisfy the constant pressure and constant radial stress boundary conditions. 
	In both the experiments and the simulations, the only differences between the six tests are the initial void ratio ($e = \phi_f / \phi_s$), the confining stresses, and the applied deformations. All six samples have the the same initial particle size distribution ($d_{50} = 200$ $\mu$m, $B_0 = 0.52$) and are simulated using the same model parameters. 
	%
	%
	
	In Figure \ref{fig:dogs_bay_triaxial}a,b, the compaction curves reported in \citet{bandini2011} are compared with the predictions of the constitutive model proposed in this work. The void ratios measured at the end of each test --- highlighted by enlarged markers in both plots --- have strikingly similar values across the  range of loading paths considered. Importantly, these final points are not fit: they are predicted using the parameters calibrated in {\ref{sec:appendix_fitting}}, following the loading paths in Table \ref{tab:dogs_bay_triaxial}. The observed similarity along these types of loading paths is unmatched by similar models proposed in the literature.
	
	
	
	In Figure \ref{fig:dogs_bay_triaxial}c, we compare the predicted pressure--shear-stress path simulated for test OR8 with the experimentally measured path reported in \citet{bandini2011}. Here, the shear stress ($q = \sqrt{(3/2) \boldsymbol{\sigma}_{s0}:\boldsymbol{\sigma}_{s0}}$) develops as the samples are sheared at constant pressure to the same final axial strain (shown in Table \ref{tab:dogs_bay_triaxial}). Although the compression curves shown in Figure \ref{fig:dogs_bay_triaxial}a,b are remarkably similar, the simulated shear-stress response generally under-predicts experimental measurements. 
	Additionally, as shown in Table \ref{tab:dogs_bay_breakage}, the relative breakage ($B$) predicted at the end of each simulation is significantly larger than the relative breakage measured at the end of the experiments ($B_\text{exp}$). This trend has been observed with similar breakage mechanics models \citep[e.g., see][]{kuwik2022} {and suggests that the assumed coupling angle $\omega$ in \eqref{eqn:y1_rates} and \eqref{eqn:friction_and_dilation} is generally too small --- over-predicting the evolution of $B$.} These differences are discussed further in Section \ref{sec:discussion}. 

	\subsection{Uniaxial Loading of Ottawa Sand}
	\label{sec:ottawa_uniaxial}
	We continue the analysis of the model by considering the behavior of a second material system: Ottawa sand. In \citet{kuwik2022}, a series of uniaxial compression tests --- or \emph{oedometer} tests --- are performed on Ottawa sand ($d_{50} = 300$ $\mu$m, $\rho_0 = 2650$ kg/m$^3$, $\phi_s$ = 0.63, $B_0 = 0.14$) to study the stress--strain--breakage response of the material. During these tests, samples were prepared in a thick-walled test cell and subjected to axial compression up to 30\%, leading to significant inelastic compaction of the bulk material and axial stresses measured up to 1.2 GPa. This series of tests allows us to evaluate the high-pressure predictions of the model in a simple geometry: starting from a stress-free state through to a fully compressed, pulverized sediment.
	
\begin{figure}[!h]
		\centering
		\includegraphics[width=0.49\linewidth]{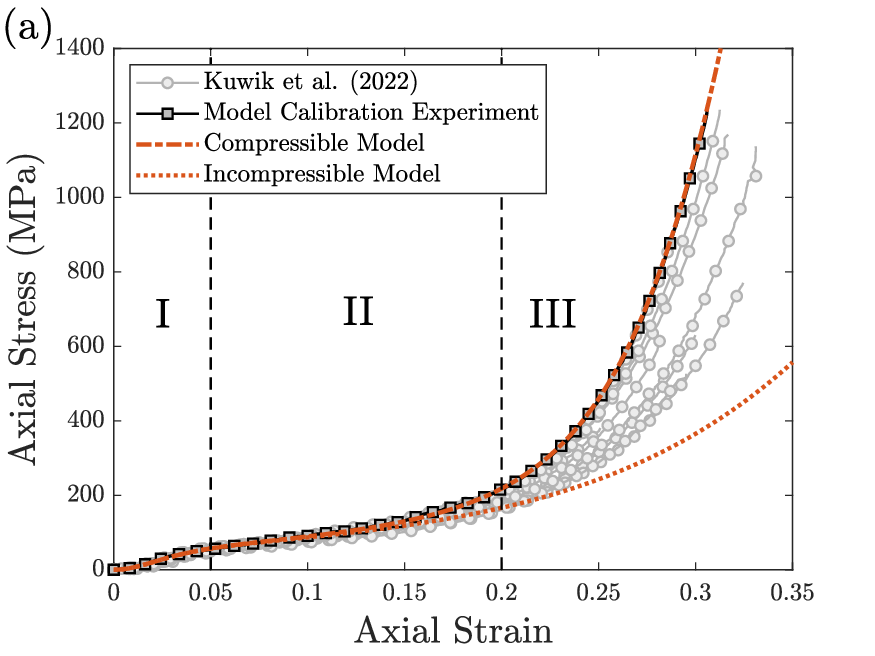}
		\includegraphics[width=0.49\linewidth]{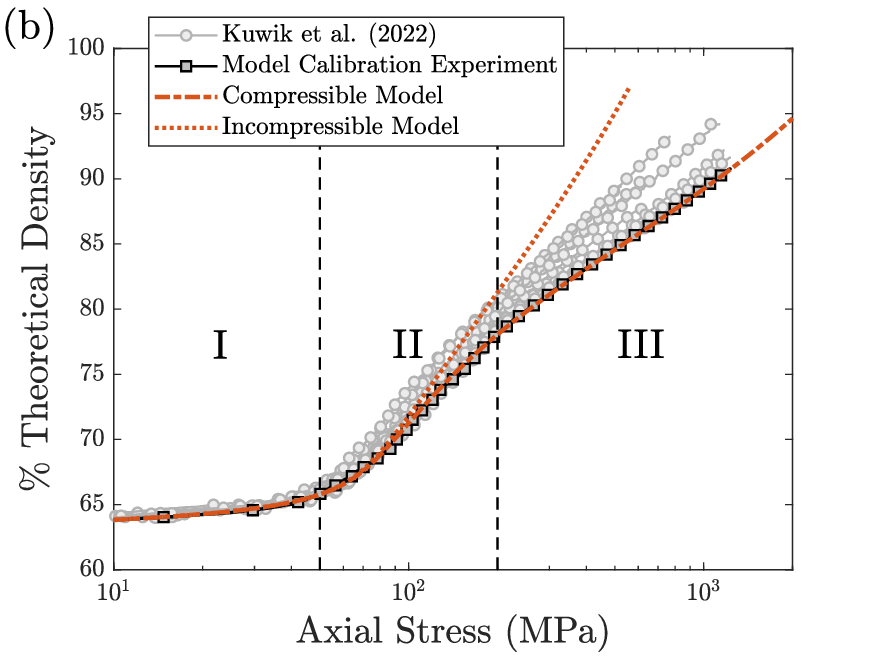}
		\newline \vskip4pt
		\includegraphics[width=0.49\linewidth]{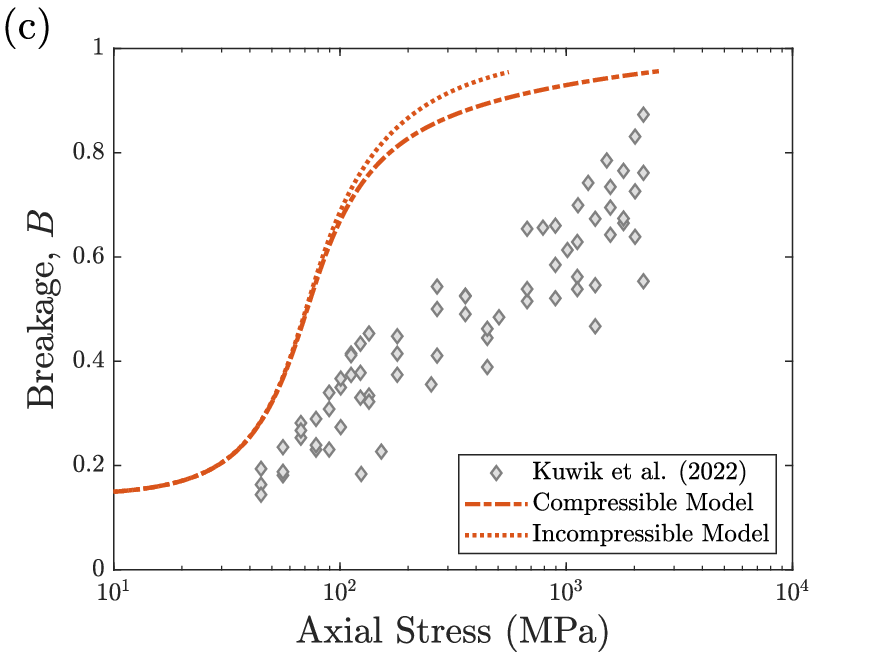}
		\caption[]{
			Comparison of predicted stress--strain--breakage response of Ottawa sand with uniaxial compression (oedometer) data reported in \citet{kuwik2022}. (a) Plot of predicted axial stress against engineering strain --- in compression --- for compressible (red, dashed line) and incompressible (red, dotted line) models, overlaid on experimental measurements. (b) Plot of predicted theoretical density ($\bar{\rho}_s/\rho_0$) against compressive axial stress highlighting three canonical stages of compaction \citep[][]{reed1995}. (c) Comparison of simulated and experimental relative breakage $B$ along uniaxial compression path.
		}
		\label{fig:ottawa_uniaxial}
	\end{figure}

	The compression data reported in \citet{kuwik2022} is some of first to highlight the three canonical stages of powder compaction \citep[Figures \ref{fig:example_uniaxial}a,b and \ref{fig:ottawa_uniaxial}a,b; see][]{reed1995} for Ottawa sand and considers strain rates from 10$^{-3}$ to 10$^{3}$ s$^{-1}$ and pressures between 10$^6$ to 10$^9$ Pa. In this work, we are particularly interested in the data collected on the MTS Criterion 43, which have complete stress--strain histories. 
	%
	From this set of experiments, we highlight the stress--strain data collected at 10$^{-1}$ s$^{-1}$ (shown in Figure \ref{fig:ottawa_uniaxial}a,b), which is used for partial model calibration. This curve is chosen to fit $\bar{K}$, $E_c$, and $b$ following the procedure discussed in {\ref{sec:appendix_fitting}}. The remaining model parameters in Table \ref{tab:parameters} are calibrated from other data sources, as listed in Section \ref{sec:methods}. 
	
	
	To simulate this experiment, we implement the stress update algorithm described in {\ref{sec:appendix_algorithm}} and model a single material point under uniaxial loading conditions --- i.e., $\boldsymbol{L}_s = -\dot{\epsilon}_{11}\ \boldsymbol{e}_1 \otimes \boldsymbol{e}_1$ with $\dot{\epsilon}_{11}$ the compressive strain-rate and $\boldsymbol{e}_1$ the unit vector aligned with the loading axis.
	Here, we again consider two possible implementations of the constitutive model: a compressible version and an incompressible version.
	%
	%
	%
	As discussed in Section \ref{sec:example_uniaxial}, below 100 MPa, the individual sand grains may be reasonably modeled as elastically incompressible. 
	However, the predictions of this simplified implementation are expected to be poor and potentially non-physical at higher pressures. Under these conditions the full compressible model should be used.
	
	In Figure \ref{fig:ottawa_uniaxial}a,b, the compaction curves reported in \citet{kuwik2022} are compared with the predictions of both the compressible and incompressible implementations described above. In both plots, the three canonical stages of compaction are labeled \citep[see][]{reed1995}. Stage I indicates the low-pressure stage, dominated by granular rearrangement and the growth of contact stresses (see Figure \ref{fig:strain_energy}a). Stage II indicates the compaction stage, dominated by particle fracture, fragmentation, and pore collapse. Finally, stage III indicates the compacted stage, dominated by elastic compression of the constituent solid (see Figure \ref{fig:strain_energy}b). In stages I and II, the predicted behavior of both model implementations closely matches the data reported in \citet{kuwik2022}; however, in stage III, the incompressible model begins to deviate, highlighting its limited application to pressures below 100 MPa.
	
	Additionally, in Figure \ref{fig:ottawa_uniaxial}c, we compare the relative breakage ($B$) predicted by both model implementations with the particle size distribution data reported in \citet{kuwik2022}. As in Section \ref{sec:dogs_bay_triaxial}, both models over-predict the measured relative breakage {while following a similar general trend. This is discussed further in Section \ref{sec:discussion} and suggests that the assumed coupling angle $\omega$ in \eqref{eqn:y1_rates} and \eqref{eqn:friction_and_dilation} should be investigated further.}
	

	\subsection{Triaxial Loading of Ottawa Sand}
	\label{sec:ottawa_triaxial}
	To validate the calibration of the model from the previous section, we analyze data from a set of experiments reported in \citet{shahin2022}.
	There, a series of triaxial compression and shearing tests are performed on Ottawa sand ($d_{50} = 175$--300 $\mu$m, $\rho_0 = 2650$ kg/m$^3$, $\phi_s = 0.65$--0.70) to study the stress--strain--dilation response of the material, including analysis of shear band formation. During these tests, samples were prepared in a high-pressure triaxial compression instrument, which is capable of applying several MPa pressures and shear stresses simultaneously. At the start of each test, samples are compressed isotropically to pressures between 10 and 45 MPa. The compressed samples are then sheared at fixed radial stresses to between 10\% and 16\% axial compression. This series of experiments allows us to evaluate the predictive capabilities of the model when no additional calibration of model parameters is performed.
	
	Here, we focus on the stress--strain--dilation data reported in \citet{shahin2022} for tests OS-10, OS-15, OS-20, OS-25, OS-30, OS-35, and OS-45, which are summarized in Table \ref{tab:ottawa_triaxial}. To simulate these seven experiments, we implement the stress update algorithm described in {\ref{sec:appendix_algorithm}} and model a single material point under drained loading conditions. The deformation rate $\boldsymbol{L}_s$ is assigned to follow the loading histories summarized in Table 2, using a numerically computed stress-gradient to satisfy the constant radial stress boundary conditions. In both the experiments and the simulations, the only differences between the seven tests are the initial porosity ($\phi_f$), the initial confining stress, and the initial particle size distribution ($d_{50}$ and $B_0$). All seven samples are simulated using the same model parameters.
	
	\begin{table}[!h]
		\centering\footnotesize
		\caption{List of simulated, triaxial compression tests for Ottawa sand, following \citet{shahin2022}.}
		\label{tab:ottawa_triaxial}
		\begin{tabular}{l l l l l}
			\noalign{\smallskip}
			\hline\noalign{\smallskip}
			Test & Initial Porosity (--) & Initial Breakage (--) & Initial $d_{50}$ ($\mu$m) & Initial Loading (MPa)\\
			\hline\noalign{\smallskip}
			OS-10 & 0.30 & 0.25 & 300 & 10 \\
			OS-15 & 0.35 & 0.125 & 175 & 15 \\
			OS-20 & 0.32 & 0.25 & 300 & 20 \\
			OS-25 & 0.34 & 0.125 & 175 & 25 \\
			OS-30 & 0.31 & 0.25 & 300 & 30 \\
			OS-35 & 0.34 & 0.125 & 175 & 35 \\
			OS-45 & 0.32 & 0.123 & 175 & 45 \\
			\hline\noalign{\smallskip}
		\end{tabular}
	\end{table}
	
\begin{figure}[!h]
		\centering
		\includegraphics[width=0.49\linewidth]{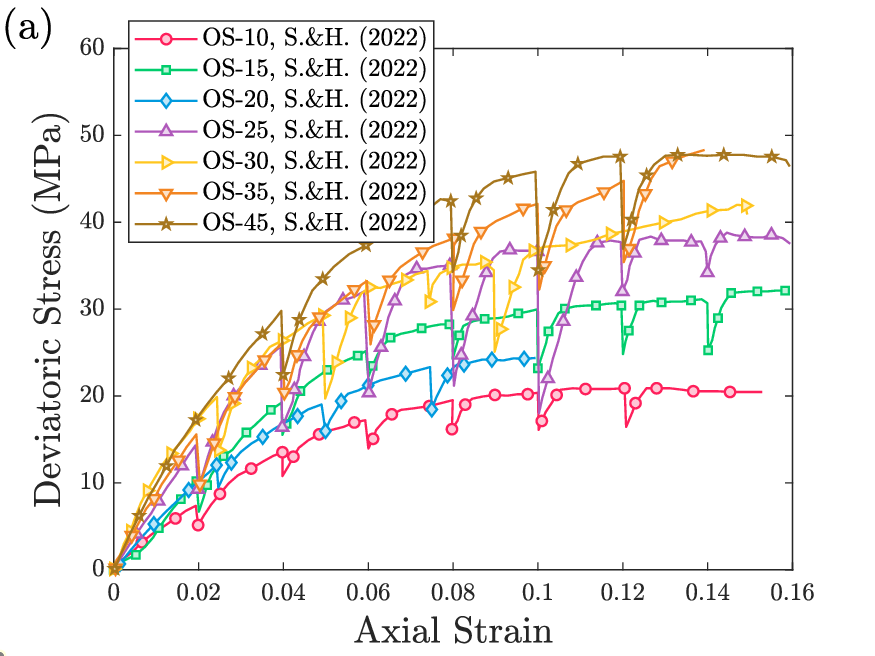}
		\includegraphics[width=0.49\linewidth]{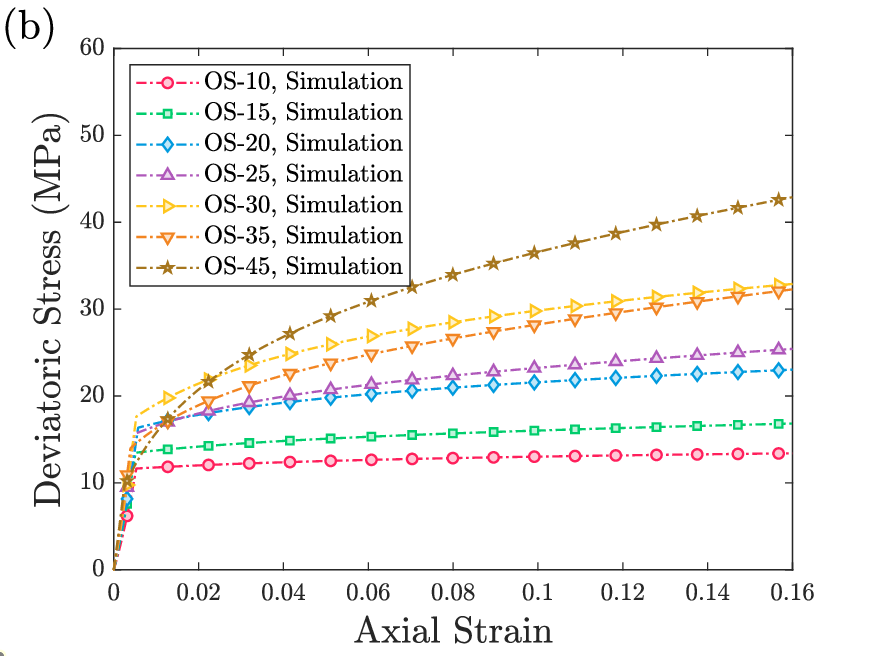}
		\includegraphics[width=0.49\linewidth]{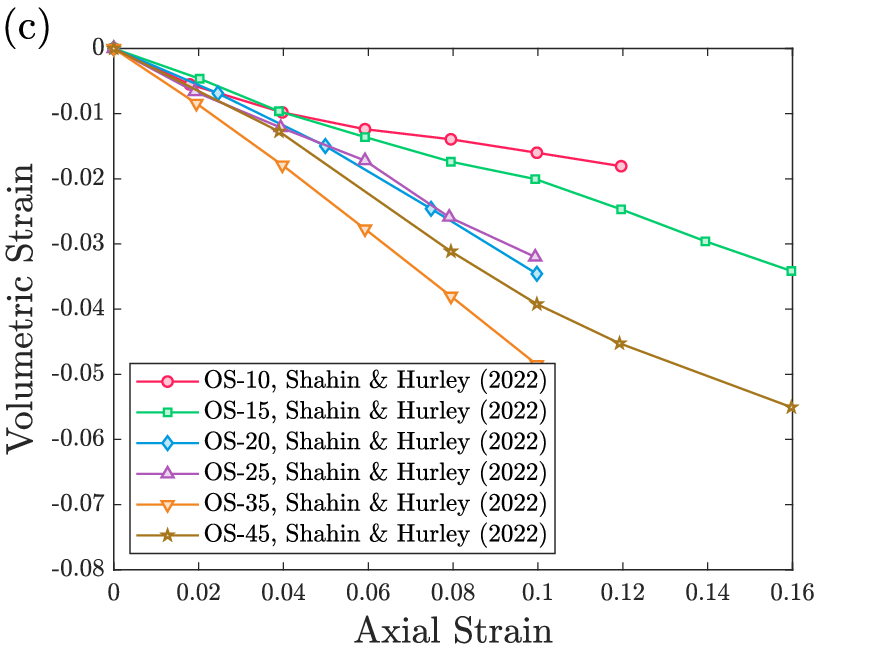}
		\includegraphics[width=0.49\linewidth]{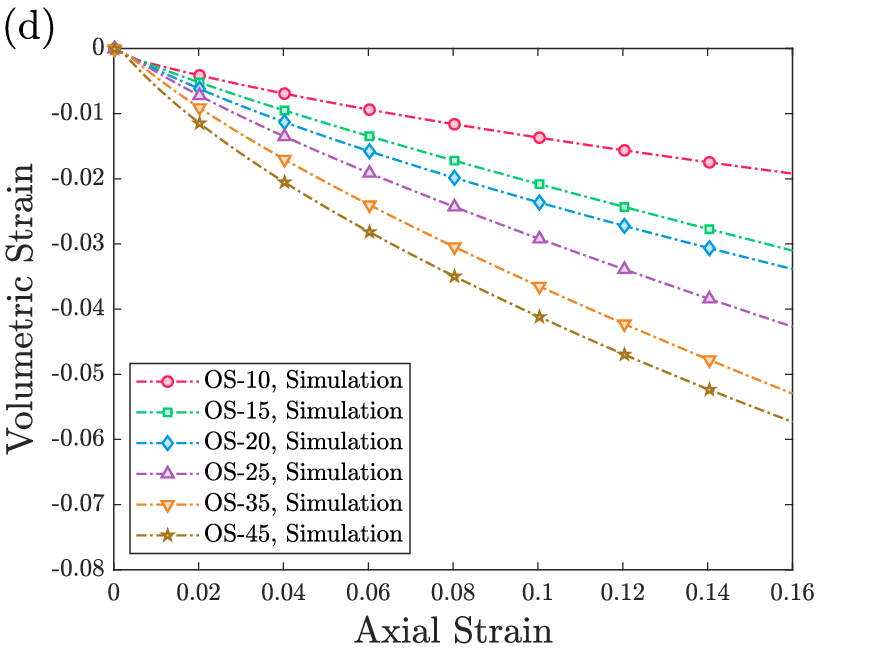}
		\caption[]{
			(a) Experimental deviatoric stress--strain curves for the Ottawa sand tests listed in Table \ref{tab:ottawa_triaxial} from \citet{shahin2022}. (b) Simulated deviatoric stress--strain curves for the same loading conditions. 
			(c) Experimental dilation curves comparing the volumetric strain to axial compression for tests reported in \citet{shahin2022}. (d) Corresponding simulated dilation curves comparing predicted volumetric strain during axial compression.
		}
		\label{fig:ottawa_triaxial}
	\end{figure}

	In Figure \ref{fig:ottawa_triaxial}a,b, the deviatoric stresses ($q = \sqrt{(3/2)\boldsymbol{\sigma}_{s0}:\boldsymbol{\sigma}_{s0}}$) reported in \citet{shahin2022} are compared with the predicted stress--strain response of the constitutive model. The stress drops observed in Figure \ref{fig:ottawa_triaxial}a are stress relaxation occurring during pauses in the experimental loading (for X-ray scanning of the samples) and are not features of the steady loading conditions simulated in Figure \ref{fig:ottawa_triaxial}b. Although the simulated stresses shown in Figure \ref{fig:ottawa_triaxial}b tend to  under-predict the measurements shown in Figure \ref{fig:ottawa_triaxial}a, similar strain-hardening and pressure-dependent yielding trends are observed (e.g., compare tests OS-10 and OS-45).
	

	Additionally, in Figure \ref{fig:ottawa_triaxial}c,d, we compare the measured volumetric strains reported in \citet{shahin2022} with the predicted volumetric strains determined from the simulations. Here, the model predictions are much closer to the experimentally measured values, with dilation angles ($\epsilon_v/\epsilon_s$) ranging from -0.15 to -0.48 for the data reported in \citet{shahin2022} and from -0.13 to -0.36 in the simulations reported here.
	
	
	\subsection{High-Velocity Impact into Ottawa Sand}
	\label{sec:ottawa_impact}
	We now consider the application of the model to the primary problem of interest in this work: prediction of projectile {dynamics and crater development} during high-velocity impact events.
	To validate these model predictions, we also report a series of tests performed on Ottawa sand ($d_{50} = 300$ $\mu$m, $\rho_0 = 2650$ kg/m$^3$, $\phi_s = 0.63$, $B_0 = 0.14$) that studies the influence of confined interstitial fluids on the cratering process and the penetration depth of high-speed projectiles. In this section, no data fitting or calibration is performed. All model predictions are determined from simulations that use the model parameters listed in Table \ref{tab:parameters}, which are calibrated following the procedure described in {\ref{sec:appendix_fitting}}.
	
	As discussed in Section \ref{sec:experimental_methods}, the experiments reported in this section were performed at the Hopkins Extreme Materials Institute (HEMI) using a two-stage light gas gun to launch 3 mm, 440C stainless steel spheres into dry and water-saturated Ottawa sand. 
	The complete set of tests performed in this work are listed in Table \ref{tab:shots} and use the experimental configuration shown in Figure \ref{fig:hyfire}. {\citep[See][for further details.]{kuwik2023}}

\begin{figure}[!h]
		\centering
		\includegraphics[width=1.0\linewidth]{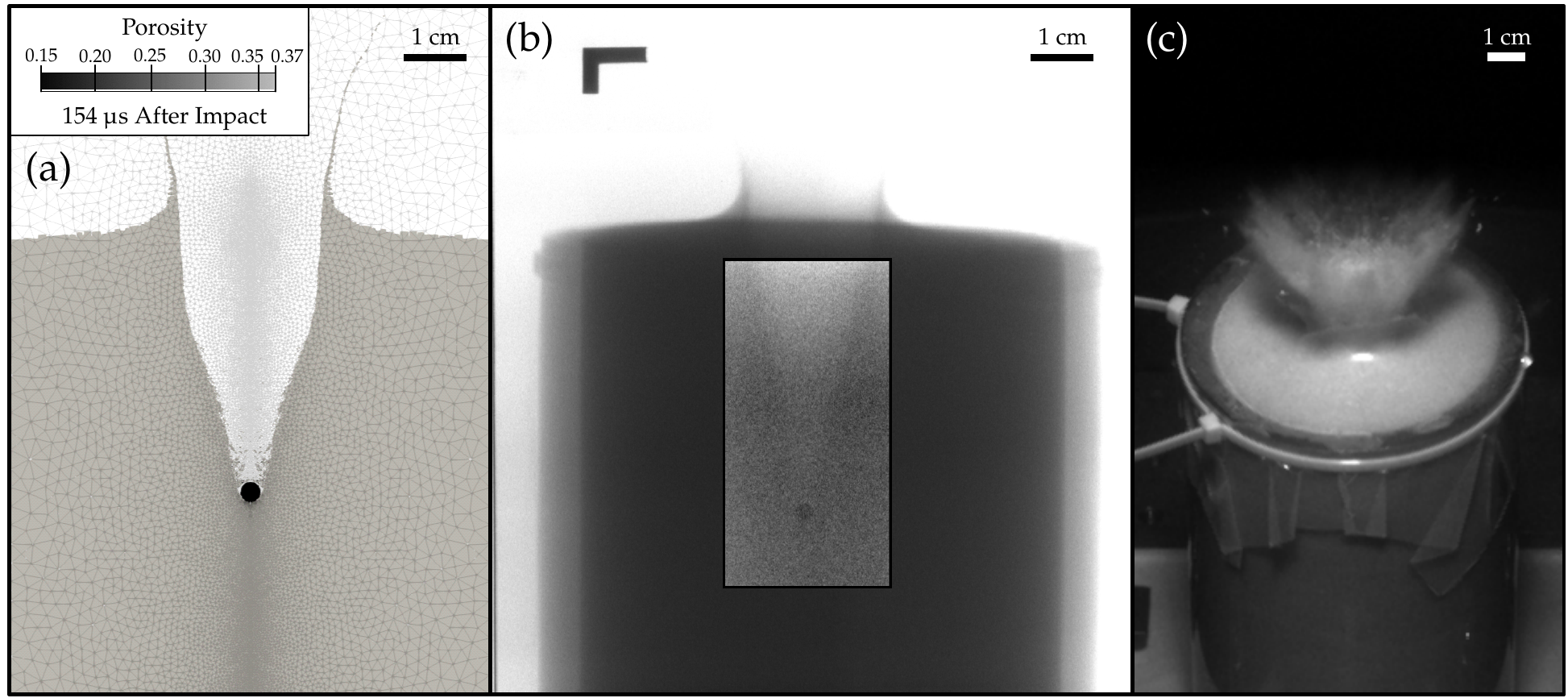}
		\caption[]{
			Comparison of high-velocity impact simulation with flash X-ray imaging and high-speed scattered light imaging of impact into water-saturated Ottawa sand. (a) Simulation snapshot taken 154 $\mu$s after impact using two-phase material point method \citep[MPM; see][]{baumgarten2019a}. (b) Scanned, in-situ X-ray image from test \#8 in Table \ref{tab:shots} captured 154 $\mu$s after impact (inset brightness and contrast adjusted using GNU Image Manipulation Program, GIMP; ``L'' indicates use of left-side X-ray flash). (c) High-speed video frame taken during test \#8 from Table \ref{tab:shots} captured 154 $\mu$s after impact. Note that the dispersion of the ejecta visible in (c) is greater than what is apparent in (a), highlighting a potential model limitation.
		}
		\label{fig:ottawa_impact}
	\end{figure}

	During each test, collected data included the initial projectile velocity; in-situ, stereo or asynchronous flash X-ray images; and high-speed scattered light images of the impact event.
	The position of the projectile within the sample was determined visually and measured using pixel counts in the GNU Image Manipulation Program (GIMP). 
	Example X-ray and scattered light images from test \#8 are shown in Figure \ref{fig:ottawa_impact}b,c.

	To simulate these impact events, we implement the stress update algorithm described in {\ref{sec:appendix_algorithm}} into the axisymmetric, two-phase material point method (MPM) from \citet{baumgarten2019a}. The simulated domain measures 3.81 cm $\times$ 20.0 cm and is discretized into 15,159 triangular elements with a minimum side length of 300 $\mu$m: 
	the $d_{50}$ of Ottawa sand. The Ottawa sand samples ($\bar{\rho}_s = 1670$ kg/m$^3$) are represented with 51,623 material points, 
	and an additional layer of 51,623 material points is added to the water-saturated simulations to represent the interstitial fluid ($\bar{\rho}_f = 370$ kg/m$^3$). The 3 mm, stainless steel projectile ($\rho = 8000$ kg/m$^3$, $E = 195$ GPa, $\nu = 0.3$) is represented by 326 material points 
	initially positioned 6 mm above the target surface. An example snapshot from a simulated impact into water-saturated Ottawa sand is shown in Figure \ref{fig:ottawa_impact}a.
	
\begin{figure}[!h]
		\centering
		\begin{minipage}{0.58\linewidth}
			\includegraphics[width=1.0\linewidth]{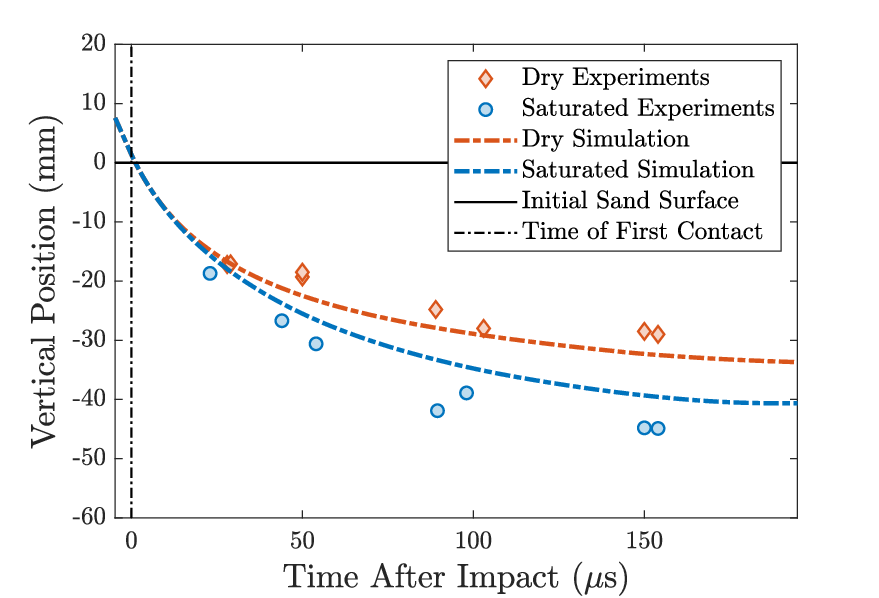}
		\end{minipage}
		\begin{minipage}{0.39\linewidth}
			\scriptsize
			\begin{tabular}{l l l}
				\hline\noalign{\smallskip}
				Test \# & Time ($\mu$s) & Position (mm)\\
				\hline\noalign{\smallskip}
				\multicolumn{3}{l}{Dry}\\
				\hline\noalign{\smallskip}
				1 & 50 & -19.5\\
				2 & 150 & -28.5\\
				3 & 50 & -18.5\\
				4 & 89 & -25.0\\
				10 & 28 & -17.5\\
				10 & 103 & -28.0\\
				11 & 29 & -17.0\\
				11 & 154 & -29.0\\
				\hline\noalign{\smallskip}
				\multicolumn{3}{l}{Saturated}\\
				\hline\noalign{\smallskip}
				5 & 44 & -26.5\\
				6 & 90 & -42.0\\
				7 & 150 & -45.0\\
				8 & 54 & -30.5\\
				8 & 154 & -45.0\\
				9 & 23 & -18.5\\
				9 & 98 & -39.0\\
				\hline
			\end{tabular}
		\end{minipage}
		\caption[]{
			Comparison of projectile position relative to initial target surface for dry and water-saturated Ottawa sand. Experimental measurements from tests listed in Table \ref{tab:shots} are marked with symbols (Dry: red diamonds; Saturated: blue circles) and are determined from in-situ, flash X-ray images. The simulated projectile positions are marked with dashed lines (Dry: red; Saturated: blue) and are determined using the two-phase material point method (MPM) algorithm developed in \citet{baumgarten2019a}.
		}
		\label{fig:ottawa_depth}
	\end{figure}

\begin{figure}[H]
		\centering
		\includegraphics[width=1.0\linewidth]{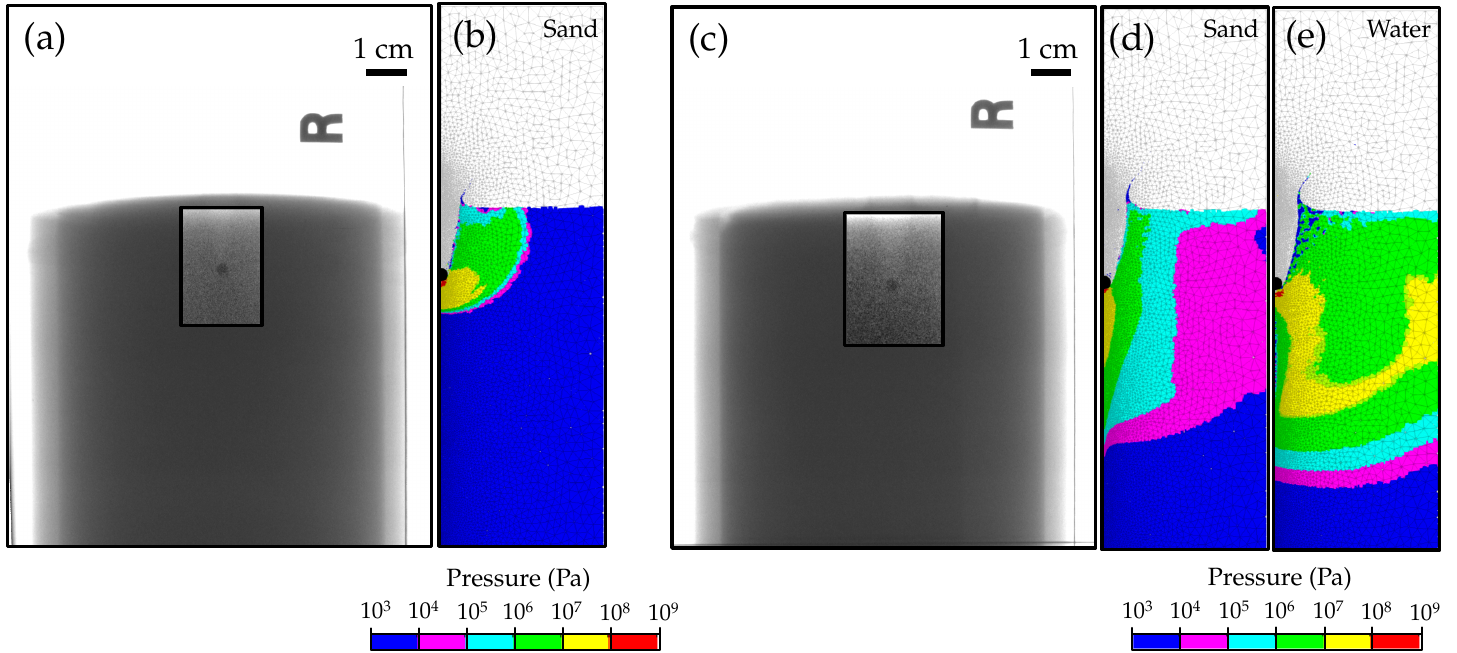}
		\caption[]{
			Comparison of crater shape, projectile position, and pressure contours approximately
			25 $\mu$s after impact of 3 mm, stainless steel sphere into dry (a,b) and water-saturated (c--e) Ottawa sand at 1.3 km/s. (a) Flash X-ray image from dry test \#10 at  28 $\mu$s (inset brightness and contrast adjusted). (b) Pressure contours from dry impact simulation at 25 $\mu$s.  (c) Flash X-ray image from saturated test \#9 at 23 $\mu$s (inset brightness and contrast adjusted). (d) Solid pressure contours, $p_s = - \text{tr}(\boldsymbol{\sigma}_s) / 3$, from saturated impact simulation at 25 $\mu$s. (e) Fluid pressure contours, $p_f$, from saturated impact simulation at 25 $\mu$s.
		}
		\label{fig:ottawa_25us}
	\end{figure}

\begin{figure}[H]
		\centering
		\includegraphics[width=1.0\linewidth]{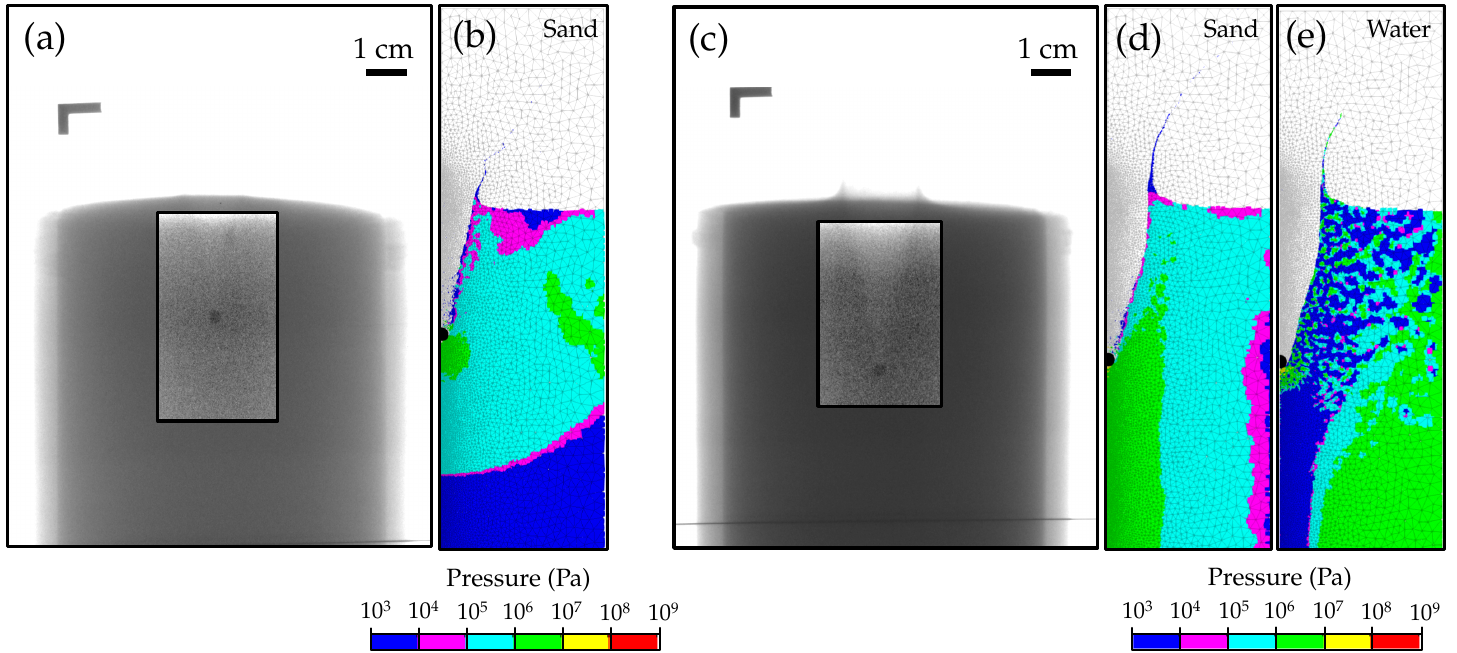}
		\caption[]{
			Comparison of crater shape, projectile position, and pressure contours approximately
			100 $\mu$s after impact of 3 mm, stainless steel sphere into dry (a,b) and water-saturated (c--e) Ottawa sand at 1.3 km/s. (a) Flash X-ray image from dry test \#10 at  103 $\mu$s (inset brightness and contrast adjusted). (b) Pressure contours from dry impact simulation at 100 $\mu$s.  (c) Flash X-ray image from saturated test \#9 at 98 $\mu$s (inset brightness and contrast adjusted). (d) Solid pressure contours, $p_s = - \text{tr}(\boldsymbol{\sigma}_s) / 3$, from saturated impact simulation at 100 $\mu$s. (e) Fluid pressure contours, $p_f$, from saturated impact simulation at 100 $\mu$s.
		}
		\label{fig:ottawa_100us}
	\end{figure}

\begin{figure}[H]
		\centering
		\includegraphics[width=1.0\linewidth]{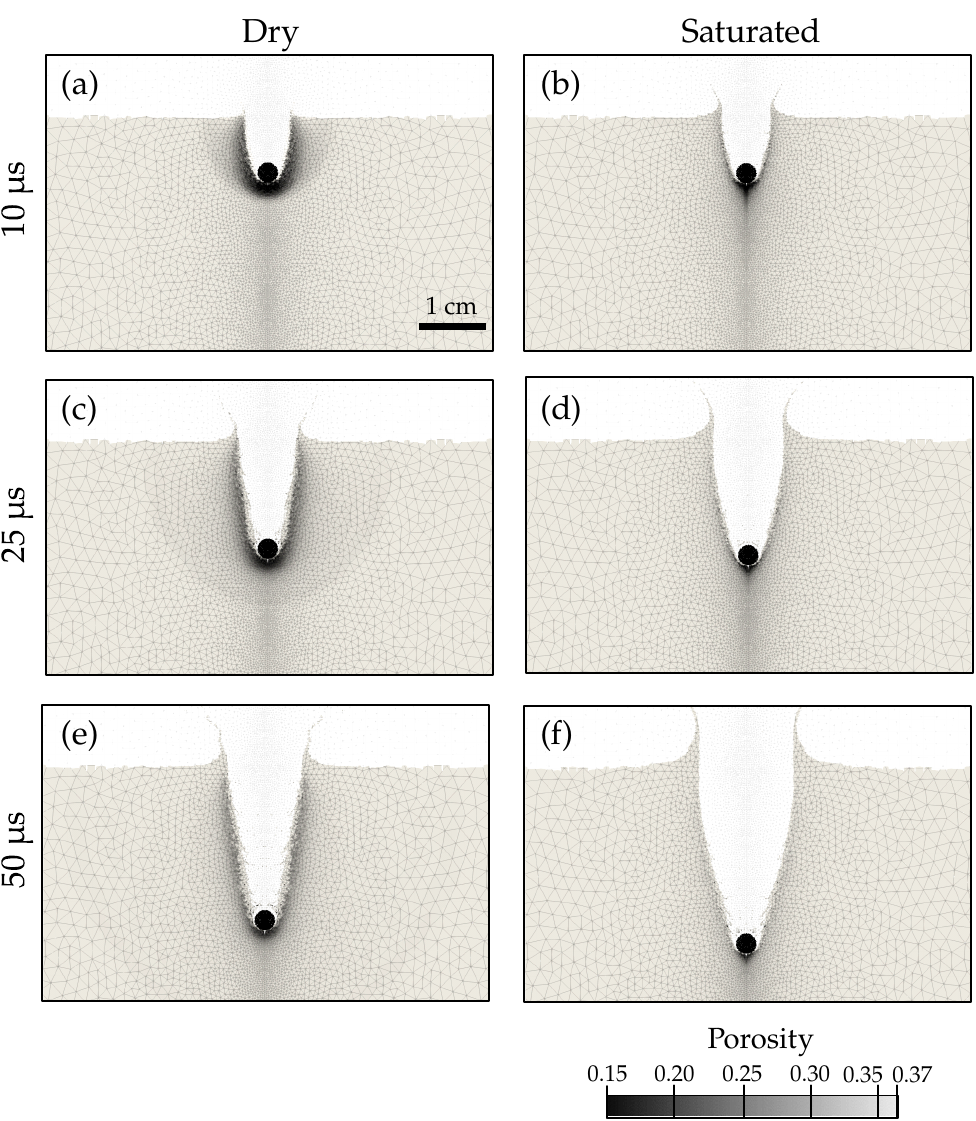}
		\caption[]{
			Comparison of crater shape, projectile position, and predicted porosity during impact of 3 mm, stainless steel sphere into dry (a,c,e) and water-saturated (b,d,f) Ottawa sand at 1.3 km/s. (a,b) Simulation snapshots at 10 $\mu$s. (c,d) Simulation snapshots at 25 $\mu$s. (e,f) Simulation snapshots at 50 $\mu$s. The presence of interstitial water in the saturated sand simulations on the right appears to inhibit pore collapse, which is apparent in the dry simulations on the left. This may explain the apparent {fluidizing} effect of the water as it inhibits the formation of the granular contacts required to develop frictional stresses.
		}
		\label{fig:ottawa_porosity}
	\end{figure}

	For this analysis, we examine two impact simulations in particular: one into dry sand and another into water-saturated sand. Both simulations consider impacts at 1.3 km/s: the average velocity for the tests reported in Table \ref{tab:shots}. The vertical position of the projectile relative to the target surface is computed during both simulations and is plotted over time in Figure \ref{fig:ottawa_depth}. These two time histories highlight an interesting result, which is also observed in the experimental data: the presence of interstitial water appears to weaken the Ottawa sand targets, allowing the projectiles to penetrate further into the samples than when the dry sand is impacted alone. This observation is further highlighted in Figures \ref{fig:ottawa_25us}--\ref{fig:ottawa_porosity}.
	
	Although the presence of water in the pore space between grains increases the combined density of the material ahead of the projectile, it also appears to reduce the sand's ability to absorb and dissipate kinetic energy. This behavior is consistent with the observed fluidization of loosely packed sediments during submarine landslides \citep[e.g., $\phi_s \leq 0.55$; see][]{pailha2009}. However, densely packed sediments (e.g., $\phi_s \geq 0.60$) and fine-particle suspensions \citep[see][]{hoffman1974,waitukaitis2012,baumgarten2019b}, generally exhibit the opposite trend, becoming stronger when saturated with water. 
	
	To understand this counter-intuitive result, we examine the cratering process and internal pressures predicted by these two simulations. In Figure \ref{fig:ottawa_25us} and Figure \ref{fig:ottawa_100us}, the predicted pressure contours and crater shapes are compared with flash X-ray images taken during tests \#9 and \#10. Figure \ref{fig:ottawa_25us} shows this comparison at 25 $\mu$s after impact, and Figure \ref{fig:ottawa_100us} shows this comparison 100 $\mu$s after impact. 
	
	At 25 $\mu$s, the granular pressures, $p_s = - \text{tr}(\boldsymbol{\sigma}_s)/3$, in both simulations exceed 100 MPa in the impact zone immediately ahead of the projectile. 
	However, in the water-saturated simulation (shown in Figure \ref{fig:ottawa_25us}d), {this region of high granular pressure is much smaller than in the dry simulation and does not extend as far laterally.}
	%
	In this second simulation, the fluid pore pressure $p_f$ appears to drive the combined response of the mixture, exceeding the granular pressure $p_s$ throughout much of the domain. This transfer of stresses from the granular material to the interstitial water appears to inhibit the dominant mechanisms of energy dissipation in the granular sediment: frictional sliding, particle fragmentation, and pore collapse.
	
	This hypothesis that the interstitial water inhibits the evolution of granular stress is further supported by the simulation snapshots shown in Figure \ref{fig:ottawa_porosity}. Here, the evolution of the porosity $\phi_f$ in the absence of pore fluid (shown in Figure \ref{fig:ottawa_porosity}a,c,e) is compared with the evolution in presence of interstitial water (shown in Figure \ref{fig:ottawa_porosity}b,d,f). In the dry simulation, the granular material is free to compact ahead of the projectile, as the individual grains are pulverized and the pore space collapses. In the water-saturated simulation, on the other hand, the presence of water in this pore space inhibits bulk compaction, reduces the stresses normally carried by the particles, and effectively fluidizes the material in the early stages of the cratering process. 


	\section{Discussion}
	\label{sec:discussion}
	The model proposed in this work combines soil mechanics, poromechanics, and shock physics into a system of governing equations and constitutive expressions that predicts the behavior of fluid-saturated sediments in extreme loading environments. Key to this modeling approach is the use of granular micromechanics in defining the mathematical descriptions of granular elasticity; granular rearrangement; particle fracture and fragmentation; and pore fluid coupling. Using this approach, the model is shown to have unique predictive capabilities. In this section, we further discuss the results presented in this work and highlight important model limitations.
	
	One advantage of the micromechanics modeling approach is that it allows us to isolate model equations and evaluate their applicability to hypothetical problems. For each of the model mechanisms listed above, we may define an associated microscopic time-scale: i.e.,
	\begin{equation}
		\label{eqn:micro_timescales}
		\tau_\sigma \propto d_{50}/C_0, \quad 
		\tau_i \propto \sqrt{\rho_s d_{50}^2/p_s}, \quad
		\tau_v \propto \eta_0 / p_s, \quad \text{and} \quad
		\tau_B \propto d_{50} / v_c.
	\end{equation}
	Here, $\tau_\sigma$ denotes the time-scale of granular elasticity and is determined by the time required for stress-waves to propagate across individual particles. This time-scale is associated with the model equations in \eqref{eqn:effective_granular_stress}, \eqref{eqn:contact_stresses}, \eqref{eqn:solid_pressure}, and \eqref{eqn:yield_stress}. Similarly, $\tau_i$ denotes the time-scale of inertia-dominated granular rearrangement and is determined by the time required to move particles around one another. This time-scale is associated with the model equations in \eqref{eqn:inelastic_deformation_rate}, \eqref{eqn:solid_dissipation}, \eqref{eqn:y1}, \eqref{subeqn:devdt}, \eqref{subeqn:desdt}, and \eqref{eqn:friction_and_dilation}. On the other hand, $\tau_v$ denotes the time-scale of viscous-dominated granular rearrangement and is associated with \eqref{eqn:y1}, \eqref{eqn:friction_and_dilation},  \eqref{eqn:effective_fluid_shear_stress}, and \eqref{eqn:interphase_drag}. Finally, $\tau_B$ denotes the time-scale of fracture and fragmentation and is determined by the time required for \emph{critical} cracks to span individual particles. This time-scale is associated with the model equations in \eqref{eqn:solid_dissipation}, \eqref{eqn:y1}, \eqref{subeqn:dBdt}, and \eqref{eqn:critical_state}. Note that $v_c$ denotes the critical crack propagation speed in the constituent solid; however, a sub-critical time-scale may also be considered. \citep[See][for additional discussion.]{jop2006,boyer2011,zhang2017}

	An implicit assumption in the model is that each micromechanism behaves similarly across the range of temperatures and strain-rates considered in this work (i.e., 180 to 1000 K and 10$^{-1}$ to 10$^{6}$ s$^{-1}$, respectively). The temperature component is a clear model limitation, as we do not consider melting or a possible brittle--ductile transitions of the constituent material. The strain-rate component, on the other hand, can be reasonably evaluated by comparing the microscopic time-scales above with a mesoscopic deformation time-scale: i.e.,
	\begin{equation}
		\label{eqn:meso_timescale}
		\tau_\epsilon \propto 1/\sqrt{\boldsymbol{D}_s : \boldsymbol{D}_s},
	\end{equation}
	with $\boldsymbol{D}_s$ the mesoscopic strain-rate from \eqref{eqn:strain_rates}. Importantly, we assume that the time-scales in \eqref{eqn:micro_timescales} are fast relative to the mesoscopic timescale in \eqref{eqn:meso_timescale}, such that strain-rate effects can be neglected in their mathematical description.

	Following such an analysis for water-saturated Ottawa sand ($d_{50} = 300$ $\mu$m; $C_0 = 3630$ m/s; $v_c \approx 1000$ m/s; $\eta_0 = 8.91\times10^{-4}$ Pa$\cdot$s) near the impact zone in Figure \ref{fig:mechanisms} ($p_s \approx 1$ GPa), we predict that $\tau_\epsilon \gg \tau_\sigma$, $\tau_i$, $\tau_v$, $\tau_B$ for mesoscopic strain-rates between 10$^{-1}$ to 10$^{5}$ s$^{-1}$. Above this rate --- very near the impacting body in Section \ref{sec:ottawa_impact} --- we anticipate that the model will likely under-predict mechanical stresses in \eqref{eqn:solid_pressure} and particle fragmentation in \eqref{eqn:y1_rates} due to dynamic fracture and the inability of particles to rearrange fast enough: $\tau_\epsilon \approx \tau_B$, $\tau_i \gg \tau_\sigma$, $\tau_v$.
	
	In addition to this time-scale evaluation, we can independently evaluate several of the model equations by considering their prediction of secondary model quantities, including the relative breakage variable, $B$. As shown in {Table \ref{tab:dogs_bay_breakage}} and Figure \ref{fig:ottawa_uniaxial}c, the model equations in \eqref{eqn:y1_rates} appear to systematically over-predict the amount of particle pulverization that is occurring within the simulated samples in Section \ref{sec:dogs_bay_triaxial} and \ref{sec:ottawa_uniaxial}. These model equations are adjusted from similar forms proposed in \citet{tengattini2016} and \citet{cil2020a}, which exhibit similar deficiencies \citep[e.g., see][]{kuwik2022}. Adjusting these equations, more carefully modeling the coupling angle $\omega$, or changing the interpretation of the variable $B$ in \eqref{eqn:y1} may be particularly interesting directions for future model development.
	
	All together, the model is robust and reasonably accurate across a range of conditions: from low-pressure testing of highly porous sediments (shown in Section \ref{sec:dogs_bay_triaxial}) to high-pressure compression of silicate sands (shown in Sections \ref{sec:ottawa_uniaxial} and \ref{sec:ottawa_triaxial}) to high-rate dynamic loading of water-saturated samples (shown in Section \ref{sec:ottawa_impact}).
	The results presented in this work highlight the importance of modeling the combined response of the solid sediment particles alongside the interstitial fluid using an appropriate mixture theory. Additionally, as shown in Section \ref{sec:ottawa_uniaxial}, the proposed decomposition of solid strain energy ($\psi_s$) into a contact component ($\psi_c$) and a compressible granular component ($\psi_g$) enhances model predictions at pressures above 100 MPa. Below this stress level, however, a more traditional treatment of the constituent solid appears sufficient for modeling the response of these materials.
	
	
	The ability of the model to capture the multi-stage loading response of laboratory samples (shown in Figures \ref{fig:dogs_bay_triaxial} and \ref{fig:ottawa_triaxial}) appears unmatched by similar models in the literature, which are usually calibrated to a single loading condition (e.g., see Figure \ref{fig:ottawa_uniaxial}). Additionally, the close agreement between the numerical simulations and experimental measurements reported in Section \ref{sec:ottawa_impact} (shown in Figure \ref{fig:ottawa_depth}) indicate that the model for Ottawa sand has a well-calibrated, predictive capability over a wide range of stresses and strain-rates. 

	\section{Conclusion}
	\label{sec:conclusion}
	In this work, we have proposed, calibrated, and validated a predictive constitutive model for fluid-saturated, brittle granular materials and used this model to study the dynamics of projectile impact into dry and water-saturated sediments. Model parameters are provided for an example carbonate sand (Dog's Bay sand), which is calibrated to low pressures, and for an example silicate sand (Ottawa sand), which is calibrated to stresses between 10$^3$ and 10$^9$ Pa. Numerical simulations of the model highlight its predictive capabilities and allow study of the apparent {fluidizing} effect of interstitial water during a 1.3 km/s impact into Ottawa sand.
	
	Application of the model is limited to conditions where the granular particles exhibit brittle, solid-like behavior and stresses have time to propagate during loading. 
	However, the model is formulated to continue making physically realistic predictions somewhat outside of this range. 

	\section{CRediT Authorship Contribution Statement}
	\textbf{A.\ S.\ Baumgarten:} Conceptualization, Methodology, Software, Formal Analysis, Investigation, Writing - Original \& Draft, Visualization.
	\textbf{J.\ Moreno:} Investigation, Resources, Data Curation, Writing - Review \& Editing.
	\textbf{B.\ Kuwik:} Methodology, Formal Analysis, Investigation, Data Curation, Writing - Review \& Editing.
	\textbf{S.\ Ghosh:} Investigation, Data Curation, Writing - Review \& Editing.
	\textbf{R.\ Hurley:} Conceptualization, Methodology, Resources, Writing - Review \& Editing, Supervision, Project Administration.
	\textbf{K.T.\ Ramesh:} Conceptualization, Methodology, Resources, Writing - Review \& Editing, Supervision, Project Administration, Funding Acquisition.
	
	\section{Acknowledgements}
	{The project or effort depicted was or is sponsored by the Department of the Defense, Defense Threat Reduction Agency under the MSEE URA, HDTRA1-20-2-0001. The content of the information does not necessarily reflect the position or the policy of the federal government, and no official endorsement should be inferred. A.\ S.\ Baumgarten, J.\ Moreno, B.\ Kuwik, S.\ Ghosh, R.\ Hurley, and K.T.\ Ramesh gratefully acknowledge helpful discussions with all participants of the Material Constitutive Laws focus area of the Material Science in Extreme Environments University Research Alliance.}

	\bibliographystyle{elsarticle-harv} 
	\bibliography{references}

	\pagebreak
	\appendix
	
	\setcounter{figure}{0}
	
	\section{Parameter Fitting}
	\label{sec:appendix_fitting}
	In this section, we discuss a procedure for determining the constitutive model parameters listed in Table \ref{tab:parameters}. This approach requires experimental data collected from between three (3) and five (5) different loading conditions --- namely, triaxial compression \citep[e.g.,][]{dakoulas1992,bandini2011}; bender element shear-wave testing \citep[e.g.,][]{robertson1995,jovicic1997}; uniaxial compression \citep[e.g.,][]{kuwik2022}; ring shear testing \citep[e.g.,][]{sadrekarimi2011}; and tap density measurements \citep[e.g.,][]{youd1973}. Additionally, this approach assumes that the constituent solid material that composes the individual particles has a well characterized elastic shock response \citep[e.g., see][]{wackerle1962,heyliger2003}.
	
	\subsection{Determining $M_0$ from Low-Pressure Shearing}
\begin{figure}[!h]
		\centering
		\includegraphics[width=0.8\linewidth]{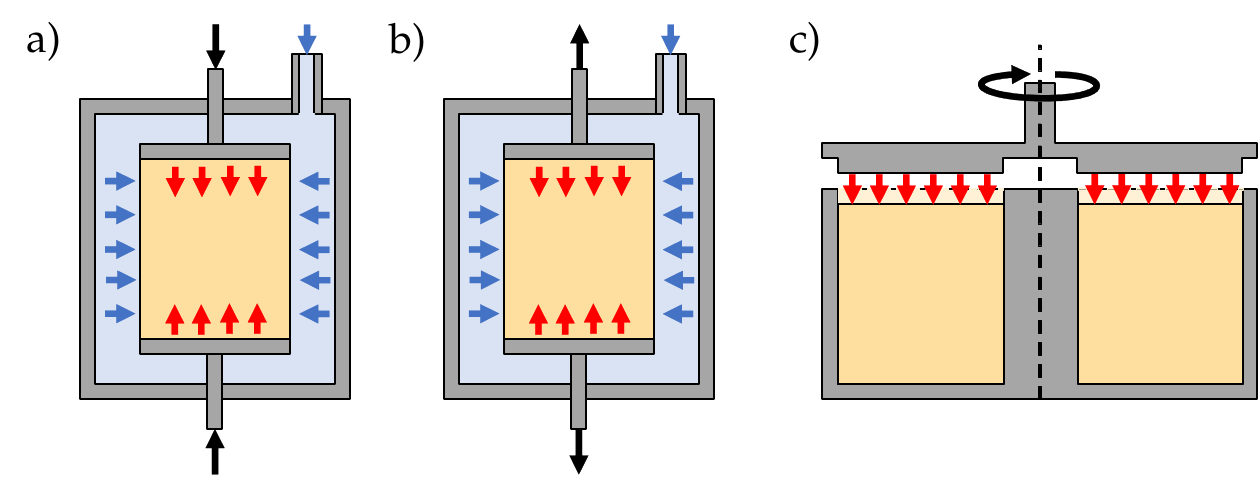}
		\caption[]{(a) Illustration of triaxial compression: the axial compression stress $\sigma_1$ (red) is \emph{greater} than the radial confining stress $\sigma_2$ (blue), leading to axial compression. (b) Illustration of triaxial extension: the axial compression stress $\sigma_1$ (red) is \emph{less} than the radial confining stress $\sigma_2$ (blue), leading to axial extension. (c) Illustration of ring shear: a normal compression stress $\sigma_n$ (red) is applied to a ring of material, and the shearing resistance to rotation $\tau$ is measured.
		}
		\label{fig:M_0_experiments}
	\end{figure}

	The critical state friction angle $M_0$ determines the inelastic response of the model during low-pressure shearing, and it can  be determined from either triaxial compression, triaxial extension, or ring shear experiments (see Figure \ref{fig:M_0_experiments}). Under these loading conditions, the model predicts the following simple relationship between the internal pressure $p_{cs}$ and internal shear stress $q_{cs}$ for steady, inelastic deformation (i.e., critical state yielding):
	\begin{equation}
		\label{eqn:low_pressure_yielding}
		p_{cs} = M_0 q_{cs}, \quad \text{with} \quad p_{cs} = -\tfrac{1}{3} \text{tr}(\boldsymbol{\sigma}_s), \quad \text{and} \quad q_{cs} = \sqrt{\tfrac{3}{2}(\boldsymbol{\sigma}_{s0}:\boldsymbol{\sigma}_{s0})}.
	\end{equation}
	In triaxial compression (see Figure \ref{fig:M_0_experiments}a), this pressure--shear stress relationship can be expressed in terms of the axial compression stress $\sigma_1$ and radial confining stress $\sigma_2$:
	\begin{equation}
		\label{eqn:M_0_triaxial_compression}
		\sigma_1 = \frac{3 + 2 M_0}{3 - M_0} \sigma_2, \quad \text{with} \quad M_0 = \frac{3(\sigma_1 - \sigma_2)}{\sigma_1 + 2 \sigma_2}.
	\end{equation}
	In triaxial extension (see Figure \ref{fig:M_0_experiments}b), a similar relationship can be found, which is expressed as follows:
	\begin{equation}
		\label{eqn:M_0_triaxial_extension}
		\sigma_1 = \frac{3 - 2 M_0}{3 + M_0} \sigma_2, \quad \text{with} \quad M_0 = \frac{3(\sigma_2 - \sigma_1)}{\sigma_1 + 2 \sigma_2}.
	\end{equation}
	Finally, in ring shear experiments (see Figure \ref{fig:M_0_experiments}c), the pressure--shear stress relationship in \eqref{eqn:low_pressure_yielding} can be expressed in terms of the applied normal stress $\sigma_n$ and the measured surface shear stress $\tau$:
	\begin{equation}
		\label{eqn:M_0_ring_shear}
		\tau = \frac{M_0 \sigma_n}{\sqrt{3}}, \quad \text{with} \quad M_0 = \sqrt{3} \bigg(\frac{\tau}{\sigma_n}\bigg).
	\end{equation}
	Together, \eqref{eqn:M_0_triaxial_compression}--\eqref{eqn:M_0_ring_shear} allow us to calibrate the critical state friction angle $M_0$ used in the model with data collected in several common experimental geometries.
	
	\begin{figure}[!h]
		\centering
		\includegraphics[width=0.48\linewidth]{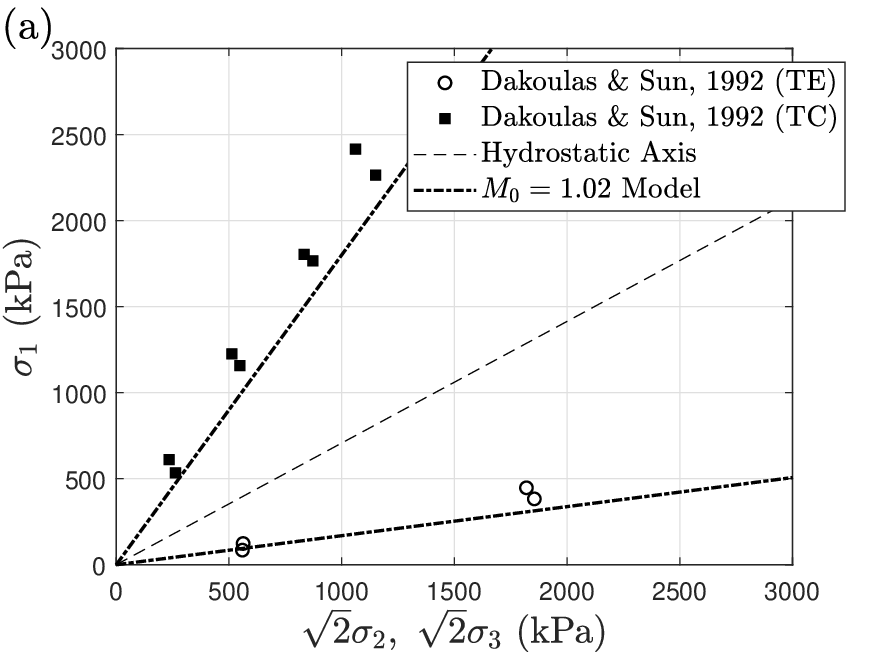}
		\includegraphics[width=0.48\linewidth]{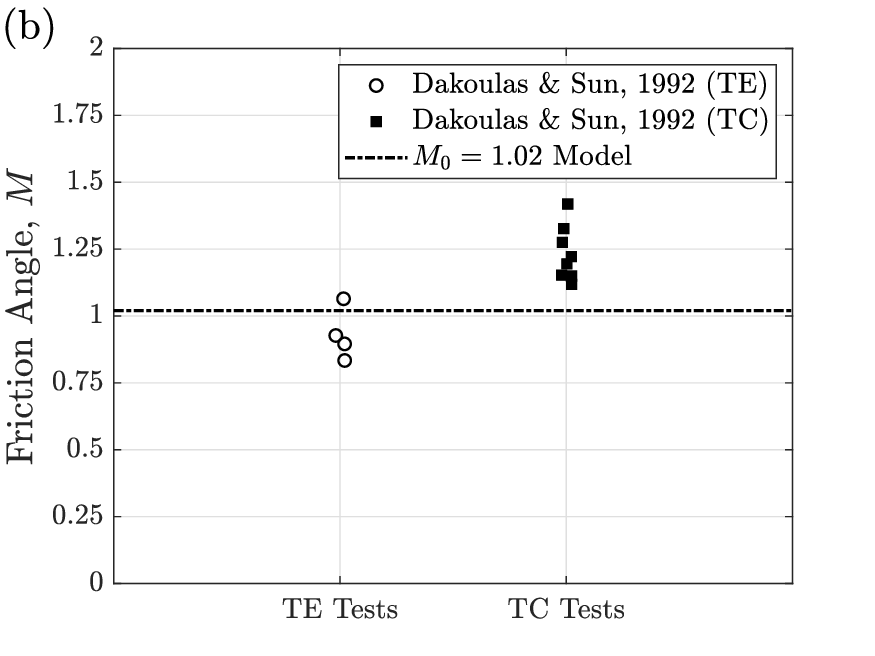}
		\caption[]{(a) Data from triaxial extension (TE) and compression (TC) tests reported in \citet{dakoulas1992} plotted in stress space and compared with the predictions of the model in \eqref{eqn:M_0_triaxial_compression} and \eqref{eqn:M_0_triaxial_extension}. (b) Plot showing the estimated values of $M_0$ using \eqref{eqn:M_0_triaxial_compression} and \eqref{eqn:M_0_triaxial_extension} for each data point in (a). 
		}
		\label{fig:dakoulas_data}
	\end{figure}
	
	For example, consider the model parameters for Ottawa sand shown in Table \ref{tab:parameters}. The value listed for $M_0$ was determined, in part, using triaxial extension and compression data reported in \citet{dakoulas1992}: shown in Figure \ref{fig:dakoulas_data}a. Using \eqref{eqn:M_0_triaxial_compression} and \eqref{eqn:M_0_triaxial_extension}, we can estimate the value of $M_0$ for each reported measurement of $\sigma_1$ and $\sigma_2$: shown in Figure \ref{fig:dakoulas_data}b. Along with data collected in \citet{wijewickreme1986}, the parameter value of 1.02 was chosen to best match the observed behavior of Ottawa sand.

	\subsection{Determining $\bar{K}$ from Isotropic or Uniaxial Compression}
\begin{figure}[!h]
		\centering
		\includegraphics[width=0.48\linewidth]{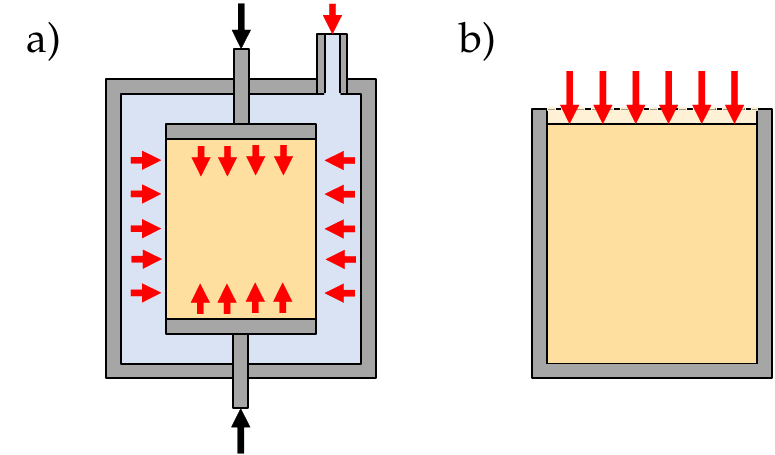}
		\caption[]{(a) Illustration of isotropic compression: the axial compression stress $\sigma_1$ is \emph{equal to} the radial confining stress $\sigma_2$, leading to volumetric compression. (b) Illustration of uniaxial compression: the axial compression stress $\sigma_1$ (red) leads to axial compression, and the thick-walled test cell generally inhibits radial expansion.
		}
		\label{fig:K_experiments}
	\end{figure}

	The reference bulk modulus, $\bar{K}$; the reference shear modulus, $\bar{G}$; and the reference pressure, $p_r$, determine the low-pressure elastic response of the model, which is characterized by the specific Helmholtz free energy function $\hat{\psi}_c(\epsilon_v^e, \epsilon_s^e, B)$ in \eqref{eqn:helmholtz_free_energy}. Parameter values for $p_r$ and $\bar{K}$ can be determined from either isotropic or uniaxial compression experiments (see Figure \ref{fig:K_experiments}). Under these loading conditions, the model predicts simple relationships between applied stresses and measured deformations.
	
	In isotropic compression, the applied hydrostatic load $p_h$ can be expressed in terms of the measured volumetric strain $\epsilon_v$ as follows:
	\begin{equation}
		\label{eqn:K_isotropic}
		p_h = \frac{\bar{\rho}_s}{\rho_0} (1 - \theta B) p_r \bigg( \frac{\bar{K}^2 \epsilon_v^2}{4} \bigg), \quad \text{with} \quad \bar{K} = \sqrt{ \frac{4 \rho_0 p_h}{\bar{\rho}_s (1 - \theta B) p_r \epsilon_v^2}}.
	\end{equation}
	Here, $\bar{\rho}_s$ is the effective mass density of the tested sample; $\rho_0$ is the mass density of the individual grains; $\theta$ is the \emph{grading index}, which has a nominal value of 0.83 \citep{tengattini2016}; $B$ is the initial relative breakage measured for the tested sample (see \ref{sec:appendix_breakage}); and $p_r$ is a nominal reference pressure (frequently 1 kPa).
	
	In uniaxial compression, on the other hand, the radial confinement of the sample produces both shear and volumetric strains. As shown in Figure \ref{fig:example_uniaxial}, this results in an initial compression (Stage I) that includes \emph{both} elastic compression and inelastic granular rearrangement. As a result, the relationship between the applied compression stress $\sigma_1$ and the measured compression strain $\epsilon_1$ depends on the critical state friction angle $M_0$:
	\begin{equation}
		\label{eqn:K_uniaxial}
		\sigma_1 = \bigg( \frac{2 M_0}{3} + 1 \bigg) \frac{\bar{\rho}_s}{\rho_0} ( 1- \theta B) p_r \bigg( \frac{\bar{K}^2 \epsilon_1^2}{4} \bigg), \quad \text{with} \quad \bar{K} = \sqrt{ \frac{12 \rho_0 \sigma_1}{(2 M_0 + 3)\bar{\rho}_s (1 - \theta B) p_r \epsilon_1^2}}.
	\end{equation}
	Here, $\bar{\rho}_s$, $\rho_0$, $\theta$, $B$, and $p_r$ are determined in the same manner as the isotropic compression case.
	
\begin{figure}[!h]
		\centering
		\includegraphics[width=0.48\linewidth]{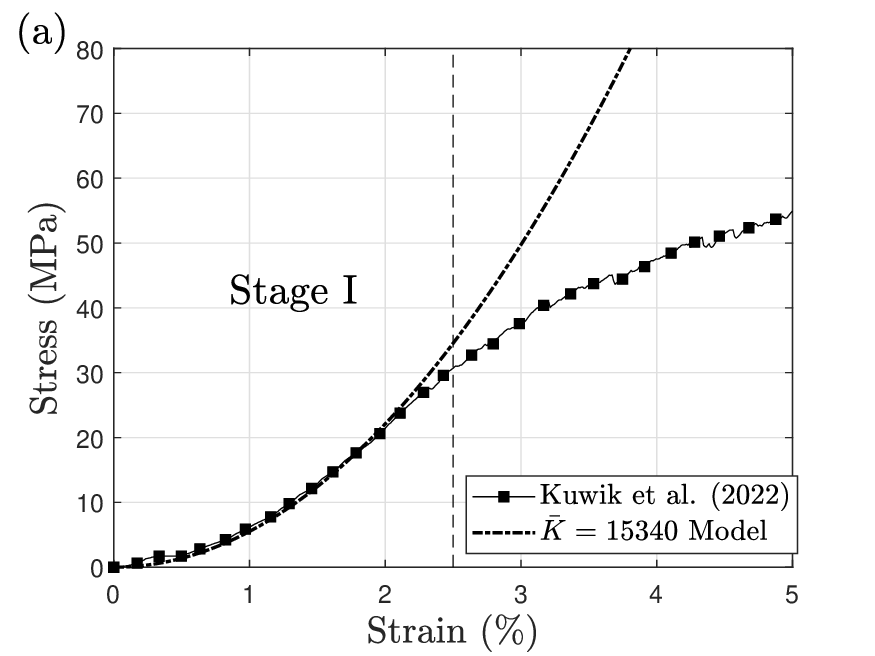}
		\includegraphics[width=0.48\linewidth]{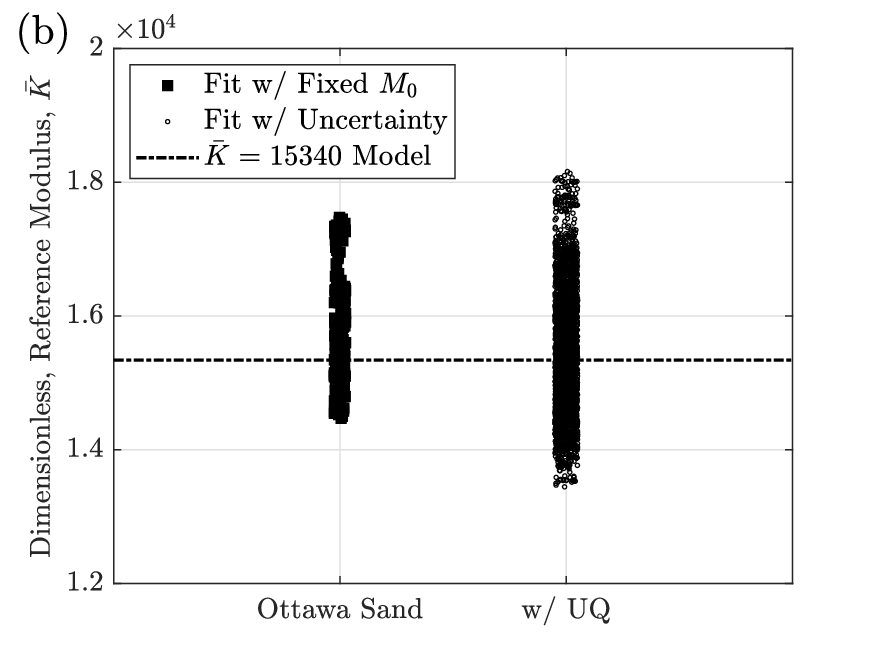}
		\caption[]{(a) Data from one uniaxial compression test reported in \citet{kuwik2022} plotted in stress--strain space and compared with the predictions of the model in \eqref{eqn:K_uniaxial}. (b) Plot showing the estimated values of $\bar{K}$ calculated using \eqref{eqn:K_uniaxial} for each data point in (a) with either a nominal value for $M_0$ (1.02) or the range of values calculated in Figure \ref{fig:dakoulas_data}b.
		}
		\label{fig:kuwik_data}
	\end{figure}

	Together, \eqref{eqn:K_isotropic} and \eqref{eqn:K_uniaxial} allow us to calibrate the reference pressure $p_r$ and the dimensionless, reference bulk modulus $\bar{K}$ used in the model with data collected in two common experimental geometries. For example, consider the model parameters for Ottawa sand in Table 1. The value listed for $\bar{K}$ was determined using the initial uniaxial compression data (Stage I) reported in \citet{kuwik2022}: shown in Section \ref{sec:ottawa_uniaxial} and in Figure \ref{fig:kuwik_data}. The parameter value of 15340 was chosen to best match the observed behavior of Ottawa sand. 
	
	\subsection{Determining $\bar{G}$ from Shear-wave Measurements}
\begin{figure}[!h]
		\centering
		\includegraphics[width=0.30\linewidth]{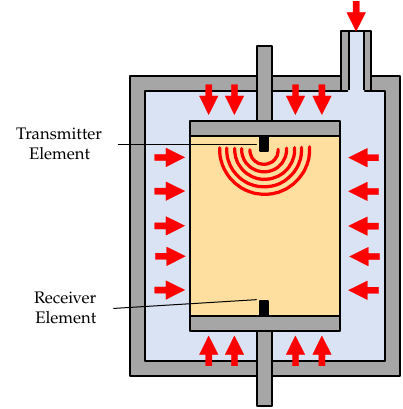}
		\caption[]{Illustration of bender element experiments. Embedded bender elements are used to measure the shear-wave speed $v_s$ within a compressed sample \citep[e.g., see][]{jovicic1997}.
		}
		\label{fig:G_experiments}
	\end{figure}

	Parameter values for $\bar{G}$, which partially determines the low-pressure response of the model, can be calculated using bender element experiments (see Figure \ref{fig:G_experiments}). In these tests, bender elements are used to measure the shear-wave speed $v_s$ in a sample at varying densities, $\bar{\rho}_s$, and pressures, $p_s$. Under these loading conditions, the model predicts the following relationship between these measured quantities:
	\begin{equation}
		\label{eqn:G_wave_speed}
		v_s^2 = \frac{\bar{G} p_r}{\bar{\rho}_s} \sqrt{\frac{\bar{\rho}_s p_s (1 - \theta B)}{\rho_0 p_r}}, \quad \text{with} \quad \bar{G} = \frac{\bar{\rho}_s v_s^2}{p_r} \sqrt{\frac{\rho_0 p_r}{\bar{\rho}_s p_s (1 - \theta B)}}.
	\end{equation}
	Here, $\bar{\rho}_s$ is the effective mass density of the tested sample; $\rho_0$ is the mass density of the individual grains; $\theta$ is the \emph{grading index}, which has a nominal value of 0.83 \citep{tengattini2016}; $B$ is the initial relative breakage measured for the tested sample (see \ref{sec:appendix_breakage}); and $p_r$ is the nominal reference pressure (frequently 1 kPa).
	
\begin{figure}[!h]
		\centering
		\includegraphics[width=0.48\linewidth]{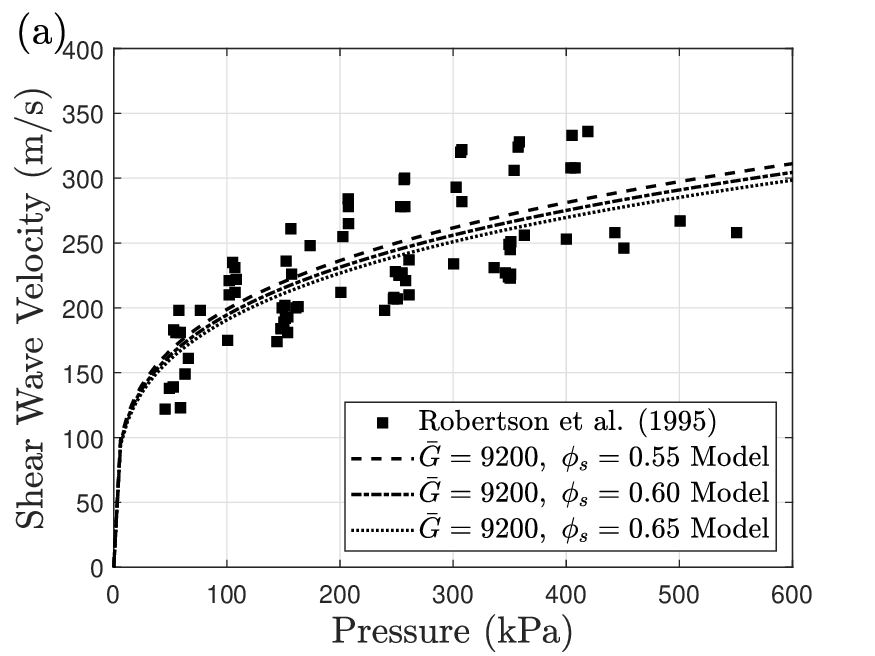}
		\includegraphics[width=0.48\linewidth]{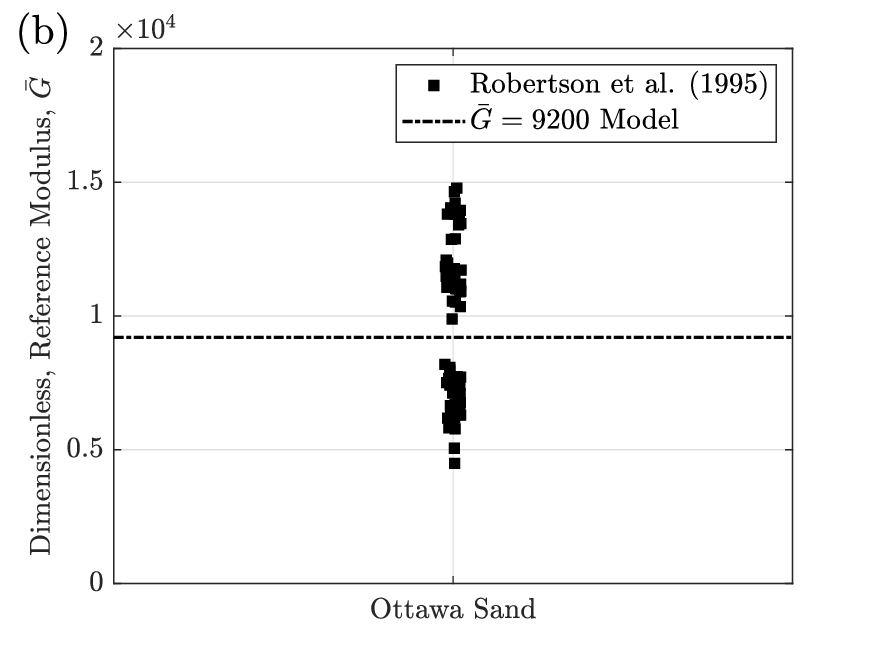}
		\caption[]{(a) Shear wave-speed data from bender element tests reported in \citet{robertson1995} plotted against confining pressure and compared with the predictions of the model in \eqref{eqn:G_wave_speed}. (b) Plot showing the estimated values of $\bar{G}$ calculated using \eqref{eqn:G_wave_speed} for each data point in (a).
		}
		\label{fig:robertson_data}
	\end{figure}

	Using \eqref{eqn:G_wave_speed}, we calibrate the dimensionless, reference shear modulus used in the model with data that is commonly available in the literature. For example, consider the parameterization of Ottawa sand shown in Table \ref{tab:parameters}. The value listed for $\bar{G}$ was determined using the bender element tests reported in \citet{robertson1995}: shown in Figure \ref{fig:robertson_data}. The parameter value of 9200 was chosen to best match the observed behavior of Ottawa sand.

	\subsection{Determining $E_c$ from Isotropic or Uniaxial Compression}
	The critical breakage energy $E_c$ determines the inelastic response of the model during high-pressure loading, and it can be determined from either isotropic or uniaxial compression experiments (see Figure \ref{fig:K_experiments}). Under these loading conditions, the model predicts that the onset of particle fragmentation will occur (approximately) when the breakage energy $E_B$ reaches a critical value --- i.e., $E_B = (1 - B)^2 E_c$. In isotropic and uniaxial compression, this transition is typically identified by a critical compression stress $\sigma_c$ or critical pressure $p_c$, respectively. Under these loading conditions, the following relationship between the breakage energy $E_B$ and the critical stress $\sigma_c$ provides a reasonable approximation of the model:
	\begin{equation}
		\label{eqn:E_c_uniaxial}
		E_B \approx \frac{\bar{\rho}_s}{\rho_0} \frac{2 \theta p_r}{3 \bar{K}} \bigg(\frac{\sigma_c}{p_r}\bigg)^{\tfrac{3}{2}} \bigg(\frac{\rho_0}{\bar{\rho}_s (1 - \theta B)}\bigg)^{\tfrac{3}{2}}, \quad \text{with} \quad E_c = \frac{E_B}{(1 - B)^2}.
	\end{equation}
	Here, $\bar{\rho}_s$ is the effective mass density of the tested sample; $\rho_0$ is the mass density of the individual grains; $\theta$ is the \emph{grading index}, which has a nominal value of 0.83 \citep{tengattini2016}; $B$ is the initial relative breakage measured for the tested sample (see \ref{sec:appendix_breakage}); and $p_r$ is the nominal reference pressure (frequently 1 kPa). It is important to note that the calibrated value of $E_c$ found using \eqref{eqn:E_c_uniaxial} depends on both experimental measurements of $\sigma_c$ \emph{and} prior calibration of the dimensionless, reference bulk modulus $\bar{K}$.

\begin{figure}[!h]
		\centering
		\includegraphics[width=0.48\linewidth]{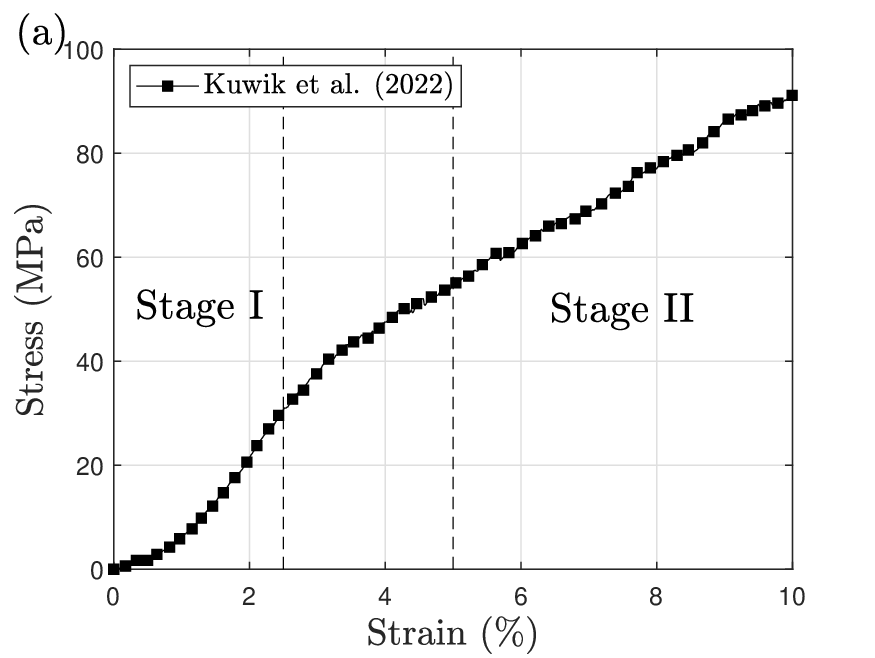}
		\includegraphics[width=0.48\linewidth]{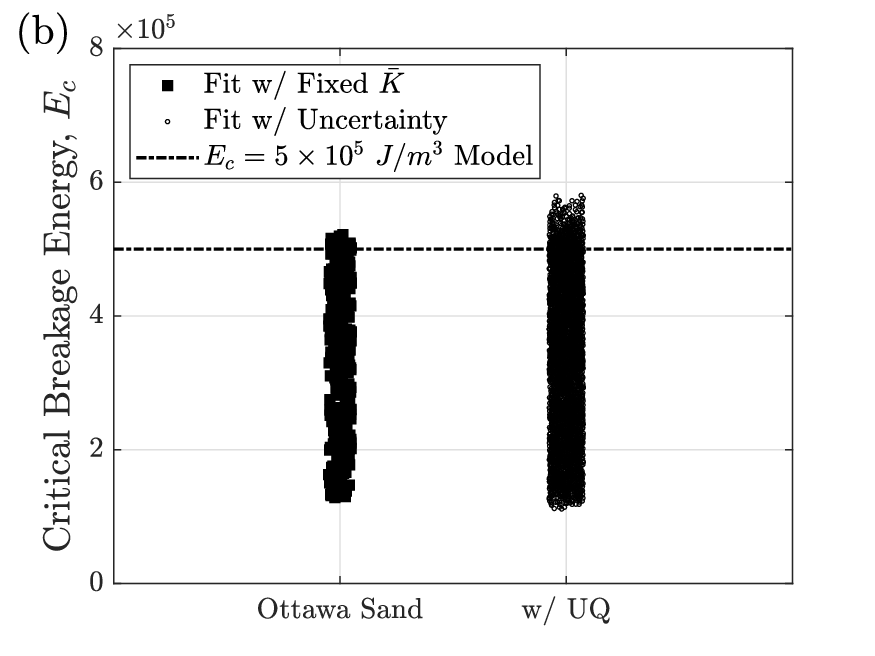}
		\caption[]{(a) Data from one uniaxial compression test reported in \citet{kuwik2022} plotted in stress--strain space highlighting transition from granular rearrangement (Stage I) to particle fragmentation (Stage II). (b) Plot showing the estimated values of $E_c$ calculated using \eqref{eqn:E_c_uniaxial} for each data point in (a) with either a nominal value for $\bar{K}$ (15340) or samples from the range of values calculated in Figure \ref{fig:kuwik_data}b.
		}
		\label{fig:kuwik_data_2}
	\end{figure}

	Using \eqref{eqn:E_c_uniaxial}, we calibrate the critical breakage energy used in the model with data collected in two common experimental geometries. For example, the model parameter $E_c$ for Ottawa  sand listed in Table \ref{tab:parameters} was determined using the uniaxial compression data (Stage I--II) reported in Kuwik et al. (2022): shown in Section \ref{sec:ottawa_uniaxial} and in Figure \ref{fig:kuwik_data_2}. The parameter value of 500 kJ/m$^3$ was chosen to best match the observed behavior of Ottawa sand.
	
	\subsection{Determining $\phi_u$, $\phi_l$, $u$, and $l$ from Tap Densities}
	The porosity parameters $\phi_u$, $\phi_l$, $u$, and $l$ determine, in part, the critical state behavior of the model at low confining stresses. The upper porosity limit $\phi_u$ represents the highest porosity at which the granular material is capable of carrying static loads (i.e., before the particles lose contact with their neighbors). The lower porosity limit $\phi_l$, on the other hand, represents the lowest porosity that the granular sediment can compact to \emph{without breaking or fracturing particles} --- typically achieved through repeated tapping or tamping of samples. The dimensionless parameters $u$ and $l$ are used in the power-law model from \citet{rubin2011} that forms a component of the model proposed in this work. In particular, these parameters are used to define $\phi_\text{min}$ and $\phi_{\text{max}}$ --- the minimum and maximum porosities at different relative breakage values --- as follows,
	\begin{equation}
		\label{eqn:phi_max_phi_min}
		\phi_\text{max} = \phi_u (1 - B)^u, \quad \text{and} \quad \phi_\text{min} = \phi_l (1 - B)^l.
	\end{equation}
	Here, B is the relative breakage measured for each tested sample (see \ref{sec:appendix_breakage}).
	
\begin{figure}[!h]
		\centering
		\includegraphics[width=0.48\linewidth]{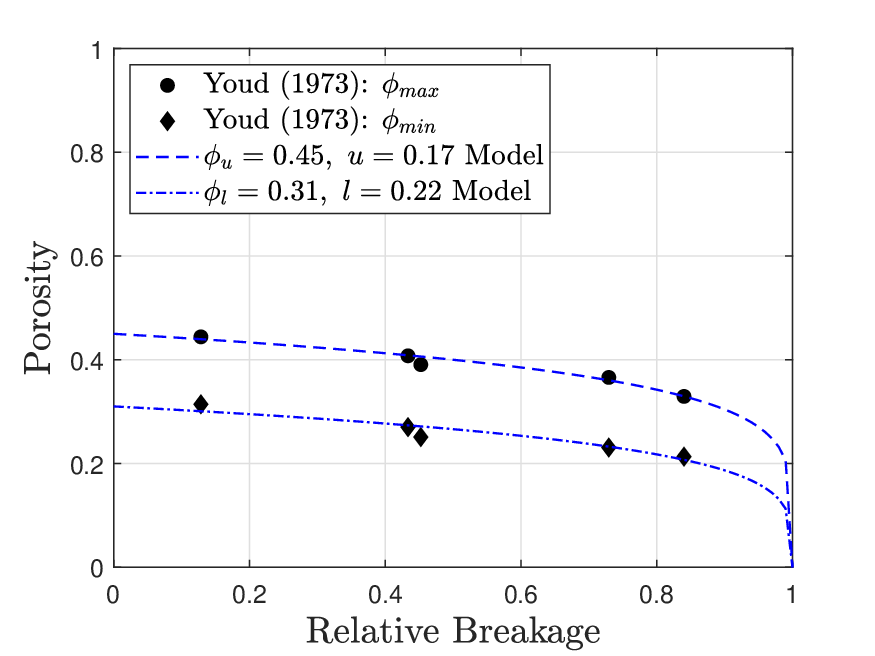}
		\caption[]{Data from tap density tests reported in \citet{youd1973} plotted in breakage--porosity space compared with the model proposed in \citet{rubin2011}.
		}
		\label{fig:youd_data}
	\end{figure}

	Using \eqref{eqn:phi_max_phi_min}, we calibrate these critical state model parameters with data commonly available in the literature. For example, the parameters $\phi_u$, $\phi_l$, $u$, and $l$ for Ottawa sand listed in Table \ref{tab:parameters} were determined using porosity data reported in \citet{youd1973}: shown in Figure \ref{fig:youd_data}. The parameter values of $\phi_u = 0.45$, $\phi_l = 0.31$, $u = 0.17$, and $l = 0.22$ were chosen to best match the observed behavior of Ottawa sand.
	
	\subsection{Determining $\gamma$ from Intermediate-Pressure Shearing}
	The dilation parameter $\gamma$ determines, in part, the low-pressure critical state behavior of the  model, controlling the critical relative density $\tau_{cs}$ in \eqref{eqn:critical_state} and the dilation angle $M_d$ in \eqref{eqn:friction_and_dilation}. It can be determined through either triaxial compression or ring shear experiments (see Figure \ref{fig:M_0_experiments}). Under these loading the conditions, the  model is reasonably approximated by the following relationships between the measured critical state porosity $\phi_{c}$ and applied mean effective stress $p_s$:
	\begin{equation}
		\label{eqn:gamma_shear}
		\begin{aligned}
			\phi_{c} &\approx \phi_\text{max} - (\phi_\text{max} - \phi_\text{min}) \frac{(1 - B)}{\gamma} \bigg( \frac{\bar{\rho}_s}{\rho_0} \frac{2 \theta p_r}{3 \bar{K}} \bigg)^{\tfrac{1}{2}} \bigg(\frac{p_s}{p_r}\bigg)^{\tfrac{3}{4}} \bigg(\frac{\rho_0}{\bar{\rho}_s (1 - \theta B)}\bigg)^{\tfrac{3}{4}},\\[1ex]
			\gamma &\approx (1 - B) \bigg(\frac{\phi_\text{max} - \phi_\text{min}}{\phi_\text{max} - \phi_f}\bigg) \bigg( \frac{\bar{\rho}_s}{\rho_0} \frac{2 \theta p_r}{3 \bar{K}} \bigg)^{\tfrac{1}{2}} \bigg(\frac{p_s}{p_r}\bigg)^{\tfrac{3}{4}} \bigg(\frac{\rho_0}{\bar{\rho}_s (1 - \theta B)}\bigg)^{\tfrac{3}{4}}.
		\end{aligned}
	\end{equation}
	Here, $\phi_\text{max}$ and $\phi_\text{min}$ are the limiting porosities from \eqref{eqn:phi_max_phi_min}, $\bar{\rho}_s$ is the effective mass density of the tested sample; $\rho_0$ is the mass density of the individual grains; $\theta$ is the \emph{grading index}, which has a nominal value of 0.83 \citep{tengattini2016}; $B$ is the initial relative breakage measured for the tested sample (see \ref{sec:appendix_breakage}); and $p_r$ is the nominal reference pressure (frequently 1 kPa).
	
\begin{figure}[!h]
		\centering
		\includegraphics[width=0.48\linewidth]{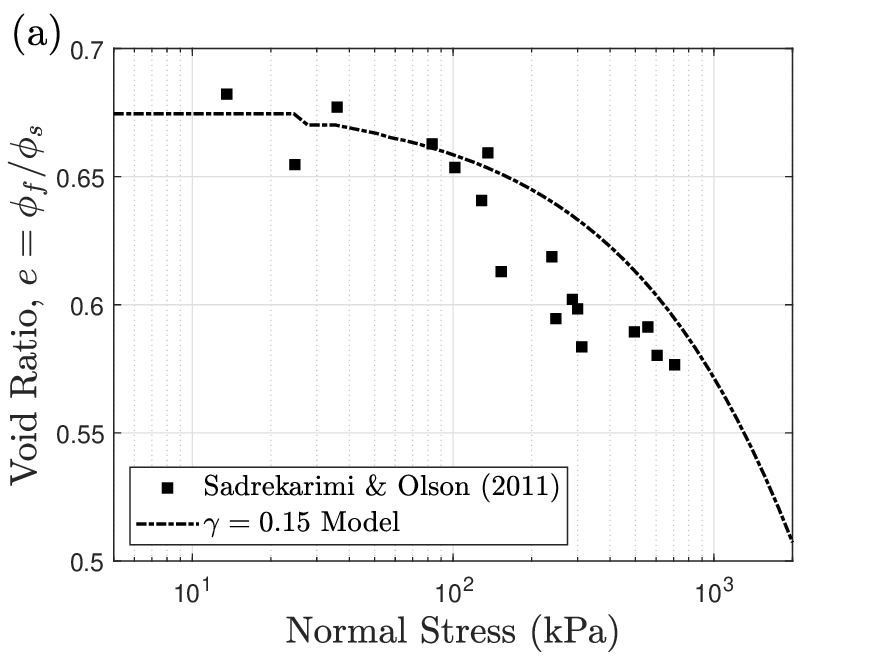}
		\includegraphics[width=0.48\linewidth]{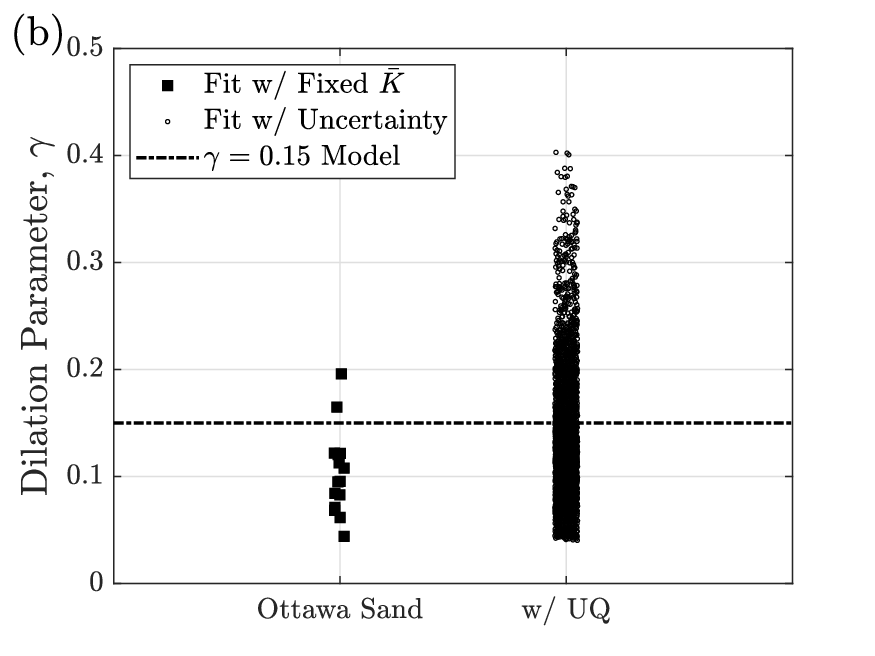}
		\caption[]{(a) Data from multiple ring shear tests reported in \citet{sadrekarimi2011} plotted in stress--void-ratio space and compared with the predictions of the model from \eqref{eqn:gamma_shear}. (b) Plot showing the estimated values of $\gamma$ calculated using \eqref{eqn:gamma_shear} for each data point in (a) with either a nominal value for $\bar{K}$ (15340) or samples from the range of values calculated in Figure \ref{fig:kuwik_data}b.
		}
		\label{fig:sadrekarimi_data}
	\end{figure}

	Using \eqref{eqn:gamma_shear}, we calibrate the dilation parameter $\gamma$ with data from two common experimental geometries. For example, consider the model parameters for Ottawa sand in Table \ref{tab:parameters}. The value listed for $\gamma$ was determined using the porosity measurements in the ring shear experiments reported in \citet{sadrekarimi2011}: shown in Figure \ref{fig:sadrekarimi_data}. The parameter value of 0.15 was chosen to best match the observed behavior of Ottawa sand.
	
	\subsection{Determining $b$ from Uniaxial Compression}
	The density--elasticity coupling parameter $b$ determines, in part, the high-pressure elastic response of the  model, which is characterized by the specific Helmholtz free energy function $\hat{\psi}_g(\rho_s, T_s)$ in \eqref{eqn:helmholtz_free_energy}. Parameter values for $b$ can be determined from either isotropic or uniaxial compression experiments (see Figure \ref{fig:K_experiments}); however, calibration of this model parameter requires experimental apparatuses that can reach and sustain stresses greater than approximately $0.1$\% of the elastic moduli (e.g., $E^*$, $G^*$, $K^*$) of the constituent solid that composes the individual grains. Above this relative stress magnitude, the Hertzian contact theory used to formulate the low-pressure elastic response, $\hat{\psi}_c(\epsilon_v^e, \epsilon_s^e, B)$, begins to significantly deviate from linear elastic behavior.
	
	Under these loading conditions, the model can be approximated by the following relationships between the measured volumetric strain $\epsilon_v$ and the applied pressure or axial compression stress $p$:
	\begin{equation}
		\label{eqn:b_uniaxial}
		\begin{aligned}
			\epsilon_0^e &\approx \sqrt{ \frac{4 \rho_0 p_0}{\bar{\rho}_s (1 - \theta) p_r \bar{K}^2}},\\[1ex]
			p &\approx  \bigg(\frac{(1 - \phi_0)(1 - \theta) p_r}{(1 - \epsilon_v) \rho_0}\bigg) \bar{K}^2 (\epsilon_0^e + \epsilon_v - \epsilon_0)^2 + \bigg(\frac{1 - \phi_0}{1- \epsilon_v}\bigg)^{2b + 1} \rho_0 C_0^2 (\epsilon_0^e + \epsilon_v - \epsilon_0),\\[1ex]
			b &\approx \frac{\displaystyle \log \bigg[ \frac{p}{\rho_0 C_0^2 (\epsilon_0^e + \epsilon_v - \epsilon_0)} - \bigg(\frac{(1 - \phi_0)(1 - \theta) p_r}{(1 - \epsilon_v) \rho_0}\bigg) \frac{\bar{K}^2}{\rho_0 C_0^2} (\epsilon_0^e + \epsilon_v - \epsilon_0)\bigg]}{\displaystyle2 \log \bigg[\frac{1 - \phi_0}{1- \epsilon_v}\bigg]} - \frac{1}{2}.
		\end{aligned}
	\end{equation}
	Here, $p_0$ and $\epsilon_0$ are the pressures and strains measured at the onset of Stage III: shown in Sections \ref{sec:example_uniaxial} and \ref{sec:ottawa_uniaxial}. These measurements are observer-dependent; however, they help isolate the high-pressure stress response predicted by the model. Additionally, $\rho_0$ is the mass density of the individual grains; $\theta$ is the \emph{grading index}, which has a nominal value of 0.83 \citep{tengattini2016}; $p_r$ is the nominal reference pressure (frequently 1 kPa); $\bar{K}$ is the dimensionless, reference bulk modulus; $\phi_0$ is the initial porosity of the sample measured before compaction; and $C_0$ is bulk wave-speed of constituent solid material.

	\begin{figure}[!h]
		\centering
		\includegraphics[width=0.48\linewidth]{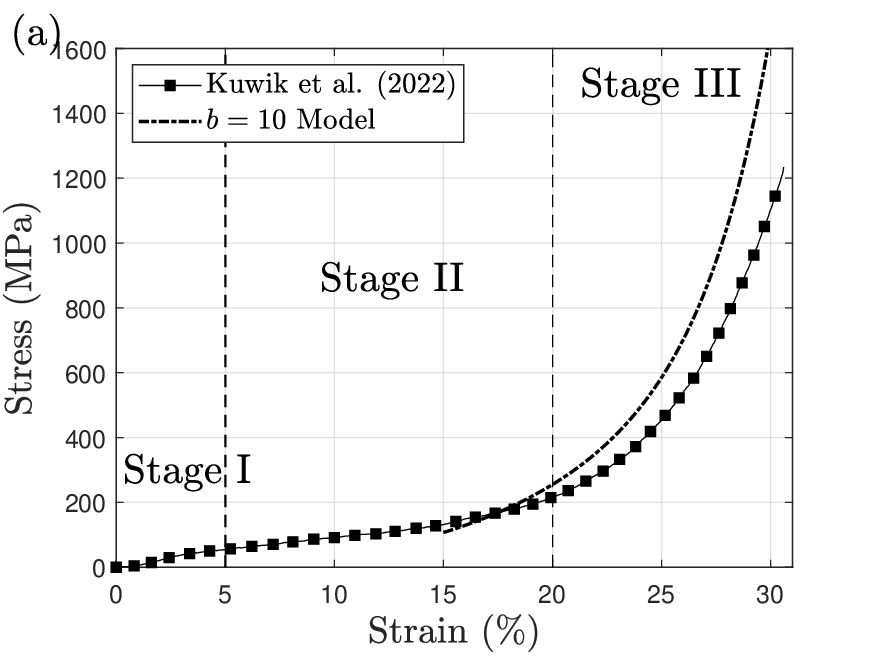}
		\includegraphics[width=0.48\linewidth]{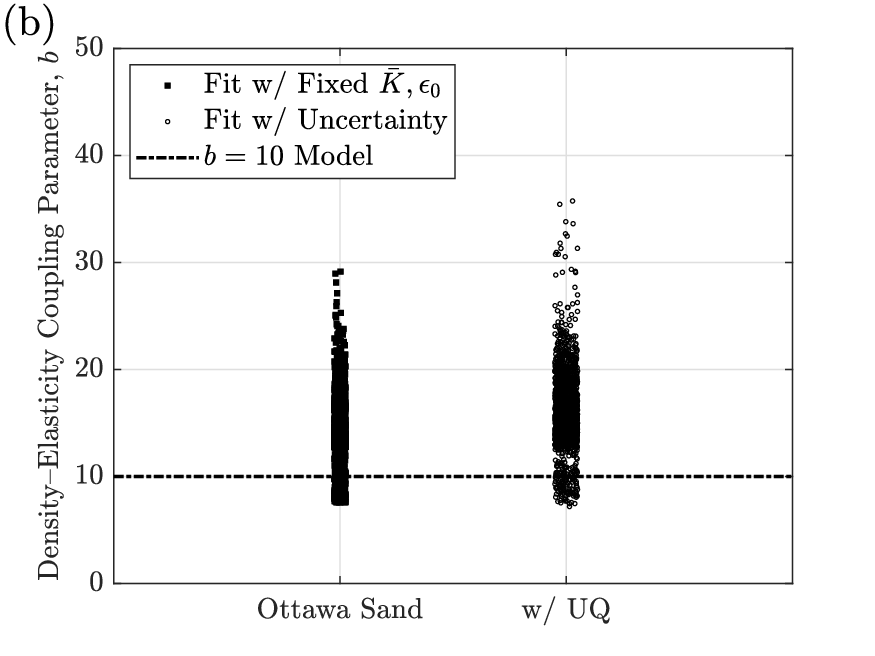}
		\caption[]{(a) Data from one uniaxial compression test reported in Kuwik et al. (2022) plotted in stress--strain space and compared with the approximation of the model in \eqref{eqn:b_uniaxial}. (b) Plot showing the estimated values for $b$ calculated using \eqref{eqn:b_uniaxial} for each data point in Stage III of (a) with either a nominal value for $\bar{K}$ (15340) and $\epsilon_0$ (20\%) or samples from the range of $\bar{K}$ values calculated in Figure \ref{fig:kuwik_data}b and initial strains between 20--25\%.
		}
		\label{fig:kuwik_data_3}
	\end{figure}
	
	Using \eqref{eqn:b_uniaxial}, we calibrate the density--elasticity coupling parameter $b$ with data from two common experimental geometries. For example, consider the model parameters for Ottawa sand shown in Table \ref{tab:parameters}. The value listed for $b$ was determined using the pressure--strain measurements from one uniaxial compression test reported in \citet{kuwik2022}: shown in Figure \ref{fig:kuwik_data_3}. The parameter value of 10 was chosen to best match the observed behavior of Ottawa sand.

	\setcounter{figure}{0}
	
	\section{Numerical Integration Procedure}
	\label{sec:appendix_algorithm}
	In this section, we briefly introduce the explicit time-integration procedure used in Section \ref{sec:example_uniaxial} and Section \ref{sec:results} to update the effective granular stress $\boldsymbol{\sigma}_s$ of a single material point, from current time $t^k$ to updated time $t^{k+1}$. As inputs, this algorithm takes in the material parameters listed in Table \ref{tab:parameters} and the relevant material point state at time $t^k$: namely,
	\begin{enumerate}[itemsep=0em,label=(\roman*)]
		\item the current relative breakage, $B^k$;
		\item the current solid-phase, effective mass density, $\bar{\rho}_s^k$;
		\item the current inelastic porosity, $\phi_p^k$;
		\item the current left, elastic Cauchy--Green tensor, $\boldsymbol{B}^{ek}$;
		\item the current solid-phase temperature $T_s^k$;
		\item the current solid-phase, specific internal energy, $\varepsilon_s^k$;
		\item and the current effective granular stress, $\boldsymbol{\sigma}_s^k$.
	\end{enumerate}
	Additionally, this algorithm requires an average deformation-rate (i.e., the velocity gradient $\boldsymbol{L}_s^k = \nabla \boldsymbol{v}_s^k$) experienced by the material during the time increment $\Delta t^k = t^{k+1} - t^k$.
	As outputs, this algorithm updates the relevant material point states and stresses: namely,
	\begin{enumerate}[itemsep=0em,label=(\roman*)]
		\item the updated relative breakage, $B^{k+1}$;
		\item the updated solid-phase, effective mass density, $\bar{\rho}_s^{k+1}$;
		\item the updated inelastic porosity, $\phi_p^{k+1}$;
		\item the updated left, elastic Cauchy--Green tensor, $\boldsymbol{B}^{ek+1}$;
		\item the updated solid-phase temperature $T_s^{k+1}$;
		\item the updated solid-phase, specific internal energy, $\varepsilon_s^{k+1}$;
		\item and the updated effective granular stress, $\boldsymbol{\sigma}_s^{k+1}$.
	\end{enumerate}
	
	The stress update algorithm proceeds in three principle steps: (a) estimate the material state and stresses at $t^{k+1}$, \emph{assuming that no inelastic deformation occurs during the time increment}; (b) check whether the material is inelastically yielding at $t^{k+1}$ (i.e., $y_1 \geq 0$, $y_2 \geq 0$, or $y_3 \geq 0$); and (c) if the material is yielding, solve for $\lambda_1$, $\lambda_2$, or $\lambda_3$ to ensure $y_1 = 0$, $y_2 = 0$, or $y_3 = 0$, respectively. This method is sometimes referred to as a prediction--correction method: (a) \emph{predicts} the material state at the end of the time increment --- sometimes called the \emph{trial state} --- (b) checks whether the trial state is valid; and (c) \emph{corrects} the material state in order to obey the model yield conditions in \eqref{eqn:y1}--\eqref{eqn:y3}. 
	A complete outline of the stress update algorithm is provided in Algorithms \ref{alg:stress_update}--\ref{alg:stress_update_3}.
	
	Using this algorithm, the entire state of a single, solid-phase material point can be forward integrate from $t^k$ to $t^{k+1}$. In Section \ref{sec:example_uniaxial} and Sections \ref{sec:dogs_bay_triaxial}--\ref{sec:ottawa_triaxial}, this stress update algorithm is implemented in MATLAB to integrate the material state over time in simple uniaxial or triaxial loading conditions. In Section \ref{sec:ottawa_impact}, this algorithm is implemented within the MPM numerical simulation framework from \citet{baumgarten2019a}. 
	
	\begin{algorithm}
		\caption{Stress update algorithm: Part I.}
		\label{alg:stress_update}
		\begin{algorithmic}[1]
			\Statex\itemsep=0.75ex
			\Function{StressUpdate}{$B^k, \bar{\rho}_s^k, \phi_p^k, \boldsymbol{B}^{ek}, T_s^k, \varepsilon_s^k, \boldsymbol{L}_s^k$}
			\LineComment{1}{Update Mass Density:}
			\State \setequal{1.5}{$\bar{\rho}_s^{k+1}$}{$\bar{\rho}_s^k (1 - \Delta t^k\ \text{tr}(\boldsymbol{L}_s^k))$}
			\Comment{from Eq.\ \eqref{eqn:mass_conservation}}
			\LineComment{1}{Estimate Elastic Deformation w/o Plasticity:}
			\State \setequal{1.5}{$\boldsymbol{B}^{e*}$}{$\boldsymbol{B}^{ek} + \Delta t^k (\boldsymbol{L}_s^k \boldsymbol{B}^{ek} + \boldsymbol{B}^{ek} \boldsymbol{L}_s^{k \top})$}
			\Comment{Note: $\boldsymbol{B}^e = \boldsymbol{F}^{e\top} \boldsymbol{F}^e$}
			\LineComment{1}{Estimate Elastic Strain Invariants:}
			\State \setequal{0.7}{$\epsilon_v^{e}$}{$\tfrac{1}{2} \text{tr}(\boldsymbol{B}^{e*}) - \tfrac{3}{2}$}
			\Comment{from Eq.\ \eqref{eqn:elastic_strain}}
			\State \setequal{0.7}{$\epsilon_s^{e}$}{$\sqrt{\tfrac{1}{6} \boldsymbol{B}^{e*}_0 : \boldsymbol{B}^{e*}_0}$}
			\Comment{$\boldsymbol{B}^{e*}_0$ denotes deviator of $\boldsymbol{B}^{e*}$}
			\State \setequal{0.7}{$J^{e}$}{$\sqrt{\text{det}(\boldsymbol{B}^{e*})}$}
			\Comment{from Eq.\ \eqref{eqn:elastic_strain}}
			\LineComment{1}{Check $y_2$ and $y_3$ Yield Functions:}
			\If{$J^{e} > 1$ \textbf{or} $\bar{\rho}_s^{k+1} < \rho_0 (1 - \phi_u(1 - B^k)^u)$}
			\LineComment{2}{Material is Either Pulling Apart ($y_2 = 0$) or Unable to Support Stresses ($y_3 = 0$)}
			\State \setequal{2.5}{$B^{k+1}$}{$B^k$}
			\State \setequal{2.5}{$\phi_p^{k+1}$}{$\phi_p^k$}
			\State \setequal{2.5}{$\boldsymbol{B}^{ek+1}$}{$\boldsymbol{1}$}
			\State \setequal{2.5}{$\varepsilon_s^{k+1}$}{$\varepsilon_s^k$}
			\State \setequal{2.5}{$T_s^{k+1}$}{$\varepsilon_s^{k+1} / c_v$}
			\State \setequal{2.5}{$\boldsymbol{\sigma}_s^{k+1}$}{$\boldsymbol{0}$}
			\State \Return $\{B^{k+1}, \bar{\rho}_s^{k+1}, \phi_p^{k+1}, \boldsymbol{B}^{ek+1}, T_s^{k+1}, \varepsilon_s^{k+1}, \boldsymbol{\sigma}_s^{k+1}\}$
			\EndIf
			\LineComment{1}{Estimate Solid Volume Fraction and Solid Mass Density:}
			\State \textbf{solve} $\phi_s = \bar{\rho}_s^{k+1}/\rho_0 - {\phi_s}^{b+1}({J^{e}}^{-1} - 1)$ \textbf{for} $\phi_s$
			\Comment{from Eq.\ \eqref{eqn:solid_volume_fraction}}
			\State \setequal{1}{$\rho_s$}{$\bar{\rho}_s^{k+1} / \phi_s$}
			\Comment{from Eq.\ \eqref{eqn:porosity}}
			\LineComment{1}{Estimate Scalar Stress Measures and Breakage Energy:}
			\State \setequal{1}{$\bar{p}$}{$(\bar{\rho}_s^{k+1}/ \rho_0) (1 - \theta B^k) p_r (\bar{K}^2 {\epsilon_v^{e}}^2 + 3 \bar{G} \bar{K} {\epsilon_s^{e}}^2) / 4$}
			\Comment{from Eq.\ \eqref{eqn:contact_stresses}}
			\State \setequal{1}{$\bar{q}$}{$(\bar{\rho}_s^{k+1}/ \rho_0) (1 - \theta B^k) 3 \bar{G} p_r (-\bar{K} \epsilon_v^{e} \epsilon_s^{e} / 2 + \sqrt{3 \bar{G} \bar{K}} {\epsilon_s^{e}}^2/4)$}
			\Comment{from Eq.\ \eqref{eqn:contact_stresses}}
			\State \setequal{1}{$p^{*}$}{$\hat{p}_H(\rho_s) [1 - \Gamma_0(1 - \rho_0/\rho_s)/2] + \rho_0 \Gamma_0 (\hat{e}_c(\rho_s) + c_v (T_s^k - T_0))$}
			\Comment{from Eq.\ \eqref{eqn:solid_pressure}}
			\State \setequal{1}{$E_B$}{$(\bar{\rho}_s^{k+1}/ \rho_0) \theta p_r (-\bar{K}^2 {\epsilon_v^{e}}^3 / 12 - 3 \bar{G} \bar{K} \epsilon_v^{e} {\epsilon_s^{e}}^2 /4 + \bar{G} \sqrt{3 \bar{G} \bar{K}} {\epsilon_s^{e}}^3 / 4)$}
			\Comment{from Eq.\ \eqref{eqn:solid_dissipation}}
			\Statex
			\Statex Continued in Algorithm \ref{alg:stress_update_2}...
			
			\algstore{myalg}
		\end{algorithmic}
	\end{algorithm}
	
	\begin{algorithm}
		\caption{Stress update algorithm: Part II.}\label{alg:stress_update_2}
		\begin{algorithmic}[1]
			\algrestore{myalg}
			\Statex \itemsep=0.75ex
			\Statex Continued from Algorithm \ref{alg:stress_update}...
			\LineComment{1}{Estimate Yield Stresses:}
			\State \setequal{0.7}{$\boldsymbol{\sigma}_y$}{$\bar{q}\boldsymbol{B}^{e*}_0 \boldsymbol{B}^{e*} / (3 \epsilon_s^{e}) - \bar{p} \boldsymbol{B}^{e*} - \phi_s p^{*} \hat{C}(\phi_s, J^{e}) \boldsymbol{1}$}
			\Comment{from Eq.\ \eqref{eqn:yield_stress}}
			\State \setequal{0.7}{$p_y$}{$-\tfrac{1}{3} \text{tr}(\boldsymbol{\sigma}_y)$}
			\State \setequal{0.7}{$q_y$}{$\sqrt{\tfrac{3}{2} \boldsymbol{\sigma}_{y0}:\boldsymbol{\sigma}_{y0}}.$}
			\Comment{$\boldsymbol{\sigma}_{y0}$ denotes deviator of $\boldsymbol{\sigma}_{y}$}
			\LineComment{1}{Check $y_1$ Yield Function:}
			\State \setequal{0.7}{$y_1$}{$E_B (1 - B^k)^2/E_c + q_y^{2}/(M p_y)^2 - 1$}
			\Comment{from Eqs.\ \eqref{eqn:y1} and \eqref{eqn:friction_and_dilation}}
			\If{$y_1 < 0$}
			\LineComment{2}{Material is Not Yielding}
			\State \setequal{2}{$B^{k+1}$}{$B^k$}
			\State \setequal{2}{$\phi_p^{k+1}$}{$\phi_p^k$}
			\State \setequal{2}{$\boldsymbol{B}^{ek+1}$}{$\boldsymbol{B}^{e*}$}
			\State \setequal{2}{$\boldsymbol{\sigma}_s^{k+1}$}{$\bar{q}\boldsymbol{B}^{e*}_0 \boldsymbol{B}^{e*} / (3 \epsilon_s^{e}) - \bar{p} \boldsymbol{B}^{e*} - \phi_s p^{*} \hat{A}(\phi_s, J^{e}) \boldsymbol{1}$}
			\Comment{from Eq.\ \eqref{eqn:effective_granular_stress}}
			\State \setequal{2}{$\varepsilon_s^{k+1}$}{$\varepsilon_s^k + \tfrac{1}{2}(\Delta t/ \bar{\rho}_s^{k+1}) (\boldsymbol{\sigma}_s^k + \boldsymbol{\sigma}_s^{k+1}) : \boldsymbol{L}_s^k$}
			\Comment{from Eq.\ \eqref{eqn:energy_conservation}}
			\State \setequal{2}{$T_s^{k+1}$}{$(\varepsilon_s^{k+1} - \hat{\psi_c}(\epsilon_v^e, \epsilon_s^e, B^{k+1}) - \hat{e}_c(\rho_s)) / c_v$}
			\Comment{from Eqs. \eqref{eqn:contact_stresses} and \eqref{eqn:solid_pressure}}
			\State \Return $\{B^{k+1}, \bar{\rho}_s^{k+1}, \phi_p^{k+1}, \boldsymbol{B}^{ek+1}, T_s^{k+1}, \varepsilon_s^{k+1}, \boldsymbol{\sigma}_s^{k+1}\}$
			\Else
			\LineComment{2}{Material is Yielding ($y_1 = 0$): Calculate $\lambda_1$ such that $y_1 = 0$.}
			\While{$y_1 > 0$}
			\LineComment{3}{Estimate $\lambda_1$ (e.g., with Bisection or Newton--Raphson Method)}
			\State \rlap{$\lambda_1$} \hspace{1.5em} $\gets$ best estimate
			\LineComment{3}{Estimate Inelastic Deformation Rates}
			\State \setequal{1.5}{$\dot{B}^k$}{$\lambda_1 \frac{(1 - B^k)^2}{E_c} \cos^2(\omega)$}
			\Comment{from Eqs.\ \eqref{eqn:y1_rates} and \eqref{eqn:friction_and_dilation}}
			\State \setequal{1.5}{$\xi_v^{pk}$}{$\lambda_1 \frac{E_B (1 - B^k)^2}{E_c} \frac{-p_y}{(p_y^2 + q_y^2)}  \text{sin}^2(\omega) + \lambda_1 M_d \frac{q_y}{(M p_y)^2}$}
			\Comment{from Eqs.\ \eqref{eqn:y1_rates} and \eqref{eqn:friction_and_dilation}}
			\State \setequal{1.5}{$\xi_s^{pk}$}{$\lambda_1 \frac{E_B (1 - B^k)^2}{E_c} \frac{q_y}{(p_y^2 + q_y^2)} \text{sin}^2(\omega) + \lambda_1 \frac{q_y}{(M p_y)^2}$}
			\Comment{from Eqs. \eqref{eqn:y1_rates} and \eqref{eqn:friction_and_dilation}}
			\State \setequal{1.5}{$\boldsymbol{\tilde{D}}^{pk}$}{$\tfrac{3}{2} \xi_s^p \boldsymbol{\sigma}_{y0} / q_y + \tfrac{1}{3} \xi_v^p \boldsymbol{1}$}
			\Comment{from Eq.\ \eqref{eqn:inelastic_deformation_rate}}
			\Statex
			\Statex Continued in Algorithm \ref{alg:stress_update_3}...
			
			\algstore{myalg2}
		\end{algorithmic}
	\end{algorithm}
	
	\begin{algorithm}
		\caption{Stress update algorithm: Part III.}\label{alg:stress_update_3}
		\begin{algorithmic}[1]
			\algrestore{myalg2}
			\Statex \itemsep=0.75ex
			\Statex Continued from Algorithm \ref{alg:stress_update_2}...
			\LineComment{3}{Update Estimated Material State}
			\State \setequal{2}{$B^{k+1}$}{$B^k + \Delta t^k \dot{B}^k$}
			\State \setequal{2}{$\phi_p^{k+1}$}{$\phi_p^k + \Delta t^k \xi_{v}^{pk} (1 - \phi_p^k)$}
			\Comment{from Eq.\ \eqref{eqn:inelastic_porosity}}
			\State \setequal{2}{$\boldsymbol{B}^{ek+1}$}{$\boldsymbol{B}^{e*} - \Delta t^k (\boldsymbol{\tilde{D}}^{pk} \boldsymbol{B}^{ek} + \boldsymbol{B}^{ek} \boldsymbol{\tilde{D}}^{pk})$}
			\State \setequal{2}{$\epsilon_v^{e}$}{$\tfrac{1}{2} \text{tr}(\boldsymbol{B}^{ek+1}) - \tfrac{3}{2}$}
			\Comment{from Eq.\ \eqref{eqn:elastic_strain}}
			\State \setequal{2}{$\epsilon_s^{e}$}{$\sqrt{\tfrac{1}{6} \boldsymbol{B}^{ek+1}_0 : \boldsymbol{B}^{ek+1}_0}$}
			\Comment{$\boldsymbol{B}^{ek+1}_0$ denotes deviator of $\boldsymbol{B}^{ek+1}$}
			\State \setequal{2}{$J^{e}$}{$\sqrt{\text{det}(\boldsymbol{B}^{ek+1})}$}
			\Comment{from Eq.\ \eqref{eqn:elastic_strain}}
			\LineComment{3}{Update Volume Fraction and Mass Density from Eqs.\ \eqref{eqn:porosity} and  \eqref{eqn:solid_volume_fraction}}
			\State \textbf{solve} $\phi_s = \bar{\rho}_s^{k+1}/\rho_0 - {\phi_s}^{b+1}({J^{e}}^{-1} - 1)$ \textbf{for} $\phi_s$
			\State \setequal{2}{$\rho_s$}{$\bar{\rho}_s^{k+1} / \phi_s$}
			\LineComment{3}{Update Scalar Stress Measures from Eqs.\ \eqref{eqn:contact_stresses}, \eqref{eqn:solid_pressure} and  \eqref{eqn:solid_dissipation}}
			\State \setequal{2}{$\bar{p}$}{$(\bar{\rho}_s^{k+1}/ \rho_0) (1 - \theta B^k) p_r (\bar{K}^2 {\epsilon_v^{e}}^2 + 3 \bar{G} \bar{K} {\epsilon_s^{e}}^2) / 4$}
			\State \setequal{2}{$\bar{q}$}{$(\bar{\rho}_s^{k+1}/ \rho_0) (1 - \theta B^k) 3 \bar{G} p_r (-\bar{K} \epsilon_v^{e} \epsilon_s^{e} / 2 + \sqrt{3 \bar{G} \bar{K}} {\epsilon_s^{e}}^2/4)$}
			\State \setequal{2}{$p^{*}$}{$\hat{p}_H(\rho_s) [1 - \Gamma_0(1 - \rho_0/\rho_s)/2] + \rho_0 \Gamma_0 (\hat{e}_c(\rho_s) + c_v (T_s^k - T_0))$}
			\State \setequal{2}{$E_B$}{$(\bar{\rho}_s^{k+1}/ \rho_0) \theta p_r (-\bar{K}^2 {\epsilon_v^{e}}^3 / 12 - 3 \bar{G} \bar{K} \epsilon_v^{e} {\epsilon_s^{e}}^2 /4 + \bar{G} \sqrt{3 \bar{G} \bar{K}} {\epsilon_s^{e}}^3 / 4)$}
			\LineComment{3}{Update Yield Stresses and $y_1$ Yield Function}
			\State \setequal{2}{$\boldsymbol{\sigma}_y$}{$\bar{q}\boldsymbol{B}^{ek+1}_0 \boldsymbol{B}^{ek+1} / (3 \epsilon_s^{e}) - \bar{p} \boldsymbol{B}^{ek+1} - \phi_s p^{*} \hat{C}(\phi_s, J^{e}) \boldsymbol{1}$}
			\Comment{from Eq.\ \eqref{eqn:yield_stress}}
			\State \setequal{2}{$p_y$}{$-\tfrac{1}{3} \text{tr}(\boldsymbol{\sigma}_y)$}
			\State \setequal{2}{$q_y$}{$\sqrt{\tfrac{3}{2} \boldsymbol{\sigma}_{y0}:\boldsymbol{\sigma}_{y0}}.$}
			\Comment{$\boldsymbol{\sigma}_{y0}$ denotes deviator of $\boldsymbol{\sigma}_{y}$}
			\State \setequal{2}{$y_1$}{$E_B (1 - B^k)^2/E_c + q_y^{2}/(M p_y)^2 - 1$}
			\Comment{from Eqs.\ \eqref{eqn:y1} and \eqref{eqn:friction_and_dilation}}
			\EndWhile
			\LineComment{2}{Update Final Stress and Energy State}
			\State \setequal{2}{$\boldsymbol{\sigma}_s^{k+1}$}{$\bar{q}\boldsymbol{B}^{ek+1}_0 \boldsymbol{B}^{ek+1} / (3 \epsilon_s^{e}) - \bar{p} \boldsymbol{B}^{ek+1} - \phi_s p^{*} \hat{A}(\phi_s, J^{e}) \boldsymbol{1}$}
			\Comment{from Eq.\ \eqref{eqn:effective_granular_stress}}
			\State \setequal{2}{$\varepsilon_s^{k+1}$}{$\varepsilon_s^k + \tfrac{1}{2}(\Delta t/ \bar{\rho}_s^{k+1}) (\boldsymbol{\sigma}_s^k + \boldsymbol{\sigma}_s^{k+1}) : \boldsymbol{L}_s^k$}
			\Comment{from Eq.\ \eqref{eqn:energy_conservation}}
			\State \setequal{2}{$T_s^{k+1}$}{$(\varepsilon_s^{k+1} - \hat{\psi_c}(\epsilon_v^e, \epsilon_s^e, B^{k+1}) - \hat{e}_c(\rho_s)) / c_v$}
			\Comment{from Eqs. \eqref{eqn:contact_stresses} and \eqref{eqn:solid_pressure}}
			\State \Return $\{B^{k+1}, \bar{\rho}_s^{k+1}, \phi_p^{k+1}, \boldsymbol{B}^{ek+1}, T_s^{k+1}, \varepsilon_s^{k+1}, \boldsymbol{\sigma}_s^{k+1}\}$
			\EndIf
			\EndFunction
		\end{algorithmic}
	\end{algorithm}

	\setcounter{figure}{0}
	
	\section{Notes on Model Theory}
	\label{sec:appendix_theory}
	In this section, we provide additional details about the thermodynamic and mechanical theories used in formulating the model. The model developed in this work is constructed within the framework of continuum mechanics, using the fluid--sediment mixture theories of \citet{bedford1983} and \citet{jackson2000}; the kinematic theory of \citet{coleman1963}; and the breakage mechanics theory of \citet{einav2007a}. Here, we provide brief overviews of these theories, including derivation of equations presented in the main body of this work.
	
	\subsection{Derivation of Governing Equations}
	\label{sec:appendix_thermodynamics}
\begin{figure}[!h]
		\centering
		\includegraphics[width=0.95\linewidth]{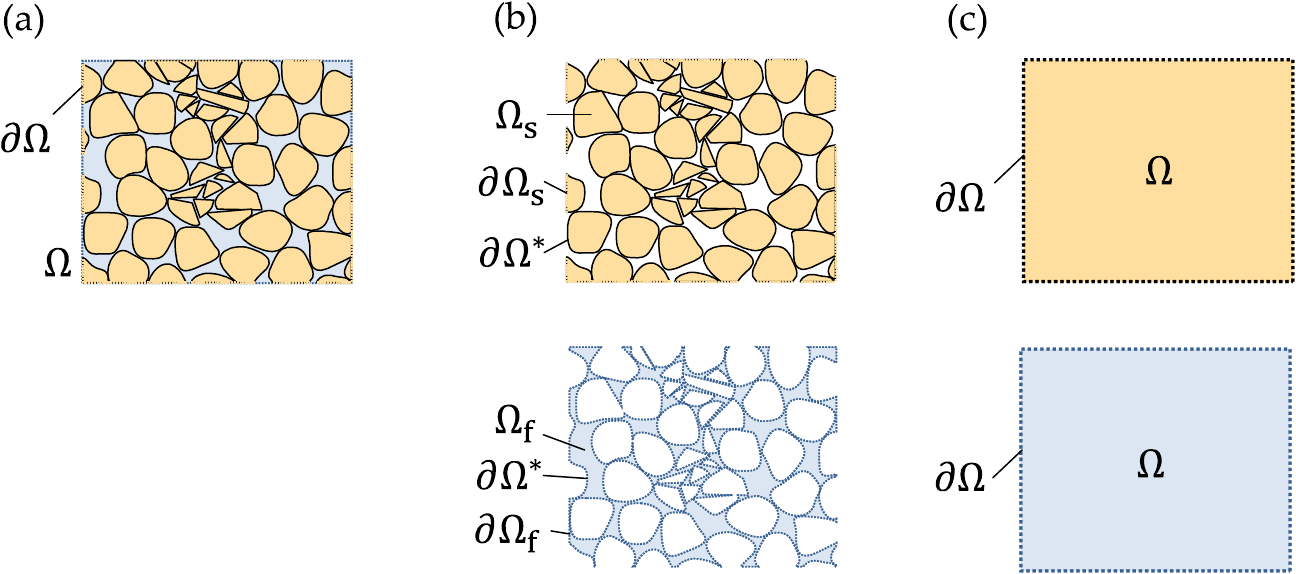}
		\caption{Illustration of a representative volume element (RVE) of fluid-saturated, granular material. The representative volume $\Omega$ shown in (a) is separable into a solid components $\Omega_s$ and a fluid component $\Omega_f$ shown in (b). These two components of the RVE, their external boundaries $\partial \Omega_s$ and $\partial \Omega_f$, and their interior surface $\partial \Omega^*$ are used to construct \emph{effective} material fields, which are continuous in the volume $\Omega$ shown in (c).}
		\label{fig:appendix_rve}
	\end{figure}

	We begin this summary by considering the system of governing equations presented in \eqref{eqn:mass_conservation}--\eqref{eqn:entropy_imbalance}. These are derived from basic conservation laws --- namely, (i) mass conservation, (ii) momentum conservation, (iii) energy conservation, and (iv) the imbalance of entropy --- as applied to mixtures of fluids and granular sediments. In this brief derivation we primarily consider the representative volume of mixed material, $\Omega$, shown in Figure \ref{fig:appendix_rve}a. We assume that the grains within this volume are composed of a solid material with \emph{true} density $\rho_s$, and we assume that the pore space between the grains is occupied by a fluid material with \emph{true} density $\rho_f$. This representative volume of material, $\Omega$, can therefore be decomposed into a solid volume, $\Omega_s$, and a fluid volume, $\Omega_f$, such that $\Omega = \Omega_s \cup \Omega_f$ (see Figure \ref{fig:appendix_rve}b). The fraction of the total volume of $\Omega$ that is occupied by solid material (i.e., $\Omega_s$) is defined as the \emph{solid volume fraction}, $\phi_s$, and the fraction occupied by fluid material (i.e., $\Omega_f$) is defined as the \emph{fluid volume fraction} or \emph{porosity}, $\phi_f$.
	
	Further, we assume that the all of the material within the volume $\Omega$ has a well-defined \emph{true} velocity, $\boldsymbol{v}^*$; specific internal energy, $\varepsilon^*$; specific entropy $s^*$; and temperature $T^*$. If we allow the true material fields to have both spatial and temporal variations, we may define:
	\begin{equation}
		\boldsymbol{v}^* = \boldsymbol{v}^*(\boldsymbol{x},t), \quad
		\varepsilon^* = \varepsilon^*(\boldsymbol{x},t), \quad
		s^* = s^*(\boldsymbol{x},t), \quad \text{and} \quad
		T^* = T^*(\boldsymbol{x},t),
	\end{equation}
	for all spatial points $\boldsymbol{x}$ in the volume $\Omega$ and all times $t$. 
	%
	From these \emph{true} material fields, we can define \emph{effective} material fields --- namely, a solid and fluid effective mass density, $\bar{\rho}_s$ and $\bar{\rho}_f$; a homogenized solid and fluid velocity, $\boldsymbol{v}_s$ and $\boldsymbol{v}_f$; homogenized specific internal energies, $\varepsilon_s$ and $\varepsilon_f$; homogenized specific entropies, $s_s$ and $s_f$; and average temperatures, $T_s$ and $T_f$ --- which also have spatial and temporal variation:
	\begin{equation}
		\bar{\rho}_s = \bar{\rho}_s(\boldsymbol{x}, t), \quad
		\boldsymbol{v}_s = \boldsymbol{v}_s(\boldsymbol{x},t), \quad
		\varepsilon_s = \varepsilon_s(\boldsymbol{x},t), \quad
		s_s = s_s(\boldsymbol{x},t), \quad \text{and} \quad
		T_s = T_s(\boldsymbol{x},t),
	\end{equation}
	and,
	\begin{equation}
		\bar{\rho}_f = \bar{\rho}_f(\boldsymbol{x}, t), \quad
		\boldsymbol{v}_f = \boldsymbol{v}_f(\boldsymbol{x},t), \quad
		\varepsilon_f = \varepsilon_f(\boldsymbol{x},t), \quad
		s_f = s_f(\boldsymbol{x},t), \quad \text{and} \quad
		T_f = T_f(\boldsymbol{x},t).
	\end{equation}
	For the problems considered in this work, it is generally infeasible to model the \emph{true} material behaviors directly. Therefore, these \emph{effective} material fields are constructed to represent volume and mass averages of conserved material quantities over an ensemble of RVEs (e.g., see Figure \ref{fig:appendix_rve}c):
	\begin{equation}
		\label{eqn:appendix_effective_fields}
		\begin{aligned}
			&\int_{\Omega} \bar{\rho}_s\ dv \equiv \int_{\Omega_s} \rho_s\ dv, \quad &&
			\int_{\Omega} \bar{\rho}_f\ dv \equiv \int_{\Omega_f} \rho_f\ dv,\\
			&\int_{\Omega} \bar{\rho}_s \boldsymbol{v}_s\ dv \equiv \int_{\Omega_s} \rho_s \boldsymbol{v}^*\ dv, \quad &&
			\int_{\Omega} \bar{\rho}_f\boldsymbol{v}_f\ dv \equiv \int_{\Omega_f} \rho_f \boldsymbol{v}^*\ dv,\\
			&\int_{\Omega} \bar{\rho}_s s_s\ dv \equiv \int_{\Omega_s} \rho_s s^*\ dv, \quad &&
			\int_{\Omega} \bar{\rho}_f s_f\ dv \equiv \int_{\Omega_f} \rho_f s^*\ dv,\\
			&\int_{\Omega} \bar{\rho}_s T_s\ dv \equiv \int_{\Omega_s} \rho_s T^*\ dv, \quad &&
			\int_{\Omega} \bar{\rho}_f T_f\ dv \equiv \int_{\Omega_f} \rho_f T^*\ dv,\\
		\end{aligned}
	\end{equation}
	and finally,
	\begin{equation*}
		\tag{\ref{eqn:appendix_effective_fields}, cont.}
		\begin{aligned}
			&\int_{\Omega} \bar{\rho}_s (\varepsilon_s + \tfrac{1}{2} \boldsymbol{v}_s \cdot \boldsymbol{v}_s)\ dv \equiv \int_{\Omega_s} \rho_s (\varepsilon^* + \tfrac{1}{2} \boldsymbol{v}^* \cdot \boldsymbol{v}^*)\ dv, \quad \text{and}\\
			&\int_{\Omega} \bar{\rho}_f (\varepsilon_f + \tfrac{1}{2} \boldsymbol{v}_f \cdot \boldsymbol{v}_f)\ dv \equiv \int_{\Omega_f} \rho_f (\varepsilon^* + \tfrac{1}{2} \boldsymbol{v}^* \cdot \boldsymbol{v}^*)\ dv.
		\end{aligned}
	\end{equation*}
	If the integrating volumes are chosen to be \emph{large} relative to the size of the grains but \emph{small} relative to the scale of any details of interest (e.g., impactor geometries), these \emph{effective} material fields should be smoothly varying, continuous, and retain important details about the properties of the deforming constituent materials.
	
	This construction of the effective material fields ensures that conservation of mass, momentum, and energy in the \emph{effective} continua will correspond to conservation of mass, momentum, and energy in the \emph{true} solid and fluid constituents of the mixture. By analogy, we also assume that this feature extends to the imbalance of entropy across scales; however, a more thorough analysis of how changing integrating volumes affects the evaluation of entropy is needed. What remains, then, is to answer the question: \emph{how do fluxes at the boundary of the volume, $\partial \Omega$, govern the change of mass, momentum, and energy in the volume, $\Omega$?}
	
	To answer this question, we need to define the boundaries of the spatial integrating volumes above --- namely, $\partial \Omega$, $\partial \Omega_s$, $\partial \Omega_f$, and $\partial \Omega^*$ in Figure \ref{fig:appendix_rve}. The exterior boundary of the spatial volume $\Omega$ is $\partial \Omega$, and it can be decomposed into a solid boundary $\partial \Omega_s$ and a fluid boundary $\partial \Omega_f$, with $\partial \Omega = \partial \Omega_s \cup \partial \Omega_f$. (Note that $\partial \Omega$, $\partial \Omega_s$, and $\partial \Omega_f$ are assumed to remain fixed in space.) The \emph{interior boundary}, $\partial \Omega^*$, denotes the boundary between the solid and fluid materials, and is assumed to be \emph{co-moving with the particles}.
	
	If we assume that the \emph{true} constituent materials are non-reactive and non-volatile (i.e, there is no combustion or vaporization at the interior grain surfaces associated with $\partial \Omega^*$), then the conservation of mass for each material within a representative volume $\Omega$ can be expressed in the following form:
	\begin{equation}
		\begin{aligned}
			&\frac{\partial}{\partial t} \int_{\Omega_s}{\rho_s\ dv} + \int_{\partial \Omega_s}{\rho_s \boldsymbol{v}^* \cdot \hat{\boldsymbol{n}}\ da} = 0, \\
			&\frac{\partial}{\partial t} \int_{\Omega_f}{\rho_f\ dv} + \int_{\partial \Omega_f}{\rho_f \boldsymbol{v}^* \cdot \hat{\boldsymbol{n}}\ da} = 0,
		\end{aligned}
	\end{equation}
	with $\hat{\boldsymbol{n}}$ the outward facing surface normal vector. This statement of mass conservation (i.e., the change of mass within the volume must be balanced by the flux of mass across its boundary) can be re-expressed in terms of the \emph{effective} material fields defined above, \emph{if we assume a similarity between the volume integral identities in \eqref{eqn:appendix_effective_fields} and the boundary integrals of sufficiently large RVEs}:
	\begin{equation}
		\begin{aligned}
			&\frac{\partial}{\partial t} \int_\Omega{\bar{\rho}_s dv} + \int_{\partial \Omega}{\bar{\rho}_s \boldsymbol{v}_s \cdot \hat{\boldsymbol{n}} da} = 0, \\
			&\frac{\partial}{\partial t} \int_\Omega{\bar{\rho}_f dv} + \int_{\partial \Omega}{\bar{\rho}_f \boldsymbol{v}_f \cdot \hat{\boldsymbol{n}} da} = 0.
		\end{aligned}
	\end{equation}
	Since these expressions should be true for any representative volume, $\Omega$, we can apply the divergence theorem and isolate the strong-form mass conservation rules:
	\begin{equation}
		\begin{aligned}
			&\frac{\partial \bar{\rho}_s}{\partial t} + \text{div} (\bar{\rho}_s \boldsymbol{v}_s) = 0,\\
			&\frac{\partial \bar{\rho}_f}{\partial t} + \text{div} (\bar{\rho}_f \boldsymbol{v}_f) = 0.
		\end{aligned}
	\end{equation}
	Together with \eqref{eqn:material_time_derivative}, these final expressions can be shown to be equivalent to \eqref{eqn:mass_conservation}.
	
	Similarly, the rate of change of the momentum for each material within a representative volume $\Omega$ can be expressed in the following form:
	\begin{equation}
		\label{eqn:appendix_momentum_integral}
		\begin{aligned}
			&\frac{\partial}{\partial t} \int_{\Omega_s}{\rho_s \boldsymbol{v}^*\ dv} + \int_{\partial \Omega_s}{\rho_s \boldsymbol{v}^* (\boldsymbol{v}^* \cdot \hat{\boldsymbol{n}})\ da} = \int_{\Omega_s}{\rho_s \boldsymbol{g} dv} - \int_{\partial\Omega^*}\boldsymbol{t}(\hat{\boldsymbol{n}})\ da + \int_{\partial\Omega_s}\boldsymbol{t}(\hat{\boldsymbol{n}})\ da,\\
			&\frac{\partial}{\partial t} \int_{\Omega_f}{\rho_f \boldsymbol{v}^*\ dv} + \int_{\partial \Omega_f}{\rho_f \boldsymbol{v}^* (\boldsymbol{v}^* \cdot \hat{\boldsymbol{n}})\ da} = \int_{\Omega_f}{\rho_f \boldsymbol{g}\ dv} + \int_{\partial\Omega^*}\boldsymbol{t}(\hat{\boldsymbol{n}})\ da + \int_{\partial\Omega_f}\boldsymbol{t}(\hat{\boldsymbol{n}})\ da,
		\end{aligned}
	\end{equation}
	with $\boldsymbol{t}(\hat{\boldsymbol{n}})$ the surface traction vector, a function of the outward pointing surface normal vector $\hat{\boldsymbol{n}}$; and $\boldsymbol{g}$ the gravitational acceleration vector. This statement of momentum conservation (i.e., the change of momentum within the volume must be balanced by flux of momentum across its boundary and the external forces acting on it) can be re-expressed in terms of the effective material fields in \eqref{eqn:appendix_effective_fields}, if we introduce \emph{effective} internal stresses and interaction forces to account for the tractions along $\partial \Omega$ and $\partial \Omega^*$, respectively:
	\begin{equation}
		\begin{aligned}
			&\int_{\partial \Omega}{\boldsymbol{\Sigma}_s\hat{\boldsymbol{n}}\ da} \equiv \int_{\partial \Omega_s}{\boldsymbol{t}(\hat{\boldsymbol{n}}) - \rho_s \boldsymbol{\delta v}_s (\boldsymbol{\delta v}_s \cdot \hat{\boldsymbol{n}})\ da},\\
			&\int_{\partial \Omega}{\boldsymbol{\Sigma}_f\hat{\boldsymbol{n}}\ da} \equiv \int_{\partial \Omega_f}{\boldsymbol{t}(\hat{\boldsymbol{n}}) - \rho_f \boldsymbol{\delta v}_f (\boldsymbol{\delta v}_f \cdot \hat{\boldsymbol{n}})\ da},\\
			&\int_\Omega (\boldsymbol{f}_b + \boldsymbol{f}_d) dv \equiv \int_{\partial \Omega^*} \boldsymbol{t}(\hat{\boldsymbol{n}}) da.
		\end{aligned}
	\end{equation}
	Here $\boldsymbol{\Sigma}_s$ is the solid Cauchy stress; $\boldsymbol{\Sigma}_f$ is the fluid Cauchy stress; $\boldsymbol{f}_b$ is the \emph{buoyant} interaction force; $\boldsymbol{f}_d$ is the \emph{drag} interaction force; and $\boldsymbol{\delta v}_s$ and $\boldsymbol{\delta v}_f$ are the \emph{fluctuational} components of the 
	\emph{true} material velocities. These velocity fluctuations are defined as $\boldsymbol{\delta v}_s \equiv \boldsymbol{v}^* - \boldsymbol{v}_s$ and $\boldsymbol{\delta v}_f \equiv \boldsymbol{v}^* - \boldsymbol{v}_s$. Importantly, these fluctuations give rise to the \emph{tortuous stresses} in the fluid and \emph{thermal stress} in the granular solid \citep[e.g., see][]{batchelor1970,haff1983,coussy2004}. Substituting these expressions --- along with the effective material identities in \eqref{eqn:appendix_effective_fields} --- into the momentum balance integrals in \eqref{eqn:appendix_momentum_integral}, we may apply the divergence theorem to find:
	\begin{equation}
		\label{eqn:appendix_effective_momentum_integral}
		\begin{aligned}
			&\frac{\partial}{\partial t} \int_\Omega \bar{\rho}_s \boldsymbol{v}_s\ dv + \int_\Omega \text{div}\big(\bar{\rho}_s\boldsymbol{v}_s \otimes \boldsymbol{v}_s \big)\ dv = \int_\Omega \big(\bar{\rho}_s \boldsymbol{g} - \boldsymbol{f_b} - \boldsymbol{f_d} + \text{div} (\boldsymbol{\Sigma}_s)\big)\ dv,\\
			&\frac{\partial}{\partial t} \int_{\Omega} \bar{\rho}_f \boldsymbol{v}_f\ dv + \int_\Omega \text{div} \big(\bar{\rho}_f\boldsymbol{v}_f 	\otimes \boldsymbol{v}_f \big)\ dv = \int_\Omega \big(\bar{\rho}_f \boldsymbol{g} + \boldsymbol{f_b} + \boldsymbol{f_d} + \text{div} (\boldsymbol{\Sigma}_f) \big)\ dv,
		\end{aligned}
	\end{equation}
	with $\otimes$ denoting the tensor product operator.
	
	Following the analyses in \citet{drumheller2000}, \citet{baumgarten2021a}, and \citet{baumgarten2021b}, we may assume that the tortuous stresses and thermal stresses mentioned above are small and that there is a smoothly varying fluid pressure, $p_f$, within the volume $\Omega$. Using these two assumptions, the Cauchy stresses and buoyant interaction force can be re-expressed as follows:
	\begin{equation}
		\boldsymbol{\Sigma}_s = \boldsymbol{\sigma}_s - \phi_s p_f \boldsymbol{1}, \quad \boldsymbol{\Sigma}_f = \boldsymbol{\tau}_f - \phi_f p_f \boldsymbol{1}, \quad \text{and} \quad \boldsymbol{f}_b = p_f \nabla \phi_f,
	\end{equation}
	where $\boldsymbol{\sigma}_s$, $\boldsymbol{\tau}_f$, and $p_f$ denote the effective granular stress tensor, the effective fluid shear stress tensor, and the fluid pore pressure described in the main text.
	Substituting these into the integral expressions in \eqref{eqn:appendix_effective_momentum_integral}, and recognizing that these expressions should be true for any representative volume $\Omega$, we can isolate the strong-form momentum conservation rules:
	\begin{equation}
		\begin{aligned}
			&\frac{\partial \bar{\rho}_s \boldsymbol{v}_s}{\partial t} + \text{div}\big(\bar{\rho}_s\boldsymbol{v}_s \otimes \boldsymbol{v}_s \big) = \bar{\rho}_s \boldsymbol{g} - \boldsymbol{f_d} + \text{div} \boldsymbol{\sigma}_s - \phi_s \nabla p_f,\\
			&\frac{\partial \bar{\rho}_f \boldsymbol{v}_f}{\partial t} + \text{div} \big(\bar{\rho}_f\boldsymbol{v}_f \otimes \boldsymbol{v}_f \big) = \bar{\rho}_f \boldsymbol{g} + \boldsymbol{f_d} + \text{div} \boldsymbol{\tau}_f - \phi_f \nabla p_f.
		\end{aligned}
	\end{equation}
	Together with \eqref{eqn:material_time_derivative} and \eqref{eqn:mass_conservation}, these expressions can be shown to be equivalent to \eqref{eqn:momentum_conservation}.
	
	The final conservation expressions that we derive in this section are for the conservation of total energy. If we assume that the \emph{true} constituent materials have a well defined internal heat flow rate, $\boldsymbol{q}^*$, and a well-defined rate of heat generation, $q^*$, associated with the absolute temperature of the materials, $T^*$, then we can express conservation of energy within a representative volume as follows:
	\begin{equation}
		\begin{aligned}
			\frac{\partial}{\partial t}\int_{\Omega_s} \rho_s \big(\varepsilon^* + \tfrac{1}{2} \boldsymbol{v}^* \cdot \boldsymbol{v}^* \big)\ dv =& - \int_{\partial \Omega_s}{\rho_s \big(\varepsilon^* + \tfrac{1}{2} \boldsymbol{v}^* \cdot \boldsymbol{v}^* \big)(\boldsymbol{v}^*\cdot\hat{\boldsymbol{n}})\ da}\\
			& -\int_{\partial\Omega_s}(\boldsymbol{q}^* \cdot \hat{\boldsymbol{n}})\ da + \int_{\partial \Omega^*} (\boldsymbol{q}^* \cdot \hat{\boldsymbol{n}})\ da\\
			& + \int_{\Omega_s} q^*\ dv \\
			& + \int_{\partial \Omega_s}(\boldsymbol{t}(\hat{\boldsymbol{n}}) \cdot \boldsymbol{v}^*)\ da -\int_{\partial \Omega^*}(\boldsymbol{t}(\hat{\boldsymbol{n}}) \cdot \boldsymbol{v}^*)\ da\\
			&+ \int_{\Omega_s}(\rho_s \boldsymbol{g} \cdot \boldsymbol{v}^*)\ dv,
		\end{aligned}
	\end{equation}
	and,
	\begin{equation}
		\label{eqn:appendix_energy_integral}
		\begin{aligned}
			\frac{\partial}{\partial t}\int_{\Omega_f} \rho_f \big(\varepsilon^* + \tfrac{1}{2} \boldsymbol{v}^* \cdot \boldsymbol{v}^* \big)\ dv =& - \int_{\partial \Omega_f}{\rho_f \big(\varepsilon^* + \tfrac{1}{2} \boldsymbol{v}^* \cdot \boldsymbol{v}^* \big)(\boldsymbol{v}^*\cdot\hat{\boldsymbol{n}})\ da}\\
			& -\int_{\partial\Omega_f}(\boldsymbol{q}^* \cdot \hat{\boldsymbol{n}})\ da - \int_{\partial \Omega^*} (\boldsymbol{q}^* \cdot \hat{\boldsymbol{n}})\ da\\
			& + \int_{\Omega_f} q^*\ dv \\
			& + \int_{\partial \Omega_f}(\boldsymbol{t}(\hat{\boldsymbol{n}}) \cdot \boldsymbol{v}^*)\ da + \int_{\partial \Omega^*}(\boldsymbol{t}(\hat{\boldsymbol{n}}) \cdot \boldsymbol{v}^*)\ da\\
			&+ \int_{\Omega_f}(\rho_f \boldsymbol{g} \cdot \boldsymbol{v}^*)\ dv.
		\end{aligned}
	\end{equation}
	These statements of energy conservation --- i.e., the change in \emph{total energy} within the volume must be balanced by flux of \emph{total energy} across its boundary, the external work acted on it, the flow of heat across its boundary, and the generation of heat within it --- can be re-expressed in terms of \emph{effective} material fields in \eqref{eqn:appendix_effective_fields}, if we introduce \emph{effective} rates of heat flow and heat generation:
	\begin{equation}
		\label{eqn:appendix_heat_flow}
		\begin{aligned}
			&\int_{\partial \Omega} (\boldsymbol{q}_s \cdot \hat{\boldsymbol{n}})\ da \equiv \int_{\partial \Omega_s} \big((\boldsymbol{q}^* \cdot \hat{\boldsymbol{n}}) + \rho_s \varepsilon^* (\boldsymbol{\delta v}_s \cdot \hat{\boldsymbol{n}})\big)\ da, \\
			&\int_{\partial \Omega} (\boldsymbol{q}_f \cdot \hat{\boldsymbol{n}})\ da \equiv \int_{\partial \Omega_f} \big((\boldsymbol{q}^* \cdot \hat{\boldsymbol{n}}) + \rho_f \varepsilon^* (\boldsymbol{\delta v}_f \cdot \hat{\boldsymbol{n}})\big)\ da,
		\end{aligned}
	\end{equation}
	and,
	\begin{equation}
		\tag{\ref{eqn:appendix_heat_flow}, cont.}
		\begin{aligned}
			&\int_\Omega q_s\ dv \equiv \int_{\Omega_s} q^*\ dv,\\
			&\int_\Omega q_f\ dv \equiv \int_{\Omega_f} q^*\ dv,\\
			&\int_\Omega q_i\ dv \equiv - \int_{\partial \Omega^*} (\boldsymbol{q}^* \cdot \hat{\boldsymbol{n}})\ da.
		\end{aligned}
	\end{equation}
	Additionally, if we assume that the magnitude of the local velocity fluctuations (i.e., $\boldsymbol{\delta v}_s$ and $\boldsymbol{\delta v}_f$) are small and that the interior grain boundaries (associated with $\partial \Omega^*$) generally move with the velocity of the solid particles, we can use the following approximation for the internal work acting on the interior boundary $\partial \Omega^*$:
	\begin{equation}
		\label{eqn:appendix_internal_work}
		\int_{\partial \Omega^*} (\boldsymbol{t}(\hat{\boldsymbol{n}}) \cdot \boldsymbol{v}^*)\ da \approx \int_\Omega \bigg( (\boldsymbol{f}_d + \boldsymbol{f}_d) \cdot \boldsymbol{v}_s - \phi_s p_f \text{div}(\boldsymbol{v}_s) + \frac{\phi_s p_f}{\rho_s} \bigg(\frac{d^s \rho_s}{dt}\bigg) \bigg)\ dv.
	\end{equation}
	(Note, the first term in this approximation derives from the definition of $\boldsymbol{f}_d$ and $\boldsymbol{f}_b$ above; the second term derives from the pressure-volume work acting on particles intersecting the boundary $\partial \Omega$; and the third term derives from the pressure-volume work acting on interior particles with changing \emph{true} densities.)
	
	Substituting the identities in \eqref{eqn:appendix_heat_flow} and the approximation in \eqref{eqn:appendix_internal_work} into the integral expressions in \eqref{eqn:appendix_energy_integral}, we may apply the divergence theorem to find \emph{effective} energy conservation expressions:
	\begin{equation}
		\begin{aligned}
			\frac{\partial}{\partial t}\int_{\Omega} \bar{\rho}_s \big(\varepsilon_s + \tfrac{1}{2} \boldsymbol{v}_s 	\cdot \boldsymbol{v}_s \big) dv =& - \int_{\Omega}{ \text{div}\big(\bar{\rho}_s \big(\varepsilon_s + \tfrac{1}{2} \boldsymbol{v}_s \cdot \boldsymbol{v}_s \big)\boldsymbol{v}_s\big)\ dv}\\
			& +\int_{\Omega} \text{div}(\boldsymbol{\sigma}_s \boldsymbol{v}_s - \boldsymbol{q}_s)\ dv\\
			& + \int_{\Omega} \bigg( q_s - q_i + (\bar{\rho}_s \boldsymbol{g} - \boldsymbol{f_d} - \phi_s \nabla p_f) \cdot \boldsymbol{v}_s - \frac{\phi_s p_f}{\rho_s} \frac{d^s \rho_s}{dt}\bigg)\ dv,
		\end{aligned}
	\end{equation}
	and,
	\begin{equation}
		\begin{aligned}
			\frac{\partial}{\partial t}\int_{\Omega} \bar{\rho}_f \big(\varepsilon_f + \tfrac{1}{2} \boldsymbol{v}_f 	\cdot \boldsymbol{v}_f \big) dv =& - \int_{\Omega}{\text{div}\big(\bar{\rho}_f \big(\varepsilon_f + \tfrac{1}{2} \boldsymbol{v}_f \cdot \boldsymbol{v}_f \big)\boldsymbol{v}_f\big)\ dv}\\
			& +\int_{\Omega} \text{div}(\boldsymbol{\tau}_f \boldsymbol{v}_f - \boldsymbol{q}_f) \ dv\\
			& + \int_{\Omega} (q_f + q_i + (\bar{\rho}_f \boldsymbol{g} - \phi_f \nabla p_f) \cdot \boldsymbol{v}_f + \boldsymbol{f_d} \cdot \boldsymbol{v}_s)\ dv\\
			& - \int_{\Omega} \bigg( \frac{\phi_s p_f}{\rho_s} \frac{d^s \rho_s}{dt} + p_f \text{div}(\phi_f \boldsymbol{v}_f + \phi_s \boldsymbol{v}_s)\bigg)\ dv.
		\end{aligned}
	\end{equation}
	Through some careful manipulation, the final integrand above can be simplified using \eqref{eqn:porosity}--\eqref{eqn:mass_conservation} to $-\phi_f p_f/\rho_f (d^f \rho_f/dt)$. Since these expressions should be valid for any representative volume $\Omega$, we can isolate the strong-form energy conservation rules:
	\begin{equation}
		\begin{aligned}
			\frac{\partial \bar{\rho}_s E_s}{\partial t} + \text{div}(\bar{\rho}_s E_s \boldsymbol{v}_s) =&\ \text{div}(\boldsymbol{\sigma}_s \boldsymbol{v}_s - \boldsymbol{q}_s) + \bar{\rho}_s \boldsymbol{g}\cdot \boldsymbol{v}_s - \boldsymbol{f}_d \cdot \boldsymbol{v}_s - \phi_s \nabla p_f \cdot \boldsymbol{v}_s - \frac{\phi_s p_f}{\rho_s} \frac{d^s \rho_s}{dt}\\
			& + q_s - q_i,\\[1em]
			\frac{\partial \bar{\rho}_f E_f}{\partial t} + \text{div}(\bar{\rho}_f E_f \boldsymbol{v}_f) =&\ \text{div}(\boldsymbol{\tau}_f \boldsymbol{v}_f - \boldsymbol{q}_f) + \bar{\rho}_f \boldsymbol{g}\cdot \boldsymbol{v}_f + \boldsymbol{f}_d \cdot \boldsymbol{v}_s - \phi_f \nabla p_f \cdot \boldsymbol{v}_f - \frac{\phi_f p_f}{\rho_f} \frac{d^f \rho_f}{dt}\\
			& + q_f + q_i,
		\end{aligned}
	\end{equation}
	with $E_s \equiv \varepsilon_s + \tfrac{1}{2} \boldsymbol{v}_s \cdot \boldsymbol{v}_s$ and $E_f \equiv \varepsilon_f + \tfrac{1}{2} \boldsymbol{v}_f \cdot \boldsymbol{v}_f$, the specific total energies for each effective material phase. Together with \eqref{eqn:material_time_derivative}, \eqref{eqn:mass_conservation}, and \eqref{eqn:momentum_conservation}, this expression can be shown to be identical to the model equations in \eqref{eqn:energy_conservation}. 
	
	This general approach to deriving the governing equations is also performed for the imbalance of entropy associated with the model expression in \eqref{eqn:entropy_imbalance}; however, we do not report this derivation here. Further (and more detailed) discussion and derivations of this system of governing equations can be found in \citet{drumheller2000}, \citet{jackson2000}, Appendix A of \citet{baumgarten2021a}, and Chapter 2 of \citet{baumgarten2021b}.

	\subsection{Derivation of Granular Kinematics}
	\label{sec:appendix_kinematics}
\begin{figure}[!h]
		\hfil True Granular Solid \hfil
		\par\vspace{1ex}
		(a)
		\begin{center}
			\includegraphics[width=0.95\linewidth]{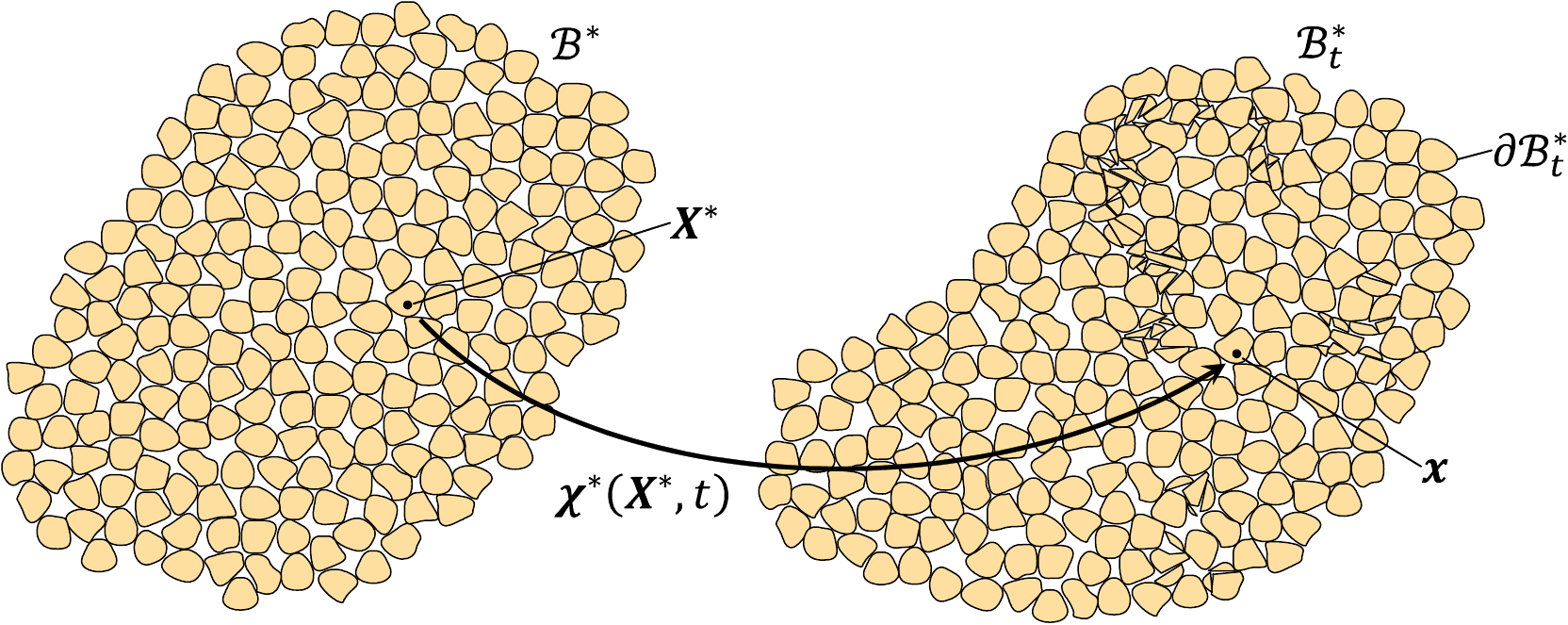}
		\end{center}
		\par\vspace{2em}
		\hfil Effective Continuum Material \hfil
		\par\vspace{1ex}
		(b)
		\begin{center}
			\includegraphics[width=0.95\linewidth]{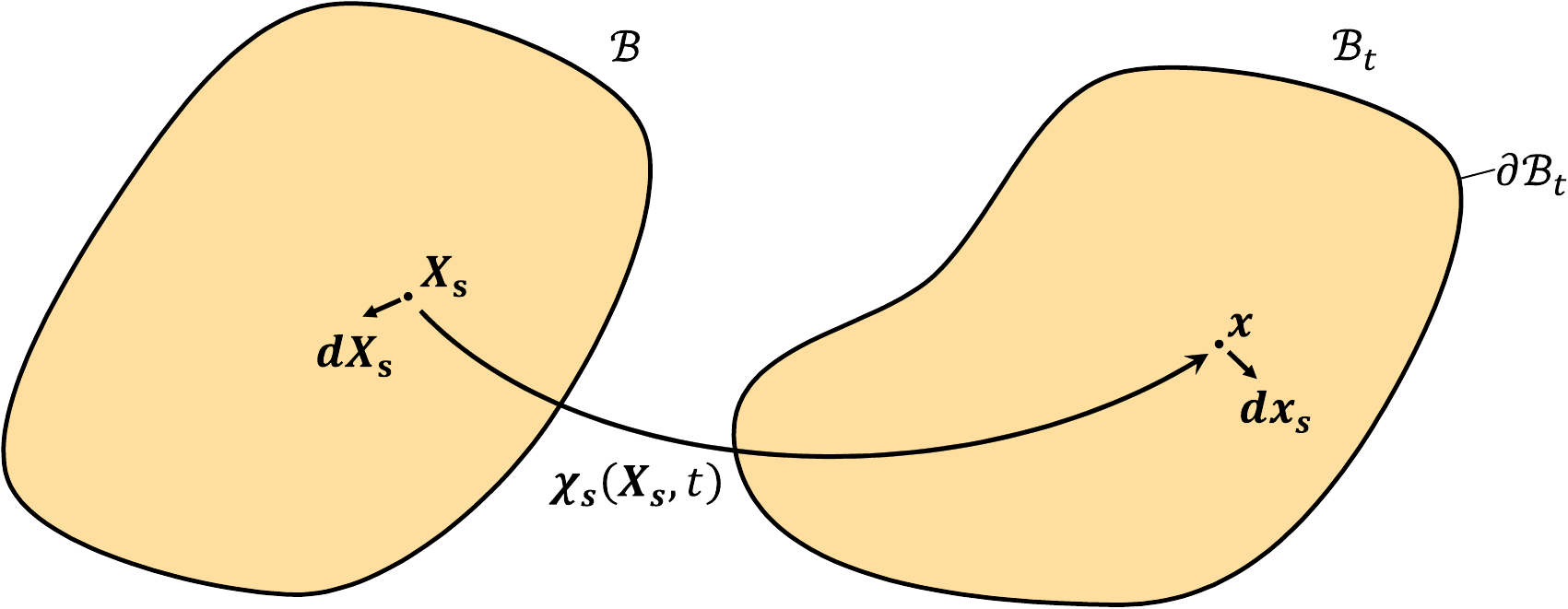}
		\end{center}
		\par\vspace{2em}
		\caption{Reference and deformed continuum bodies associated with (a) the true granular solid and (b) the effective continuum material that we use to represent it. The motion functions $\boldsymbol{\chi}^*(\boldsymbol{X}^*,t)$ and $\boldsymbol{\chi}_s(\boldsymbol{X}_s,t)$ map points in the true reference body ($\boldsymbol{X}^*$) and effective reference body ($\boldsymbol{X}_s$) to spatial points ($x$) in the true and effective deformed bodies, respectively. In the effective continuum material, it is useful to describe the deformation gradient, $\boldsymbol{F}$, which maps an infinitesimal vector $\boldsymbol{dX}_s$ in the effective reference body $\mathcal{B}$ to a vector $\boldsymbol{dx}_s$ in the effective deformed body $\mathcal{B}_t$.}
		\label{fig:appendix_kinematics}
	\end{figure}

	Here, we continue the discussion of the proposed model theory by considering the physical connection between material motions and deformations within granular materials, following the basic kinematic principles described in \citet{gurtin2010}. In this analysis, we consider both the true granular solid (shown in Figure \ref{fig:appendix_kinematics}a) alongside the effective continuum material that we use to represent it (shown in Figure \ref{fig:appendix_kinematics}b). The purpose of this discussion is to clarify the elastic-plastic decomposition of material deformations used in the formulation of the model.

	First, consider the \emph{true} granular material shown in Figure \ref{fig:appendix_kinematics}a. We identify the region of space occupied by a \emph{reference configuration} of this material as the \emph{true reference body}, $\mathcal{B}^*$. This reference configuration is commonly taken to be a hypothetical \emph{undeformed} configuration of the material that is later moved into a \emph{deformed} configuration. A position within this region of space is referred to as a \emph{true material point}, which is denoted by $\boldsymbol{X}^*$ and represents the location of a \emph{material particle}.
	
	Similarly, we identify the region of space that is actually occupied by the \emph{true} granular material --- at some time $t$ --- as the \emph{true deformed body}, $\mathcal{B}^*_t$. Each material point $\boldsymbol{X}^*$ in $\mathcal{B}^*$ can be mapped to a unique spatial point $\boldsymbol{x}$ in $\mathcal{B}^*_t$ using the \emph{motion function} $\boldsymbol{\chi}^*(\boldsymbol{X}^*, t)$. The time rate of change of the motion function --- i.e., the rate of change of the position $\boldsymbol{x}$ of a material point $\boldsymbol{X}^*$ --- is the \emph{true} velocity of the granular solid, $\boldsymbol{v}^*$:
	\begin{equation}
		\boldsymbol{v}^* = \frac{\partial \boldsymbol{\chi}^*(\boldsymbol{X}^*,t)}{\partial t}.
	\end{equation}
	
	We also consider the effective continuum material, shown in Figure \ref{fig:appendix_kinematics}b, that we use to represent the true granular solid. Using the true reference body $\mathcal{B}^*$, we construct an \emph{effective reference body} $\mathcal{B}$ through the homogenization procedure described in the previous section. We refer to a position within this region as an \emph{effective} material point, $\boldsymbol{X}_s$, which denotes the location of an \emph{effective} material particle. 
	
	Ideally, the motion and deformation of the effective continuum body $\mathcal{B}$ should represent the aggregate motion of the true continuum body $\mathcal{B}^*$. This can be accomplished by introducing a second motion function: the \emph{effective} motion function $\boldsymbol{\chi}_s(\boldsymbol{X}_s, t)$, which maps an effective material point $\boldsymbol{X}_s$ in $\mathcal{B}$ to a unique spatial point $\boldsymbol{x}$ in the \emph{effective deformed body} $\mathcal{B}_t$. Since we are interested in the aggregate motion of the material in the \emph{neighborhood} of $\boldsymbol{X}_s$, we identify the time rate of change of this motion function --- i.e., the rate of change of the position $\boldsymbol{x}$ mapped to by a particular effective material point $\boldsymbol{X}_s$ --- as the homogenized solid velocity field, $\boldsymbol{v}_s$:
	\begin{equation}
		\boldsymbol{v}_s = \frac{\partial \boldsymbol{\chi}_s(\boldsymbol{X}_s, t)}{\partial t}.
	\end{equation}
	
	Importantly, the \emph{effective} material point $\boldsymbol{X}_s$ \emph{does not} identify a true material point (denoted by $\boldsymbol{X}^*$). Rather, $\boldsymbol{X}_s$ denotes a reference position in $\mathcal{B}$ that we use to follow the aggregate motion and deformation of the material in the \emph{neighborhood} of $\boldsymbol{X}_s$. This distinction is important. Since the velocity of the true granular material $\boldsymbol{v}^*$ and the homogenized velocity field $\boldsymbol{v}_s$ are \emph{not} identical, there can be significant differences between the local deformations of individual solid grains (in $\mathcal{B}^*_t$) and the effective deformation of \emph{material neighborhoods} (in $\mathcal{B}_t$). Additionally, it is possible that \emph{all} of the \emph{true} material that is initially in the neighborhood of the effective material point $\boldsymbol{X}_s$ may be exchanged for an equivalent mass of material from outside of this neighborhood, through regular granular rearrangement \citep[e.g., see particle size-segregation phenomena in][]{liu2023}.
	
	\emph{How, then, can we predict the evolution of the effective granular stress, $\boldsymbol{\sigma}_s$, without direct knowledge about the microscopic deformations and motions of the individual grains that compose the granular solid?} In solid mechanics, it is common to consider the fundamental connection between material strains --- a measure of the local material deformation --- and material stresses. This connection arises due to the physical link between the motion of atoms and molecules in the material and the position-dependent nature of inter-atomic and inter-molecular forces. Since we have chosen to represent the true granular solid with an effective continuum material, we cannot use specific information about the true velocity field $\boldsymbol{v}^*(\boldsymbol{x},t)$ or the motion function $\boldsymbol{\chi}^*(\boldsymbol{X}^*,t)$ to determine these true, microscopic deformations. Here, we are limited to describing material deformations in aggregate using the homogenized velocity field $\boldsymbol{v}_s(\boldsymbol{x},t)$, the effective motion function $\boldsymbol{\chi}_s(\boldsymbol{X}_s,t)$, and their respective gradients.
	
	To characterize this aggregate deformation, we consider the motion of the effective continuum material in the \emph{neighborhood} of an effective material point $\boldsymbol{X}_s$. To do this, we introduce an infinitesimal vector $\boldsymbol{dX}_s$ in the effective reference body $\mathcal{B}$, shown in Figure \ref{fig:appendix_kinematics}b, which points from the effective material point $\boldsymbol{X}_s$ to an arbitrary material point within its neighborhood: $\boldsymbol{X}_s + \boldsymbol{dX}_s$. In principle, this second material point can be mapped to a spatial position $\boldsymbol{x}$ using the motion function $\boldsymbol{\chi}_s(\boldsymbol{X}_s + \boldsymbol{dX}_s, t)$. The deformation of this material neighborhood can therefore be characterized by relating a \emph{relative position vector} $\boldsymbol{dx}_s$ (in $\mathcal{B}_t$) to the infinitesimal vector $\boldsymbol{dX}_s$ (in $\mathcal{B}$), where:
	\begin{equation}
		\boldsymbol{dx}_s = \boldsymbol{\chi}_s(\boldsymbol{X}_s + \boldsymbol{dX}_s, t) + \boldsymbol{\chi}_s(\boldsymbol{X}_s, t).
	\end{equation}
	As in standard solid mechanics, we relate these vector quantities using the \emph{effective deformation gradient} $\boldsymbol{F}$:
	\begin{equation}
		\boldsymbol{F} = \frac{\partial \boldsymbol{\chi}_s(\boldsymbol{X}_s,t)}{\partial \boldsymbol{X}_s}, \quad \text{and} \quad \boldsymbol{dx}_s = \boldsymbol{F}\ \boldsymbol{dX}_s.
	\end{equation}
	Following the derivation in \citet{gurtin2010}, the time rate of change of the deformation gradient can be expressed as follows:
	\begin{equation}
		\frac{d^s \boldsymbol{F}}{dt} = \boldsymbol{L}_s \boldsymbol{F}, \quad \text{with} \quad \boldsymbol{L}_s \equiv \nabla \boldsymbol{v}_s.
	\end{equation}
	Conceptually, the deformation gradient tensor $\boldsymbol{F}$ maps the vector between neighboring effective material points ($\boldsymbol{dX}_s$) in the reference body $\mathcal{B}$ to a relative position vectos ($\boldsymbol{dx}_s$) in the deformed body $\mathcal{B}_t$. If $\boldsymbol{F}$ is the identity tensor $\boldsymbol{1}$, then the effective material in this neighborhood would be \emph{effectively unstretched} and \emph{undeformed}.
	
	Recall, however, that $\boldsymbol{X}_s$ and its associated mapping function $\boldsymbol{\chi}_s(\boldsymbol{X}_s,t)$ \emph{do not} identify the motion of the \emph{true} material that composes the individual grains. Similarly, $\boldsymbol{F}$ describes how the material in the neighborhood of $\boldsymbol{X}_s$ deforms when subjected to the aggregate velocity field $\boldsymbol{v}_s(\boldsymbol{x},t)$. Since the \emph{true} velocity $\boldsymbol{v}^*$ of the granular solid and the \emph{effective} velocity $\boldsymbol{v}_s$ of the aggregate material are \emph{not} identical, there can be significant differences between the local deformations of individual solid grains and the effective deformations described by $\boldsymbol{F}$. In particular, since the grains in the material are not fixed to their neigbors, it is possible that the material around $\boldsymbol{X}_s$ might exhibit significant distortion and strain (i.e., $\boldsymbol{F} \neq \boldsymbol{1}$) \emph{without any corresponding deformation within the individual grains that compose this neighborhood}. In other words, \emph{except for particular arrangements and loading histories}, the tensor $\boldsymbol{F}$ is generally \emph{insufficient} for characterizing the deformation of the true granular solid --- which, in turn, gives rise to the effective granular stress, $\boldsymbol{\sigma}_s$.
	
	To overcome this limitation, we consider the Kroner--Lee decomposition of the deformation gradient into an \emph{inelastic} or \emph{plastic} component, $\boldsymbol{F}^p$, and an \emph{elastic} component $\boldsymbol{F}^e$.
	\begin{equation}
		\boldsymbol{F} = \boldsymbol{F}^e \boldsymbol{F}^p.
	\end{equation}
	%
	In plasticity theory, this decomposition of material deformation represents a conceptual mapping of the material in the neighborhood of an effective material point $\boldsymbol{X}_s$ first into an \emph{intermediate space} with the \emph{plastic distortion} tensor $\boldsymbol{F}^p$ (Figure \ref{fig:kinematics}c,d) and then into the final \emph{deformed space} with the \emph{elastic distortion} tensor $\boldsymbol{F}^e$ (Figure \ref{fig:kinematics}d,e). Here we use $\boldsymbol{F}^p$ to represent \emph{any} aggregate granular motion that does not result in \emph{elastic} deformations within the individual grains (e.g., granular rearrangement, grain fragments sliding along fracture planes, sub-grain plasticity, etc.). $\boldsymbol{F}^e$, on the other hand, describes \emph{any} aggregate granular motion associated with microscopic, elastic deformations of the grains. Since the effective deformation gradients $\boldsymbol{F}$ is already disconnected from the \emph{true} physical motion of the individual grains (through the homogenization procedure described in the previous section), we must assume that this type of decomposition of $\boldsymbol{F}$ is admissible. 
	
	Additionally, we assume that $\boldsymbol{F}^e$ and $\boldsymbol{F}^p$ follow a similar evolution rule to $\boldsymbol{F}$:
	\begin{equation}
		\label{eqn:appendix_strain_rates}
		\frac{d^s \boldsymbol{F}^e}{dt} = \boldsymbol{L}^e \boldsymbol{F}^e, \quad \text{and} \quad
		\frac{d^s \boldsymbol{F}^p}{dt} = \boldsymbol{L}^p \boldsymbol{F}^p, \quad \text{with} \quad
		\boldsymbol{L}_s = \boldsymbol{L}^e + \boldsymbol{F}^e \boldsymbol{L}^p \boldsymbol{F}^{e-1} = \boldsymbol{L}^e + \boldsymbol{\tilde{D}}^p.
	\end{equation}
	Here we decompose the effective velocity gradient $\boldsymbol{L}_s$ into an \emph{elastic distortion rate} $\boldsymbol{L}^e$ and a \emph{inelastic distortion rate} $\boldsymbol{L}^p$. 
	{However, since the decomposition of $\boldsymbol{F}$ is multiplicative, we may assume that $\boldsymbol{L}^p$ is symmetric and transform it into the same space as $\boldsymbol{L}_s$, which we refer to as the \emph{effective inelastic deformation rate} $\boldsymbol{\tilde{D}}^p$:
	\begin{equation}
		\boldsymbol{\tilde{D}}^p = \boldsymbol{F}^e \boldsymbol{L}^p \boldsymbol{F}^{e-1}.
	\end{equation}
	In this way, the decomposition of $\boldsymbol{L}_s$ in \eqref{eqn:appendix_strain_rates} becomes \emph{additive}.}
	
	Together, $\boldsymbol{F}$, $\boldsymbol{F}^e$, and $\boldsymbol{F}^p$ characterize the aggregate deformation of the effective continuum material shown in Figure \ref{fig:appendix_kinematics}b. $\boldsymbol{F}^e$ represents the elastic deformations within the \emph{true} granular material that give rise to the effective granular stress $\boldsymbol{\sigma}_s$, and $\boldsymbol{F}^p$ represents all other aggregate deformations (e.g., granular rearrangement, fragmentation, sub-grain plasticity, etc.). However, this picture of material deformation is incomplete without some model for $\boldsymbol{L}^e$ or $\boldsymbol{\tilde{D}}^p$, which govern the evolution of these two deformation measures. Although the material deformation rate $\boldsymbol{L}_s$ is determined purely by aggregate material motion, $\boldsymbol{L}^e$ and $\boldsymbol{\tilde{D}}^p$ must be predicted in some other way.

	\subsection{Breakage Mechanics Theory}
	\label{sec:appendix_breakage}
	In addition to the elastic-plastic deformation theory above, the constitutive model proposed in this work also uses the theory of breakage mechanics \citep[][]{einav2007a}. This modeling approach incorporates the micro-mechanics of particle fragmentation through the auxiliary variable $B$, called the \emph{relative breakage}. This internal variable provides additional information about the \emph{true} granular solid \emph{that is not otherwise captured} by the deformation gradient $\boldsymbol{F}$ or its components $\boldsymbol{F}^e$ and $\boldsymbol{F}^p$.

\begin{figure}[h!]
		\centering
		\includegraphics[width=0.65\linewidth]{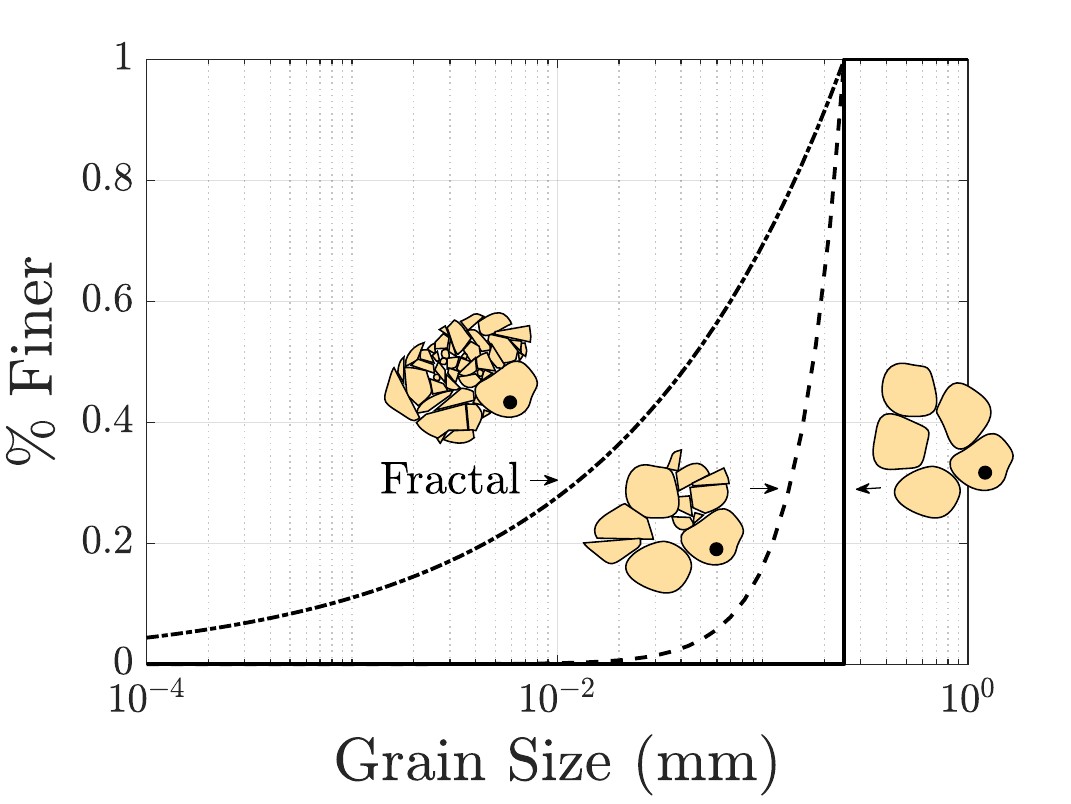}
		\caption{Illustration of cumulative size distribution function, $F(d)$, for an idealized, mono-disperse granular material (right); a partially-pulverized granular material (center); and a hypothetical, fully-pulverized granular material (right). The hypothetical distribution for a fully pulverized sediment is taken to be a fractal distribution: $F(d) = (d/d_{\text{max}})^{0.4}$, \citep[see][]{einav2007a}.}
		\label{fig:appendix_breakage}
	\end{figure}

	Originally introduced in \citet{hardin1985}, $B$ measures pulverization in granular sediments through changes in their aggregate \emph{particle size distribution} --- i.e., crushing generally causes larger grains to fracture into smaller particles. This particle size distribution is frequently reported using the \emph{cumulative size distribution function}, $F(d)$, which describes the mass fraction of material that is finer than a given sieve diameter $d$. An illustration of several cumulative size distributions are shown in Figure \ref{fig:appendix_breakage}. Since $F(d)$ represents the mass fraction that would pass through a particular sieve size, it is easily measured experimentally.
	
	In order to characterize a distribution of particle sizes using the single internal variable $B$, \citet{einav2007a} proposed the following reduction:
	\begin{equation}
		B \equiv \frac{\int_{d_\text{min}}^{d_\text{max}} \big(F(x) - F_0(x)\big) x^{-1} dx}{\int_{d_\text{min}}^{d_\text{max}} \big(F_u(x) - F_0(x)\big) x^{-1} dx}, \quad \text{with}\quad B\in[0,1].
	\end{equation}
	Here $d_\text{min}$ and $d_\text{max}$ denote the minimum and maximum grain diameters in a given sample, respectively; $F_0(d)$ denotes the \emph{initial} cumulative size distribution of a sample before any particles have fragmented; and $F_u(d)$ denotes the \emph{hypothetical} fractal distribution that the pulverized sediment approaches as more and more particles fragment. In this work, we take the \emph{initial} cumulative size distribution to be a \emph{hypothetical}, mono-disperse distribution at $d = d_\text{max}$:
	\begin{equation}
		F_0(d) = H(d - d_\text{max}),
	\end{equation}
	with $H(x)$ the Heaviside function. Additionally, following \citet{einav2007a}, we let the \emph{ultimate} cumulative size distribution be the \emph{hypothetical} fractal distribution beginning at $d = d_\text{max}$:
	\begin{equation}
		F_u(d) = \bigg(\frac{d}{d_\text{max}}\bigg)^{(3 - \alpha)}, \quad \text{with} \quad \alpha = 2.6.
	\end{equation}
	Using this approach, information about the distribution of grain sizes can be incorporated into the model through the Helmholtz free energy function, $\psi_s$, in \eqref{eqn:helmholtz_free_energy}; the compression yield function, $y_1$, in \eqref{eqn:y1}; the inelastic deformation rates in \eqref{eqn:y1_rates}; and the limiting porosities, $\phi_\text{max}$ and $\phi_\text{min}$ from \citet{rubin2011}. Further (and more detailed) discussion of this modeling theory can be found in \citet{einav2007a}, \citet{nguyen2009}, and \citet{rubin2011}. 

	\subsection{Coleman--Noll Procedure and Expressions for $\hat{A}$ and $\hat{C}$}
	\label{sec:appendix_coleman_noll}
	We now turn to thermodynamic restrictions that are specific to the model proposed in this work. In particular, we consider the model restrictions imposed by the selection of specific functional forms for the Helmholtz free energies, $\psi_s$ and $\psi_f$ in \eqref{eqn:helmholtz_free_energy}. This process is frequently referred to as the Coleman--Noll procedure \citep[see][]{coleman1963} and it determines the thermodynamic requirements of the model shown in \eqref{eqn:heat_flow}, \eqref{eqn:effective_granular_stress}, \eqref{eqn:yield_stress}, and \eqref{eqn:effective_fluid_stresses}. In this section, we provide a brief derivation of these model terms, including explicit forms of the functions $\hat{A}(\phi_s, J^e)$ and $\hat{C}(\phi_s, J^e)$ used in the model --- shown in \eqref{eqn:appendix_A_and_C}.
	
	We begin this derivation by recalling the definition of the \emph{effective} Helmholtz free energies:
	\begin{equation}
		\label{eqn:appendix_free_energy}
		\psi_s \equiv \varepsilon_s - T_s s_s, \quad \text{and} \quad \psi_f \equiv \varepsilon_f - T_f s_f,
	\end{equation}
	with $\varepsilon_s$, $\varepsilon_f$, $T_s$, $T_f$, $s_s$, and $s_f$ defined in \eqref{eqn:appendix_effective_fields}. If we take the governing equations in \eqref{eqn:energy_conservation} and \eqref{eqn:entropy_imbalance} as given, then the evolution rule for the free energies in \eqref{eqn:appendix_free_energy} can expressed in the following form:
	\begin{equation}
		\label{eqn:appendix_free_energy_imbalance}
		\begin{aligned}
			\bar{\rho}_s \frac{d^s \psi_s}{dt} \leq \boldsymbol{\sigma}_s : \nabla \boldsymbol{v}_s + \frac{\phi_s p_f}{\rho_s} \bigg(\frac{d^s \rho_s}{dt}\bigg) - \bar{\rho}_s s_s \bigg(\frac{d^s T_s}{dt}\bigg) + \frac{(T_s - T_f)q_i}{T_f} - \frac{\boldsymbol{q}_s \cdot \nabla T_s}{T_s}&,\\[1ex]
			\bar{\rho}_f \frac{d^f \psi_f}{dt} \leq \boldsymbol{\tau}_f : \nabla \boldsymbol{v}_f + \frac{\phi_f p_f}{\rho_f} \bigg(\frac{d^f \rho_f}{dt}\bigg) -  \bar{\rho}_f s_f \bigg(\frac{d^f T_f}{dt}\bigg) + \frac{(T_s - T_f)q_i}{T_s} - \frac{\boldsymbol{q}_f \cdot \nabla T_f}{T_f}&\\
			+\boldsymbol{f}_d \cdot (\boldsymbol{v}_s - \boldsymbol{v}_f).
		\end{aligned}
	\end{equation}
	In the first step of the Coleman--Noll procedure, we assert that since \eqref{eqn:energy_conservation} and \eqref{eqn:entropy_imbalance} must be true for mixtures with \emph{any arbitrary} temperature fields, $T_s$ and $T_f$, the inequalities in \eqref{eqn:appendix_free_energy_imbalance} can only be satisfied if the following inequalities are also satisfied \emph{independently}:
	\begin{equation*}
		\boldsymbol{q}_s \cdot \nabla T_s \leq 0, \quad \boldsymbol{q}_f \cdot \nabla T_f \leq 0, \quad \text{and} \quad q_i (T_s - T_f) \geq 0.
	\end{equation*}
	Additionally, a similar analysis of \eqref{eqn:appendix_free_energy_imbalance} with respect to \emph{any arbitrary} difference in material velocities, $\boldsymbol{v}_s - \boldsymbol{v}_f$, requires that the final part of \eqref{eqn:heat_flow} also be satisfied \emph{independently}:
	\begin{equation*}
		\boldsymbol{f}_d \cdot (\boldsymbol{v}_s - \boldsymbol{v}_f) \geq 0.
	\end{equation*}
	Substituting \eqref{eqn:heat_flow} into \eqref{eqn:appendix_free_energy_imbalance} yields the model equations presented in \eqref{eqn:free_energy_imbalance}:
	\begin{equation*}
		\begin{aligned}
			\bar{\rho}_s \frac{d^s \psi_s}{dt} &= \boldsymbol{\sigma}_s : \nabla \boldsymbol{v}_s + \frac{\phi_s p_f}{\rho_s} \bigg(\frac{d^s \rho_s}{dt}\bigg) - \bar{\rho}_s s_s \bigg(\frac{d^s T_s}{dt}\bigg) - D_s,\\
			\bar{\rho}_f \frac{d^f \psi_f}{dt} &= \boldsymbol{\tau}_f : \nabla \boldsymbol{v}_f + \frac{\phi_f p_f}{\rho_f} \bigg(\frac{d^f \rho_f}{dt}\bigg) -  \bar{\rho}_f s_f \bigg(\frac{d^f T_f}{dt}\bigg) - D_f.
		\end{aligned}
	\end{equation*}
	with $D_s \geq 0$ and $D_f \geq 0$ the positive rates of \emph{mechanical dissipation}.
	
	We continue this analysis by recalling the functional forms for $\psi_s$ and $\psi_f$ in \eqref{eqn:helmholtz_free_energy} --- namely, $\psi_s = \hat{\psi}_c(\epsilon_v^e, \epsilon_s^e, B) + \hat{\psi}_g(\rho_s, T_s)$ and $\psi_f = \hat{\psi}_f(\rho_f, T_s)$. If we substitute these expressions into \eqref{eqn:free_energy_imbalance}, apply the chain rule, and group terms on the right-hand side we find:
	\begin{equation}
		\label{eqn:appendix_mechanical_free_energy_imbalance}
		\begin{aligned}
			0 = &-\bar{\rho}_s \bigg( \frac{\partial \hat{\psi}_c}{\partial \epsilon_v^e} \frac{d^s \epsilon_v^e}{dt} + \frac{\partial \hat{\psi}_c}{\partial \epsilon_s^e} \frac{d^s \epsilon_s^e}{dt} + \frac{\partial \hat{\psi}_c}{\partial B} \frac{d^s B}{dt} + \frac{\partial \hat{\psi}_g}{\partial \rho_s} \frac{d^s \rho_s}{dt} + \frac{\partial \hat{\psi}_g}{\partial T_s} \frac{d^s T_s}{dt}\bigg) +\boldsymbol{\sigma}_s : \nabla \boldsymbol{v}_s \\
			&+ \frac{\phi_s p_f}{\rho_s} \bigg(\frac{d^s \rho_s}{dt}\bigg) - \bar{\rho}_s s_s \bigg(\frac{d^s T_s}{dt}\bigg) - D_s,\\[1em]
			0 = &-\bar{\rho}_f \bigg( \frac{\partial \psi_f}{\partial \rho_f} \frac{d^f \rho_f}{dt} + \frac{\partial \psi_f}{\partial T_f} \frac{d^f T_f}{dt}\bigg) +\boldsymbol{\tau}_f : \nabla \boldsymbol{v}_f + \frac{\phi_f p_f}{\rho_f} \bigg(\frac{d^f \rho_f}{dt}\bigg) -  \bar{\rho}_f s_f \bigg(\frac{d^f T_f}{dt}\bigg) - D_f.
		\end{aligned}
	\end{equation}
	Additionally, recalling the elastic strain definitions in \eqref{eqn:elastic_strain} --- namely, $\boldsymbol{E}^e = \tfrac{1}{2} (\boldsymbol{F}^{e\top} \boldsymbol{F}^e - \boldsymbol{1})$, $\epsilon_v^e = \text{tr}(\boldsymbol{E}^e)$, and $\epsilon_s^e = \sqrt{\tfrac{2}{3} \boldsymbol{E}_0 : \boldsymbol{E}_0}$ --- along with the definition of the left Cuachy--Green tensor ($\boldsymbol{B}^e = \boldsymbol{F}^e \boldsymbol{F}^{e\top}$) we can use \eqref{eqn:appendix_strain_rates} to re-express \eqref{eqn:appendix_mechanical_free_energy_imbalance} as follows:
	\begin{equation}
		\label{eqn:appendix_mechanical_free_energy_imbalance_v2}
		\begin{aligned}
			0 = &-\bar{\rho}_s \frac{\partial \hat{\psi}_c}{\partial \epsilon_v^e} (\boldsymbol{B}^e : \boldsymbol{L}^e)  
			-\frac{\bar{\rho}_s}{3 \epsilon_s^e} \frac{\partial \hat{\psi}_c}{\partial \epsilon_s^e} (\boldsymbol{B}^e_0 \boldsymbol{B}^e : \boldsymbol{L}^e)
			-\bar{\rho}_s \frac{\partial \hat{\psi}_c}{\partial B} \frac{d^s B}{dt}\\
			& -\bar{\rho}_s \frac{\partial \hat{\psi}_g}{\partial \rho_s} \frac{d^s \rho_s}{dt} - \bar{\rho}_s \frac{d^s T_s}{dt} \bigg(\frac{\partial \hat{\psi}_g}{\partial T_s} + s_s\bigg) +\boldsymbol{\sigma}_s : \nabla \boldsymbol{v}_s + \frac{\phi_s p_f}{\rho_s} \bigg(\frac{d^s \rho_s}{dt}\bigg) - D_s,\\[1em]
			0 = & - \frac{\phi_f}{\rho_f} \frac{d^f \rho_f}{dt} \bigg(\rho_f^2 \frac{\partial \psi_f}{\partial \rho_f} - p_f\bigg)  - \bar{\rho}_f \frac{d^f T_f}{dt}\bigg(\frac{\partial \psi_f}{\partial T_f} + s_f \bigg) +\boldsymbol{\tau}_f : \nabla \boldsymbol{v}_f  - D_f.
		\end{aligned}
	\end{equation}
	At this stage, we make a \emph{significant simplifying assumption}: when the \emph{true} solid phase density is changing (i.e., $d^s \rho_s / dt \neq 0$) the contribution of the pressure-volume work exerted by the fluid on the solid is \emph{negligible} (i.e., $\rho_s^2\ \partial \hat{\psi}_g/\partial \rho_s \gg p_f$). Using this assumption, we therefore isolate two \emph{independent} expressions for the imbalance of free energy:
	\begin{equation}
		\label{eqn:appendix_solid_free_energy_imbalance}
		\begin{aligned}
			0 = &-\bar{\rho}_s \frac{\partial \hat{\psi}_c}{\partial \epsilon_v^e} (\boldsymbol{B}^e : \boldsymbol{L}^e)  
			-\frac{\bar{\rho}_s}{3 \epsilon_s^e} \frac{\partial \hat{\psi}_c}{\partial \epsilon_s^e} (\boldsymbol{B}^e_0 \boldsymbol{B}^e : \boldsymbol{L}^e)
			-\bar{\rho}_s \frac{\partial \hat{\psi}_c}{\partial B} \frac{d^s B}{dt}
			-\bar{\rho}_s \frac{\partial \hat{\psi}_g}{\partial \rho_s} \frac{d^s \rho_s}{dt}\\
			& - \bar{\rho}_s \frac{d^s T_s}{dt} \bigg(\frac{\partial \hat{\psi}_g}{\partial T_s} + s_s\bigg) +\boldsymbol{\sigma}_s : \nabla \boldsymbol{v}_s - D_s,
		\end{aligned}
	\end{equation}
	and,
	\begin{equation}
		\label{eqn:appendix_fluid_free_energy_imbalance}
		0 =  - \frac{\phi_f}{\rho_f} \frac{d^f \rho_f}{dt} \bigg(\rho_f^2 \frac{\partial \psi_f}{\partial \rho_f} - p_f\bigg)  - \bar{\rho}_f \frac{d^f T_f}{dt}\bigg(\frac{\partial \psi_f}{\partial T_f} + s_f \bigg) +\boldsymbol{\tau}_f : \nabla \boldsymbol{v}_f  - D_f.
	\end{equation}
	
	Let us continue this analysis by first focusing on \eqref{eqn:appendix_solid_free_energy_imbalance}. In particular, we may substitute $d^s \rho_s/dt$ using \eqref{eqn:elastic_strain} and \eqref{eqn:appendix_strain_rates} by recalling the \emph{mechanical model} for the solid density in \eqref{eqn:solid_density}: $\rho_s = \rho_0 (1 + \hat{\alpha}(\phi_s)(J^{e-1} - 1))$. Through careful manipulation of this equation, the time-rate of change of the \emph{true} solid mass density can be expressed as,
	\begin{equation}
		\label{eqn:appendix_solid_density_rate}
		\frac{d^s \rho_s}{dt} = -\rho_s \bigg(\frac{\hat{\alpha}(\phi_s) \boldsymbol{L}^e : \boldsymbol{1} + \phi_s \frac{\partial \hat{\alpha}}{\partial \phi_s} (1 - J^e) \boldsymbol{L}_s:\boldsymbol{1}}{J^e + \hat{\alpha}(\phi_s)(1 - J^e) + \phi_s \frac{\partial \hat{\alpha}}{\phi_s}(1 - J^e)}\bigg), \quad \text{with} \quad \hat{\alpha}(\phi_s) = \phi_s^b.
	\end{equation}
	Further, if we introduce $\hat{A}(\phi_s, J^e)$ and $\hat{C}(\phi_s, J^e)$ as follows,
	\begin{equation}
		\label{eqn:appendix_A_and_C}
		\hat{A}(\phi_s, J^e) \equiv \frac{\phi_s^b  + b \phi_s^b (1 - J^e)}{J^e + (b+1)\phi_s^b(1 - J^e)},
		\quad \text{and} \quad
		\hat{C}(\phi_s, J^e) \equiv \frac{\phi_s^b}{J^e + (b+1)\phi_s^b(1 - J^e)},
	\end{equation}
	then \eqref{eqn:appendix_solid_free_energy_imbalance} can be re-expressed as,
	\begin{equation}
		\label{eqn:appendix_solid_free_energy_imbalance_v2}
		\begin{aligned}
			0 = &\bigg(\boldsymbol{\sigma}_s - \bar{\rho}_s \frac{\partial \hat{\psi}_c}{\partial \epsilon_v^e} \boldsymbol{B}^e - \frac{\bar{\rho}_s}{3 \epsilon_s^e} \frac{\partial \hat{\psi}_c}{\partial \epsilon_s^e} \boldsymbol{B}^e_0 \boldsymbol{B}^e + \phi_s \rho_s^2 \frac{\partial \hat{\psi}_g}{\partial \rho_s} \hat{A}(\phi_s, J^e) \boldsymbol{1} \bigg) : \boldsymbol{L}^e\\
			& + \bigg( \boldsymbol{\sigma}_s + \phi_s \rho_s^2 \frac{\partial \hat{\psi}_g}{\partial \rho_s} \big( \hat{A}(\phi_s,J^e) - \hat{C}(\phi_s, J^e)\big) \boldsymbol{1}\bigg) : \boldsymbol{\tilde{D}}^p
			- \bar{\rho}_s \frac{\partial \hat{\psi}_c}{\partial B} \frac{d^s B}{dt}
			- D_s\\
			& - \bar{\rho}_s \frac{d^s T_s}{dt} \bigg(\frac{\partial \hat{\psi}_g}{\partial T_s} + s_s\bigg).
		\end{aligned}
	\end{equation}
	Here, a significant number of algebraic steps have been omitted; however, for brevity, we believe these omissions are necessary.
	
	Now, we may conclude the Coleman--Noll procedure by asserting that \eqref{eqn:appendix_fluid_free_energy_imbalance} and \eqref{eqn:appendix_solid_free_energy_imbalance_v2} must remain true for \emph{any arbitrary} densities, temperatures, and rates of deformation. This necessitates that \emph{each} of the following expressions must be true \emph{independently}:
	\begin{equation}
		\label{eqn:appendix_thermodynamic_requirements}
		\begin{aligned}
			&\boldsymbol{\sigma}_s = \frac{\bar{q}}{3 \epsilon_s^e}  \boldsymbol{B}^e_0 \boldsymbol{B}^e -\bar{p} \boldsymbol{B}^e - \phi_s p^* \hat{A}(\phi_s, J^e) \boldsymbol{1},\\
			&\quad \text{with} \quad \bar{p} = -\bar{\rho}_s \frac{\partial \hat{\psi}_c}{\partial \epsilon_v^e},
			\quad \bar{q} = \bar{\rho}_s \frac{\partial \hat{\psi}_c}{\partial \epsilon_s^e},
			\quad p^* = \rho_s^2 \frac{\partial \hat{\psi}_g}{\partial \rho_s};\\[1ex]
			& D_s = \boldsymbol{\sigma}_y:\boldsymbol{\tilde{D}}^p + E_B \frac{d^s B}{dt} \geq 0,\\
			& \quad \text{with} \quad E_B = - \rho_s \frac{\partial \hat{\psi}_c}{\partial B},
			\quad \boldsymbol{\sigma}_y = \frac{\bar{q}}{3 \epsilon_s^e}  \boldsymbol{B}^e_0 \boldsymbol{B}^e -\bar{p} \boldsymbol{B}^e - \phi_s p^* \hat{C}(\phi_s, J^e) \boldsymbol{1}; \\[1ex]
			& p_f = \rho_f^2 \frac{\partial \psi_f}{\partial \rho_f};\\[1ex]
			& D_f = \boldsymbol{\tau}_f : \nabla \boldsymbol{v}_f \geq 0; \\[1ex]
			& s_s = -\frac{\partial \hat{\psi}_g}{\partial T_s};
			\quad \text{and} \quad s_f = -\frac{\partial \psi_f}{\partial T_f}.
		\end{aligned}
	\end{equation}
	In \eqref{eqn:appendix_thermodynamic_requirements}, we see that the definition of the effective stresses, $\boldsymbol{\sigma}_s$, $\boldsymbol{\tau}_f$, and $p_f$ \emph{are not} arbitrary, they are a direct consequence of the specific choice of Helmholtz free energy functions in \eqref{eqn:helmholtz_free_energy} and enforcement of the first and second laws of thermodynamics.

	\subsection{Proof of Model Dissipation}
	\label{sec:appendix_dissipation}
	The final component of the modeling theory that we discuss in this section concerns proving that the model obeys the thermodynamic requirements listed in \eqref{eqn:appendix_thermodynamic_requirements}. In particular, we are interested in showing that,
	\begin{equation*}
		D_s = \boldsymbol{\sigma}_y:\boldsymbol{\tilde{D}}^p + E_B \frac{d^s B}{dt} \geq 0.
	\end{equation*} 
	The other components of \eqref{eqn:appendix_thermodynamic_requirements} are designed to be true by construction. However, because the model is \emph{non-associative} --- i.e., the inelastic yield functions in \eqref{eqn:y1}---\eqref{eqn:y3} \emph{do not} uniquely determine the inelastic deformation rate, $\boldsymbol{\tilde{D}}^p$ --- it is important to validate that the model is thermodynamically sound by substituting \eqref{eqn:inelastic_deformation_rate} into \eqref{eqn:appendix_thermodynamic_requirements} as follows:
	\begin{equation}
		\label{eqn:appendix_dissipation_1}
		D_s = \boldsymbol{\sigma}_y:\bigg(\frac{3 \xi_s^p}{2 q_y} \boldsymbol{\sigma}_{y0} + \tfrac{1}{3}\big(\xi_v^p + \xi_2^p + \xi_3^p\big) \boldsymbol{1}\bigg) + E_B \frac{d^s B}{dt}.
	\end{equation}
	Here $q_y = \sqrt{\tfrac{3}{2} \boldsymbol{\sigma}_{y0}:\boldsymbol{\sigma}_{y0}}$, and $\xi_s^p$, $\xi_v^p$, $\xi_2^p$, and $\xi_3^p$ are defined in \eqref{eqn:y1_rates}--\eqref{eqn:y3_rates}. We can re-express \eqref{eqn:appendix_dissipation_1} by applying the associative property of the tensor contraction operator (:), as follows:
	\begin{equation}
		\label{eqn:appendix_dissipation_2}
		D_s = \frac{3 \xi_s^p}{2 q_y} \boldsymbol{\sigma}_{y0} : \boldsymbol{\sigma}_{y0} + \tfrac{1}{3}\big(\xi_v^p + \xi_2^p + \xi_3^p\big) \text{tr}(\boldsymbol{\sigma}_y) + E_B \frac{d^s B}{dt}.
	\end{equation}
	Recalling that $p_y = -\tfrac{1}{3}\text{tr}(\boldsymbol{\sigma}_y)$, we arrive at the simple expression for mechanical dissipation from \eqref{eqn:solid_dissipation}:
	\begin{equation}
		\label{eqn:appendix_dissipation_3}
		D_s = q_y \xi_s^p - p_y (\xi_v^p + \xi_2^p + \xi_3^p) + E_B \frac{d^s B}{dt}.
	\end{equation}
	
	To prove that this expression for $D_s$ satisfies the condition that $D_s \geq 0$, we consider four independent conditions:
	\begin{enumerate}[itemsep=0em,label=(\roman*)]
		\item no yielding (i.e., $y_1 < 0$, $y_2 < 0$, and $y_3 < 0$);
		\item dense compression yielding (i.e., $y_1 = 0$, $y_2 < 0$, $y_3 < 0$);
		\item dense tensile yielding (i.e., $y_1 < 0$, $y_2 = 0$, $y_3 < 0$);
		\item and dilute compression yielding (i.e., $y_1 < 0$, $y_2 < 0$, $y_3 = 0$). 
	\end{enumerate}
	The first case above is trivial: in the absence of yielding \eqref{eqn:y1}--\eqref{eqn:y3} restrict $\xi_s^p$, $\xi_v^p$, $\xi_2^p$, and $\xi_3^p$ to zero, along with $d^s B / dt = 0$. Together, this ensures that $D_s = 0$. For the other three conditions, however, we must substitute in the yield equations in \eqref{eqn:y1}--\eqref{eqn:y3} along with the flow rules in \eqref{eqn:y1_rates}--\eqref{eqn:y3_rates} into \eqref{eqn:appendix_dissipation_3} directly.
	
	We begin with case (ii): $y_1 = 0$. If $y_1 = 0$, then it can be shown that $y_2 < 0$ and $y_3 < 0$, and accordingly $\xi_2^p = \xi_3^p = 0$. Following \eqref{eqn:y1}, $y_1 = 0$ also implies that,
	\begin{equation*}
		\frac{E_B (1 - B)^2}{E_c} + \frac{q_y^2}{(M p_y)^2} = 1.
	\end{equation*}
	Returning to \eqref{eqn:appendix_dissipation_3}, we can substitute the flow equations in \eqref{eqn:y1_rates}, along with $\xi_s^p = \xi_3^p = 0$, to find:
	\begin{equation*}
		\begin{aligned}
			D_s = \lambda_1 \bigg(&\frac{E_B (1 - B)^2}{E_c} \frac{q_y^2}{(p_y^2 + q_y^2)} \sin^2(\omega) + \frac{q_y^2}{(M p_y^2)} \\
			&+ \frac{E_B(1 - B)^2}{E_c} \frac{p_y^2}{(p_y^2 + q_y^2)} \sin^2(\omega) - M_d \frac{q_y}{M^2 p_y}\\
			& + \frac{E_B (1 - B)^2}{E_c} \cos^2(\omega)\bigg).
		\end{aligned}
	\end{equation*}
	Together with \eqref{eqn:y1}, this expression simplifies to:
	\begin{equation}
		\label{eqn:appendix_dissipation_4}
		D_s = \lambda_1 \bigg(1 - \frac{M_d q_y}{M^2 p_y}\bigg) \quad \text{if} \quad y_1 = 0.
	\end{equation}
	Since, in general, the model requires that $q_y < M p_y$ and $M_d < M$ --- see \eqref{eqn:friction_and_dilation} --- it is simple to show that \eqref{eqn:appendix_dissipation_4} satisfies the requirement that $D_s \geq 0$.
	
	Next, we consider case (iii) above: $y_2 = 0$. If $y_2 = 0$, then it can be shown that $y_1 < 0$ and $y_3 < 0$, and accordingly $\xi_s^p = \xi_v^p = \xi_3^p = 0$ and $d^s B / dt = 0$. Following \eqref{eqn:y2}, $y_2 = 0$ also implies that $p_y = 0$. Substitution of these equalities into \eqref{eqn:appendix_dissipation_3} yields the trivial result: $D_s = 0$ when $y_2 = 0$.
	
	Finally, we consider case (iv) above: $y_3 = 0$. If $y_3 = 0$, then it can be shown that $y_1 < 0$ and $y_2 < 0$, and accordingly $\xi_s^p = \xi_v^p = \xi_2^p = 0$ and $d^s B / dt = 0$. Following \eqref{eqn:y3}, $y_3 = 0$ also implies that $\lambda_3 \leq 0$. If we substitute the flow equation in \eqref{eqn:y3_rates} into \eqref{eqn:appendix_dissipation_3}, we find:
	\begin{equation}
		\label{eqn:appendix_dissipation_5}
		D_s = \lambda_3 p_y, \quad \text{if} \quad y_3 = 0.
	\end{equation}
	Since the model requires that $\lambda_3 \geq 0$ and $p_y \geq 0$ --- by \eqref{eqn:y2} --- it is simple to show that \eqref{eqn:appendix_dissipation_5} satisfies the requirement that $D_s \geq 0$.

	\setcounter{figure}{0}
	
	\section{Explicit Expressions for Model EOS Functions}
	\label{sec:appendix_explicit_eos}
	In this section, we provide additional details about the model equations used for the solid pressure term, $p^*$, in \eqref{eqn:effective_granular_stress} and \eqref{eqn:yield_stress} and the fluid pore pressure, $p_f$. The purpose of this section is to provide brief descriptions of these model equations alongside the abbreviated explicit equations of state. The equations selected for these two components of the model derive from existing models in the literature. For this reason, we direct the reader to other sources for further reading \citep[e.g.,][]{mie1903,gruneisen1912,tillotson1962,brundage2013,tonge2016a}.
	
	\subsection{Mie--Gr{\"u}neisen EOS}
	\label{sec:appendix_mie_gruneisen}
	In the model proposed in this work, the solid pressure term, $p^*$, is used to represent the response of a granular sediment in compression \emph{when approaching zero porosity} --- i.e., $\phi_f = 0$ and $\phi_s = 1$. From \eqref{eqn:effective_granular_stress}, we have defined $p^*$ as follows:
	\begin{equation*}
		p^* = \rho_s^2 \frac{\partial \hat{\psi}_g}{\partial \rho_s}.
	\end{equation*}
	At this stage, we may define $p^*$ using \emph{any thermodynamically consistent} model --- i.e., a model with either an explicit or implicit equation for $\hat{\psi}_g(\rho_s, T_s)$ in \eqref{eqn:helmholtz_free_energy}. In this work, we have chosen the Mie--Gr{\"u}neisen EOS to define this relationship.
	
	The Mie--Gr{\"u}neisen EOS is a common model for solids in the shock-compression literature \citep[e.g., see][]{heuze2012, tonge2016a}, which utilizes a simple pressure-energy relationship:
	\begin{equation}
		p^* - p_0 = \rho_s \Gamma (\varepsilon^* - \varepsilon_0). 
	\end{equation} 
	Here, $\varepsilon^*$ denotes the \emph{true} specific internal energy of the material and generally has two contributions: a thermal component ($c_v T^*$) and a cold component ($\hat{e}_c(\rho_s)$). For the model proposed in this work, we adopt a similar definition for the granular internal energy associated with $p^*$:
	\begin{equation}
		\varepsilon_g = \hat{e}_c(\rho_s) + c_v T_s.
	\end{equation}
	Following the work of \citet{drumheller1998}, we let the specific entropy associated with $p^*$ take the following form:
	\begin{equation}
		s_g = c_v \ln (T_s/T_0) - c_v \Gamma_0 (1 - \rho_0/\rho_s) + s_0,
	\end{equation}
	where $T_0$ and $\rho_0$ are reference temperatures and densities at zero pressure and $s_0$ is the non-zero reference entropy. Together, these define the specific Helmholtz free energy function, $\hat{\psi}_g(\rho_s, T_s)$:
	\begin{equation}
		\label{eqn:appendix_solid_free_energy_explicit}
		\hat{\psi}_g(\rho_s, T_s) = \hat{e}_c(\rho_s) + c_v T_s - T_s \big(c_v \ln (T_s/T_0) - c_v \Gamma_0 (1 - \rho_0/\rho_s) + s_0\big).
	\end{equation}
	
	The final component of the Mie--Gr{\"u}neisen EOS is the specific functional form of $\hat{e}_c(\rho_s)$, which we adopt from \citet{tonge2016a}:
	\begin{equation}
		\hat{e}_c(\rho_s) = e^{\Gamma_0(1 - J)} \big(c_v \hat{T}_H(J) - c_v T_0 \big) + c_v T_0, \quad \text{with} \quad J = \rho_0/\rho_s.
	\end{equation}
	Here, $\hat{T}_H(J)$ is the temperature Hugoniot, as a function of the specific volume ratio, $\rho_0/\rho_s$. If the corresponding pressure Hugoniot function, $\hat{p}_H(J)$, is known, then $\hat{T}_H(J)$ can be expressed as,
	\begin{equation}
		c_v \hat{T}_H(J) = - \int_1^J \frac{\hat{p}_H(J')}{\rho_s} \bigg[1 - \frac{\Gamma_0}{2}(1 - J')\bigg] e^{-\Gamma_0(1 - J')} dJ'.
	\end{equation}
	For the first-order Mie--Gr{\"u}neisen EOS, the pressure Hugoniot, $\hat{p}_H(J)$, is given by the following piece-wise function:
	\begin{equation}
		\hat{p}_H(J) = 
		\begin{cases}
			\displaystyle
			\frac{\rho_0 C_0^2 (1 - J)}{(1 - S_0 (1 - J))^2} & \quad \text{if} \quad J < 1,\\[1em]
			\rho_0 C_0^2 (1 - J) & \quad \text{if} \quad J \geq 1,
		\end{cases}
		\quad \text{with} \quad J = \rho_0/\rho_s.
	\end{equation}
	Substituting these explicit functions into \eqref{eqn:appendix_solid_free_energy_explicit}, we arrive at the expression for $p^*$ from \eqref{eqn:solid_pressure}
	\begin{equation*}
		p^* = \rho_s^2 \frac{\partial \hat{\psi}_g}{\partial \rho_s} = \hat{p}_H\bigg(\frac{\rho_0}{\rho_s}\bigg) \bigg[1 - \frac{\Gamma_0}{2} \bigg(1 - \frac{\rho_0}{\rho_s}\bigg)\bigg] + \rho_0 \Gamma_0 \big(\hat{e}_c(\rho_s) + c_v (T_s - T_0)\big)
	\end{equation*}

	\subsection{Tillotson EOS}
	\label{sec:appendix_tillotson}
	In the model proposed in this work, the fluid pore pressure, $p_f$, describes the response of the pore fluid to compression and tension. Not only does $p_f$ partially determine the motion of the pore fluid, it also drives the motion of the solid material, through normal tractions along the boundary $\partial \Omega^*$ in Figure \ref{fig:appendix_rve}. From \eqref{eqn:effective_fluid_stresses}, we have defined $p_f$ as follows:
	\begin{equation*}
		p_f = \rho_f^2 \frac{\partial \psi_f}{\partial \rho_f}.
	\end{equation*}
	At this stage, we may define $p_f$ using \emph{any thermodynamically consistent} model, and here we choose the Tillotson EOS.
	
	The Tillotson EOS \citep{tillotson1962} is another common model for solids and liquids in the shock-compression literature \citep[e.g., see][]{brundage2013}, which is slightly more sophisticated than the Mie--Gr{\"u}neisen EOS. In addition to the compressive and thermal stresses, the Tillotson EOS also incorporates vaporization, which may occur in the pore fluid at sufficiently high impact velocities. As with the Mie--Gr{\"u}neisen EOS, the specific internal energy of the fluid $\varepsilon_f$ generally has two contributions: a thermal component ($c_{vf} (T_f - T_0)$) and a cold component ($\hat{e}_{cf}(\rho_f)$). For the model proposed in this work, we follow the implementation of \citet{brundage2013}:
	\begin{equation}
		\varepsilon_f = \hat{e}_{cf}(\rho_f) + c_{vf} (T_f - T_0).
	\end{equation}
	However, unlike the Mie--Gr{\"u}neisen EOS, the Tillotson EOS \emph{does not} have a simple closed form expression for the entropy of the pore fluid. Rather, the Tillotson EOS defines the pressure $p_f$ as a function of $\rho_f$ and $\varepsilon_f$, with the entropy implicitly determined by these equations.
	
	Additional details about the Tillotson EOS may be found in \citet{tillotson1962} and \citet{brundage2013}. Here we provide the full form of the model equations used in this work:
	\begin{equation}
		\hat{p}_{f1}(\rho_f, \varepsilon_f) = \bigg[a_f + \frac{b_f}{\varepsilon_f/(E_0 \eta^2) + 1}\bigg] \rho_f \varepsilon_f + A_f \mu + B_f \mu^2,
	\end{equation}
	\begin{equation}
		\hat{p}_{f2}(\rho_f, \varepsilon_f) = a_f \rho_f \varepsilon_f + \bigg[\frac{b_f \rho_f \varepsilon_f}{\varepsilon_f/(E_0 \eta^2) + 1} + A_f \mu e^{-\beta_f (\rho_0/\rho_f - 1)}\bigg]e^{-\alpha_f(\rho_0/\rho_f - 1)^2},
	\end{equation}
	and,
	\begin{equation}
		\hat{p}_{f3}(\rho_f, \varepsilon_f) = \bigg[a_f + \frac{b_f}{\varepsilon_f/(E_0 \eta^2) + 1}\bigg] \rho_f \varepsilon_f + A_f \mu,
	\end{equation}
	with $\eta = \rho_f / \rho_0$ and $\mu = \eta - 1$.
	From $\hat{p}_{f1}$, $\hat{p}_{f2}$, and $\hat{p}_{f3}$, we construct the fluid pore pressure function $\hat{p}_f(\rho_f, \varepsilon_f)$ as follows:
	\begin{equation}
		\hat{p}_f(\rho_f, \varepsilon_f) = 
		\begin{cases}
			\hat{p}_{f1}(\rho_f, \varepsilon_f), & \text{if} \quad \rho_f \geq \rho_{f0},\\[1em]
			\hat{p}_{f1}(\rho_f, \varepsilon_f), & \text{if} \quad \rho_{f0} > \rho_f \geq \rho_\text{IV},\ \varepsilon_f \leq E_\text{IV},\\[1em]
			\displaystyle
			\frac{(\varepsilon_f - E_\text{IV})\hat{p}_{f1} + (E_\text{CV} - \varepsilon_f)\hat{p}_{f2}}{E_\text{CV} - E_\text{IV}}, & \text{if} \quad \rho_{f0} > \rho_f \geq \rho_\text{IV},\ E_\text{CV} > \varepsilon_f \geq E_\text{IV},\\[1em]
			\hat{p}_{f2}(\rho_f, \varepsilon_f), & \text{if} \quad \rho_0 > \rho_f,\ \varepsilon_f \geq E_\text{CV},\\[1em]
			\hat{p}_{f3}(\rho_f, \varepsilon_f), & \text{if} \quad \rho_\text{IV} > \rho_f,\ E_\text{CV} > \varepsilon_f.
		\end{cases}
	\end{equation}

\end{document}